\documentclass[twocolumn]{aastex63}

\usepackage{amsmath}

\newcommand{\kms}{\,km\,s$^{-1}$}
\newcommand{\HI}{\mbox{H\,{\sc i}}}
\newcommand{\OI}{\mbox{O\,{\sc i}}}
\newcommand{\sqcm}{\,cm$^{-2}$}
\newcommand{\msun}{\,M$_\odot$}
\newcommand{\msy}{\,M$_\odot$\,yr$^{-1}$}
\newcommand{\tm}{\tablenotemark} 
\newcommand{\tn}{\tablenotetext}

\newcommand{\I}{\,{\sc i}}
\newcommand{\II}{\,{\sc ii}}
\newcommand{\III}{\,{\sc iii}}

\newcommand{\V}{\,{\sc v}}

\newcommand{\fuse}{\emph{FUSE}}
\newcommand{\hst}{\emph{HST}}

\newcommand{\allbeta}{$-0.94_{-0.06}^{+0.06}$}
\newcommand{\ismbeta}{$-0.68_{-0.20}^{+0.17}$}
\newcommand{\ivcbeta}{$-1.01_{-0.14}^{+0.15}$}
\newcommand{\hvcbeta}{$-1.05_{-0.06}^{+0.07}$}
\newcommand{\lowbeta}{$-0.62_{-0.10}^{+0.11}$}
\newcommand{\highbeta}{$-0.72_{-0.07}^{+0.10}$}
\newcommand{\midbeta}{$-1.78_{-0.42}^{+0.29}$}
\newcommand{\lowbetahalo}{$-0.64_{-0.11}^{+0.11}$}
\newcommand{\highbetahalo}{$-0.99_{-0.11}^{+0.11}$}
\newcommand{\midbetahalo}{$-1.71_{-0.43}^{+0.33}$}
\newcommand{\lowbreak}{$16.6$}
\newcommand{\highbreak}{$17.8$}
\newcommand{\lowbreakhalo}{$16.6$}
\newcommand{\highbreakhalo}{$17.7$}

\graphicspath{{./}{figures/}}

\shorttitle{The \HI\ Column Density Distribution of the Galactic Disk and Halo}
\shortauthors{French et al.}

\begin{document}

\title{The \HI\ Column Density Distribution of the Galactic Disk and Halo}

\correspondingauthor{David M. French}
\email{dfrench@stsci.edu}

\author[0000-0003-3681-0016]{David M. French}
\affil{Space Telescope Science Institute, 3700 San Martin Drive, Baltimore, MD 21218, USA}

\author[0000-0003-0724-4115]{Andrew J. Fox}
\affil{AURA for ESA, Space Telescope Science Institute, 3700 San Martin Drive, Baltimore, MD 21218, USA}

\author[0000-0002-0507-7096]{Bart P. Wakker}
\affil{Department of Astronomy, University of Wisconsin - Madison, Madison, WI 53703, USA}

\author[0000-0002-5222-5717]{Colin Norman}
\affil{Space Telescope Science Institute, 3700 San Martin Drive, Baltimore, MD 21218, USA}
\affil{Center for Astrophysical Sciences, Department of Physics \& Astronomy, The Johns Hopkins University, Baltimore, MD 21218, USA}

\author[0000-0001-9158-0829]{Nicolas Lehner}
\affil{Department of Physics, University of Notre Dame, 225 Nieuwland Science Hall, Notre Dame, IN 46556, USA}

\author[0000-0002-2591-3792]{J. Christopher Howk}
\affil{Department of Physics, University of Notre Dame, 225 Nieuwland Science Hall, Notre Dame, IN 46556, USA}

\author[0000-0001-8016-6980]{Blair D. Savage}
\affil{Department of Astronomy, University of Wisconsin - Madison, Madison, WI 53703, USA}

\author[0000-0002-1188-1435]{Philipp Richter}
\affil{Institut für Physik und Astronomie, Universität Potsdam, Haus 28, Karl-Liebknecht-Str. 24/25, D-14476 Golm (Potsdam), Germany}

\author[0000-0002-7893-1054]{John O'Meara}
\affil{W.M. Keck Observatory, 65-1120 Mamalahoa Highway, Kamuela, HI 96743}

\author[0000-0002-2724-8298]{Sanchayeeta Borthakur}
\affil{School of Earth and Space Exploration, Arizona State University, 781 Terrace Mall, Tempe, AZ 85287, USA}

\author[0000-0001-6670-6370]{Timothy Heckman}
\affil{Center for Astrophysical Sciences, Department of Physics \& Astronomy, The Johns Hopkins University, Baltimore, MD 21218, USA}

\accepted{August 5, 2021}

\begin{abstract}

We present a census of neutral gas in the Milky Way disk and halo down to limiting column densities of $N$(HI)$\sim10^{14}$\,cm$^{-2}$ using measurements of \HI\ Lyman-series absorption from the \emph{Far Ultraviolet Spectroscopic Explorer.} Our results are drawn from an analysis of 25 AGN sightlines spread evenly across the sky with Galactic latitude  $|b|\gtrsim20\degr$. By simultaneously fitting multi-component Voigt profiles to 11 Lyman-series absorption transitions covered by \emph{FUSE} (Ly$\beta$--Ly$\mu$) plus \emph{HST} measurements of Ly$\alpha$, we derive the kinematics and column densities of a sample of 152 \HI\ absorption components. While saturation prevents accurate measurements of many components with column densities 
17$\lesssim$log\,$N$(\HI)$\lesssim$19, we derive robust measurements at log\,$N$(\HI)$\lesssim$17
and log\,$N$(\HI)$\gtrsim$19.
We derive the first ultraviolet \HI\ column density distribution function (CDDF) of the Milky Way, both globally and for low-velocity (ISM), intermediate-velocity clouds (IVCs), and high-velocity clouds (HVCs). We find that IVCs and HVCs show statistically indistinguishable CDDF slopes, with $\beta_{\rm IVC}=$\ivcbeta\ and $\beta_{\rm HVC}=$\hvcbeta. Overall, the CDDF of the Galactic disk and halo appears shallower than that found by comparable extragalactic surveys, suggesting a relative abundance of high-column density gas in the Galactic halo.
We derive the sky covering fractions as a function of \HI\ column density, finding an  enhancement of IVC gas in the northern hemisphere compared to the south.  We also find evidence for an excess of inflowing \HI\ over outflowing \HI, with $-$0.88$\pm$0.40 \msy\ of HVC inflow versus $\approx$0.20$\pm$0.10 \msy\ of HVC outflow,
confirming an excess of inflowing HVCs seen in UV metal lines.

\end{abstract}


\section{Introduction} \label{sec:intro}
The extended gaseous halos of galaxies (the circumgalactic medium; CGM) play essential roles in 
the life cycles of galaxies. These  vast reservoirs of gas, 
extending out to hundreds of kpc around galaxy disks, contain at least as much baryonic 
matter as the galaxy disks themselves (e.g., \citealt{peeples2014, werk2014, keeney2017}) and are 
fed by a number of sources, including infalling intergalactic matter and outflowing feedback 
material from their host galaxies. As a consequence they are complex, turbulent, and multiphase. 
The cool phase\footnote{In this paper we use the IGM/CGM definition of temperature where cool,
warm, and hot denote log ($T_{\rm gas}$/K) = 4--5, 5--6, and $\>$6, respectively.}
in particular represents star-formation fuel that has yet to be 
accreted. Understanding how this gas is distributed in galaxy halos is a key step toward 
understanding how galaxies grow and evolve across cosmic time 
\citep[see reviews by][]{putman2012, tumlinson2017}.

CGM studies have largely been separated into distinct Galactic and extragalactic camps 
due to their unique viewpoints and associated observational challenges.
Sitting inside our own Galaxy, we can achieve full sky-coverage and thus see the entire 
halo at once, both using absorption and emission.
The cool gas in the Milky Way disk a halo has been extensively studied 
via \HI\ 21 cm emission and UV metal-line absorption for decades. The resulting gas distribution has been 
historically separated by LSR velocity, with the assumption that low-velocity 
($|v_{\rm LSR}|\la40$\kms) gas 
traces the Galactic ISM (disk), and higher $v_{\rm LSR}$ gas 
traces the Galactic halo via 
intermediate-velocity ($40 \leq |v_{\rm LSR}|<100$\kms) and high-velocity 
($|v_{\rm LSR}|\ge100$\kms) clouds 
\citep[IVCs/HVCs; see reviews by][]{wakker1997, putman2012, richter2017}. 

This velocity-based division between disk and halo is generally supported by
the available distance information. 
The \HI\ disk has a scale height of $\approx$200--300 pc \citep{dickey1990}, whereas
IVCs are found at distances of $d\approx 0.5 - 2.0$ kpc \citep{richter2017} from the Sun, and 
HVCs are at $|d| \approx 5-15$ kpc 
\citep{wakker2001, wakker2008, thom2008, lehner2011, smoker2011, richter2015}.
However, the HVCs with good distance constraints are 21\,cm-bright clouds, which 
only uncover the tip of the Galactic-halo iceberg because current 21\,cm surveys 
are insensitive to \HI\ column densities below $\sim$10$^{18}$\sqcm. 
We follow the velocity-based
distinction between disk and halo in this paper, given the lack of a better alternative
for a sample of absorbers with no direct distance information.
However, this approach is inherently imprecise, as it relies
on the assumption that halo gas exists exclusively at high velocity. 
Simulations show that a large fraction of halo gas may be hidden at low velocity, 
where it blends with the disk (e.g., \citealt{zheng2020}).

One key measure of the gas distribution in the CGM is the column density distribution 
function (CDDF), which was developed to characterize the distribution drawn from blind 
extragalactic surveys. The CDDF describes the relative number density of absorbers
within a column density interval, $dN$, normalized by the redshift path, $dX$, over which 
an absorption survey has been completed. \citet{tytler1987} was the first to show that 
the distribution of extragalactic Ly$\alpha$ absorbers at $z$=0.2--3.5 can be well fit 
by a single power-law over seven orders of magnitude of \HI\ column, from 
log\,$N$(\HI)=14--21, with $f(N)=C\,N^{\beta}$. The slope, $\beta$, of this power-law 
encodes fundamental information about the mass distribution of the gas clouds and the 
ionizing radiation field incident on the gas. More recent studies have found evidence 
for at least one break in the power-law distribution for the extragalactic \HI\ CDDF 
(e.g., \citealt{kim2013}, \citealt{lehner2007}, \citealt{prochaska2014}, \citealt{ribaudo2011},
\citealt{omeara2013}, and \citealt{rudie2013}).

Our own Galaxy provides the unique point of view necessary to construct a complete CDDF 
from a single galaxy. However, observing Galactic \HI\ presents its own set of difficulties. 
Absorption from the Galactic disk completely dominates Ly$\alpha$ within a $\sim$2000\kms\ 
window, masking the presence of any weaker components from the halo. 
\HI\ 21\,cm radio observations have the velocity resolution to separate individual 
components, but lack the sensitivity to probe gas below $\sim10^{18}$\sqcm, and such 
gas is possibly lost near $v$=0\kms. The best solution to this issue is to observe 
the higher-order Lyman series lines, which become progressively less saturated as 
one moves up the series, and allow one to probe a much wider range of \HI\ column densities.

For this reason, we are conducting a survey of the Galactic halo using archival 
\emph{Far Ultraviolet Spectroscopic Explorer} (\fuse) sightlines toward background AGN. 
\fuse\ produced the largest and most sensitive sample of Galactic halo sightlines 
covering the Lyman series in absorption to date.
The \fuse\ wavelength coverage allows for the simultaneous fitting of Galactic Lyman 
series absorption down to Ly$\mu\,917.1805$ (the 12th transition of the Lyman series), 
allowing us to reliably probe column densities down to log\,$N$(\HI)$\lesssim$14. 
Although these measurements are technically challenging, owing to saturation, blending 
with H$_2$ and intergalactic absorption, and low S/N, there is enough useful information 
in the Lyman series to reliably measure key \HI\ properties including column densities 
and kinematics. In particular, even though the lower order lines are saturated for 
typical HVC column densities, the higher-order lines are often unsaturated and therefore 
on the linear part of the curve-of-growth, allowing more accurate column density 
measurements. This technique has been demonstrated via Lyman-series absorption 
studies of a number of individual sightlines 
\citep{sembach2002, collins2004, fox2005, ganguly2005, zech2008, richter2009}
and as part of an \ion{O}{6} survey \citep{fox2006}, but never in a blind \HI\ survey.

This paper is organized as follows. In Section~\ref{sec:methods} we discuss the sample, 
data reduction, and Voigt-profile fitting procedures. In Section~\ref{section:results} 
we discuss the observational properties of the \HI\ sample, including their basic properties, 
CDDF, sky covering fraction, and breakdown into inflow and outflow. We present a Summary in
Section~\ref{sec:summary}. Finally, in Appendix~\ref{section:lya} we present an analysis 
of the effect of including Ly$\alpha$ in the fits, and in Appendix \ref{section:fits} 
we present our Voigt profile fits for each target.

\section{Methods} \label{sec:methods}

\subsection{The Lyman Series of Neutral Hydrogen}{\label{subsec:lyman}}

\HI\ is the only directly observable ionization stage of 
the dominant element in the Universe.
The most sensitive lines of \HI\ available at any wavelength 
are the Lyman series, representing the resonance transitions out 
of the ground state ($n$=1). Absorption in the Lyman series 
probes \HI\ in the UV to very sensitive levels, 
provided that FUV-bright background sources are identified.
Since \fuse\ covered the wavelength range 905--1195\,\AA, its bandpass 
encompasses the entire \HI\ Lyman series except Ly$\alpha$. This makes
\fuse\ a well-suited instrument to conduct a sensitive \HI\ survey.

\begin{deluxetable}{lll cc}[!t]
\tablewidth{0pt}
\tablecaption{Atomic Data for Lyman Series}
\tablehead{Line & $\lambda_0$ & $f$-value & \multicolumn{2}{c}{\underline{~~~~log\,$N$($\tau_0$=1)\tablenotemark{a}~~~~}}\\
                &           &           & $b$=15 & $b$=30} 
\startdata
   Ly$\alpha$ & 1215.6700 & 0.416400 & 13.30 & 13.60 \\
    Ly$\beta$ & 1025.7223 & 0.079120 & 14.09 & 14.39 \\
   Ly$\gamma$ &  972.5368 & 0.029000 & 14.55 & 14.85 \\
   Ly$\delta$ &  949.7431 & 0.013940 & 14.88 & 15.18 \\
 Ly$\epsilon$ &  937.8035 & 0.007799 & 15.14 & 15.44 \\
    Ly$\zeta$ &  930.7483 & 0.004814 & 15.35 & 15.65 \\
     Ly$\eta$ &  926.2257 & 0.003183 & 15.53 & 15.83 \\
   Ly$\theta$ &  923.1504 & 0.002216 & 15.69 & 15.99 \\
    Ly$\iota$ &  920.9631 & 0.001605 & 15.83 & 16.13 \\
   Ly$\kappa$ &  919.3514 & 0.001200 & 15.96 & 16.26 \\
  Ly$\lambda$ &  918.1294 & 0.000921 & 16.07 & 16.37 \\
      Ly$\mu$ &  917.1806 & 0.000723 & 16.18 & 16.48 \\
\enddata 
\tablecomments{Atomic data for the Lyman series transitions under study are taken from \citet{jitrik2004}. These values are close to but slightly different than those in \citet{morton2003}.}
\tablenotetext{a}{\HI\ column density at which the optical depth at line center is unity, for two representative values of the line width $b$ (in \kms). This indicates the regime where each line is unsaturated, and therefore amenable to precise measurement.}
\label{tab:atomic}
\end{deluxetable}

The basic atomic data -- rest wavelengths ($\lambda_0$) and oscillator strengths ($f$) 
-- for the Lyman series transitions are given in Table~\ref{tab:atomic}. 
This table includes a calculation of the \HI\ column density at which each line 
reaches an optical depth at line center of 1, 
$N$($\tau_0$=1)=$m_ec b/(\sqrt\pi e^2f\lambda$), where $m_e$ is the electron mass, 
$c$ is the speed of light, $e$ is the electronic charge, and $b$ is the Doppler line width. We consider 
two different values of $b$, 15\kms\ and 30\kms, chosen to be representative of \HI\ at $|b| > 20^{\circ}$.
The purpose of calculating $N$($\tau_0$)=1 is that it gives an indication of the 
column density where each line is unsaturated and therefore amenable to precise measurement. 
The twelve Lyman lines covered by our data cover 3\,dex in log\,$N$($\tau_0$), 
from $\approx$13.3 for Ly$\alpha$ to $\approx$16.2 for Ly$\mu$, revealing the diagnostic 
power of the Lyman series for characterizing diffuse \HI\ over a wide range of conditions. 

\subsection{Sample Selection} \label{subsec:sample}

\begin{figure*}[ht]
  \vspace{10pt}
    \centering
    \includegraphics[width=0.99\linewidth]{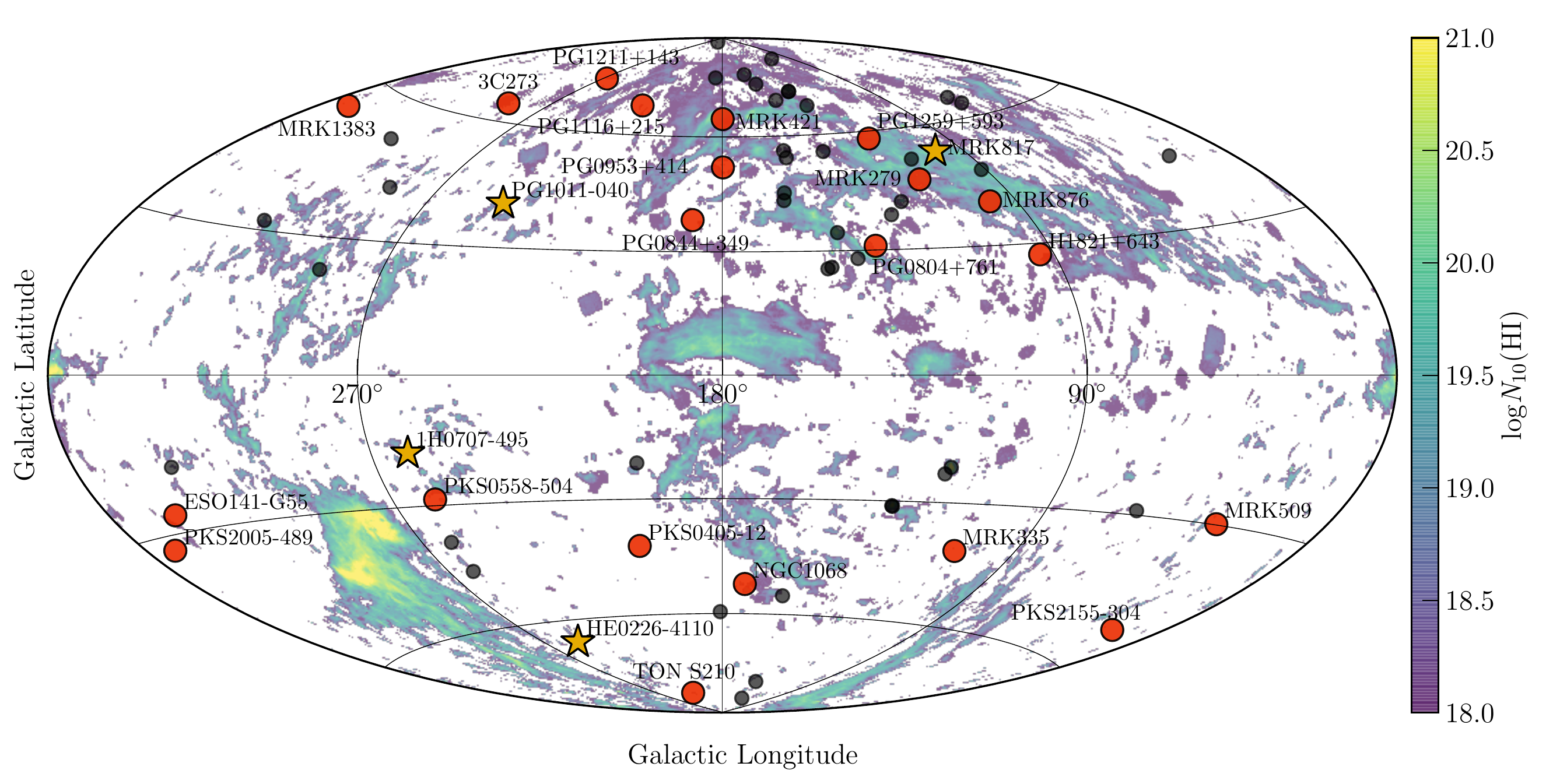}
    \caption{\small{All-sky map in Galactic coordinates showing the locations of the survey sightlines (red circles), with the four sightlines shown as examples in Figure \ref{fig:overview} highlighted as yellow stars. Black circles indicate the locations of 
    rejected sightline candidates. The map is centered on the Galactic anti-center and includes the {\it high-velocity} 21\,cm data from \citet{westmeier2018} color-coded by \HI\ column density. Note that our \fuse\ sample is split across the northern and southern hemispheres, and across the east and west.}}
  \vspace{10pt}
\label{fig:map}
\end{figure*}

We selected our sample from a preliminary pool of the 67 \fuse\ AGN spectra with 
a signal-to-noise ratio per resolution element at 977\AA\ of $\rm (S/N)_{977} \ge4$. 
Many of these proved to be unusable due to 
their low redshifts, which resulted in strong blending of intrinsic AGN emission 
or absorption. Several others were simply too low (S/N) to be useful. 
After careful inspection of each 
spectrum and removing the unusable cases, we arrived at a final sample of 25 
sightlines for which we could produce fits, 
all with $\rm (S/N)_{977}\ge$6 (see Table~\ref{table:fuse_results}). 
The on-sky locations of both our final sample and the rejected sightlines are 
shown in Figure \ref{fig:map}.

Of these 25 sightlines, 19 have \HI\ measurements presented in \citet{fox2006}; 
however, that paper only measured one Lyman series line per sightline using
apparent optical depth integrations; here we model the entire series 
using a full Voigt-profile fitting analysis.
In addition, the majority of the sightlines were analyzed
as part of the \fuse\ \ion{O}{6} survey \citep{savage2003, wakker2003, sembach2003}.
The majority also have \emph{Hubble Space Telescope}/Cosmic Origins Spectrograph
data covering UV metal lines \citep{richteretal2017, fox2020}, though 
(with the exception of the Ly$\alpha$) these are outside the scope of this analysis.

\subsection{Data Reduction} \label{subsec:reduction}
For each sightline we reduced the \fuse\ data with the reduction pipeline 
{\tt CALFUSE} version 2.4 or higher \citep{dixon2007}. 
This pipeline provides flux-calibrated, wavelength-calibrated spectra from the 
eight detector segments (LiF1A, LiF1B, LiF2A, LiF2B, SiC1A, SiC1B, SiC2A, SiC2B)
with a pixel size of 2\kms\ and velocity resolution of 20\kms\ (FWHM). 
Details of the \fuse\ satellite and its on-orbit performance are given in 
\citet{moos2000} and \citet{sahnow2000}.

All of our \fuse\ sightlines except for NGC 1068 and 1H0707-495 have also been 
observed by the Cosmic Origins Spectrograph (COS) and/or the 
Space Telescope Imaging Spectrograph (STIS) on board \hst. 
For these targets we obtained their G130M (COS) or E140M (STIS) spectra from the 
Mikulski Archive for Space Telescopes (MAST) in order to include the Ly$\alpha$ 
transition in our fitting procedure. 
From an analysis of four sightlines at varying signal-to-noise, we found that 
excluding Ly$\alpha$ from the fits (i.e. fitting the \fuse\ data alone) can 
lead to under-predicting the column density of the strongest components, 
in some cases by orders of magnitude. See Appendix \ref{section:lya} for a 
detailed discussion of this effect. Therefore we include Ly$\alpha$ in the 
fits whenever possible, even though this means including data taken by a different 
instrument. These Ly$\alpha$ data have been independently fit by \citet{wakker2011},
and we use their results as an independent check on the total \HI\ column densities.

\subsection{Voigt-Profile Fitting Procedure} \label{subsec:fitting}

\begin{figure*}[ht!]
  \vspace{10pt}
\centering
 \includegraphics[width=0.99\linewidth]{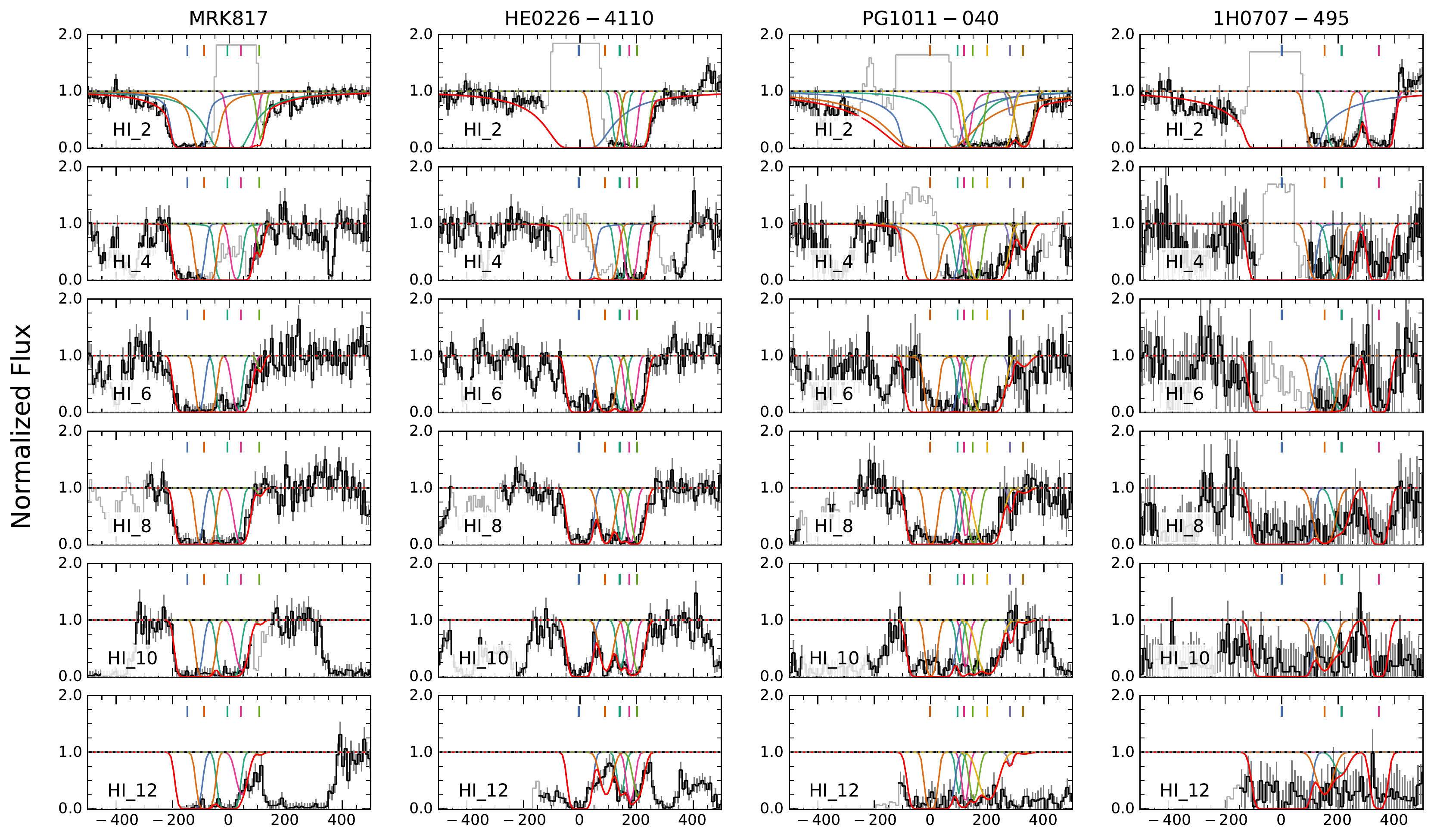}
 \includegraphics[width=0.99\linewidth]{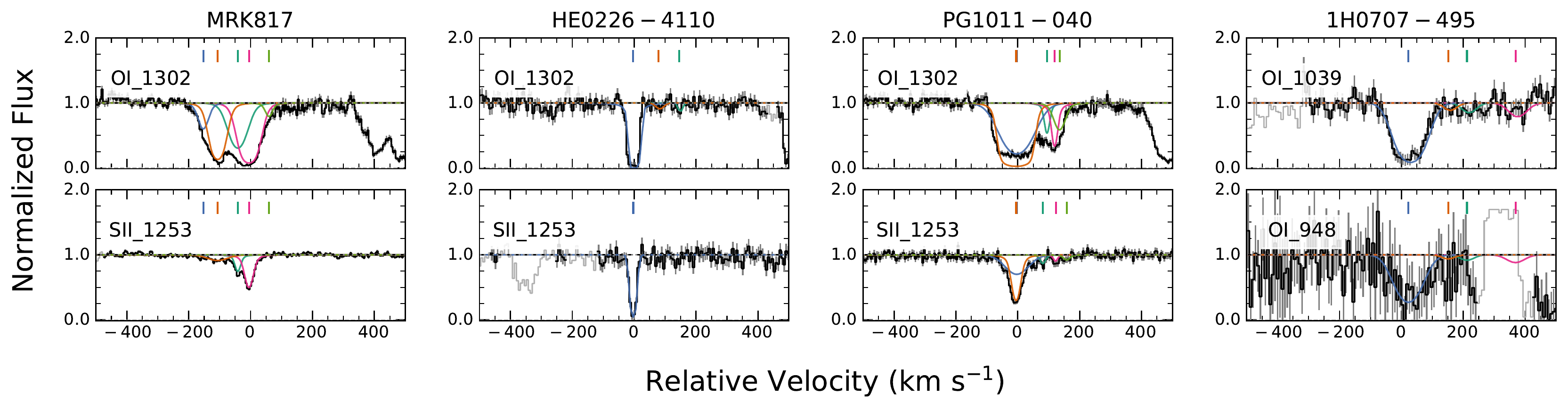}
  \caption{\small{{\bf Top:} Four example \fuse\ spectra from our survey spanning a range in sky location and signal-to-noise. Each panel shows the normalized flux vs LSR velocity for a different Lyman series absorption line, with the best-fit Voigt model shown in red, and each individual component plotted in a different color. {\bf Bottom: } Metal-line absorption fits toward the same four sightlines using COS, STIS and FUSE data. The matching colored tick marks and Voigt profiles illustrate the component structure of these unsaturated \HI\ tracers which help guide the placement of \HI\ components in the FUSE Lyman series fits.}}
  \vspace{10pt}
\label{fig:overview}
\end{figure*}

We used the Voigt-profile fitting software \emph{VoigtFit} \citep{voigtfit} to perform 
all of our Lyman series fits, with atomic data sourced from \cite{morton2003}. The code 
allows for simultaneous $\chi^2$-minimization fitting of multiple components across all twelve 
Lyman series transitions observed by both COS/STIS  ($\rm Ly\alpha$) and \fuse\ ($\rm Ly\beta - Ly\mu$). 
The \fuse\ and STIS/E140M line-spread functions (LSFs) were approximated as Gaussians, 
with FWHM=20.0\kms\ for \fuse\ and FWHM=6.5\kms\ for STIS/E140M. For COS spectra, 
we adopted the LSF tables made available by STScI for the appropriate central wavelength setting, 
wavelength, and detector lifetime position. These tabulated COS LSFs account for extended 
non-Gaussian wings in the line profiles. The LSFs are provided as inputs to \emph{VoigtFit} 
for convolution with the model components when matching the data. Our detailed fitting 
procedure is described in the following six steps.

First, all spectra were binned by a factor of two, resulting in rebinned pixels 
of 4\kms\ for \fuse, 5\kms\ for COS, and 4\kms\ for STIS. 

Second, we searched for absorption components in the interval ${|v_{\rm LSR}|}\lesssim500$\kms. 
This interval was chosen after inspection of all the spectra, since all components were found 
to lie within this window. It includes both strong absorption from the Milky Way ISM as well as 
strong \HI\ airglow emission down to approximately Ly$\rm \epsilon$ $\lambda 937.803$. These 
airglow regions are manually masked out before fitting. We endeavored to keep additional masking 
to a minimum, although it is occasionally necessary due to intervening higher-redshift 
absorption systems. 

Third, we fitted continua around each line of interest using low-order Legendre polynomials, 
and applied an $\rm H_2$ model from the \fuse\ data based on results from \cite{wakker2006}. 
This $\rm H_2$ correction sometimes results in ``ghost-flux" -- a spurious signal artificially 
pulled out from a saturated profile -- or a simple over-correction resulting in too much flux
being added back to a profile. We mask these regions only if they appear to affect the 
quality or convergence of a fit. 

Fourth, we isolated and fitted the O\I\ absorption seen at 
$\lambda 1302, 1039, 988, 971, 950, 948, 936, 929, 924$, and S\II\ absorption at 
$\lambda 1250, 1253, 1259$ using \emph{VoigtFit}, to use as an initial-guess model for the \HI\ component structure. 
This required a combination of \fuse\ data and \hst/STIS or COS data covering the O\I~$\lambda$1302 
and S\II\ $\lambda$1250, 1253, 1259 lines. 
The primary challenge for fitting the \HI\ profiles is the velocity structure, as it can be 
nearly impossible to unambiguously recover multiple absorbers overlapping in velocity space. 
The low velocity, ISM region ($|v_{\rm LSR}|<40$ \kms) is the most saturated and suffers the most. 
However, \OI\ and S\II\ are abundant and not highly depleted onto dust grains, and thus 
are excellent tracers of \HI. \OI\ is a particularly strong tracer of \HI\ because the two are closely coupled by resonant charge exchange \citep{Field1971}. We thus use these \OI\ and S\II\ fits to inform the initial 
guesses for the \HI\ component structure. These metal lines complement each other effectively, 
as the more sensitive \OI\ transitions can help reveal weak \HI\ lines, and the weaker and 
narrower S\II\ lines provides clues to the velocity structure of closely blended components at low velocities. The combination of \OI\ and S\II\ can trace \HI\ down to roughly log\,$N$(\HI)$\sim$17 at $20\%$ solar metallicity. Additionally, all our targets have been observed in 21 cm emission, so where necessary we 
consulted these emission profiles for help untangling the \HI\ component structure.

Fifth, the actual fits were then performed simultaneously to Ly$\alpha$--Ly$\mu$. 
The \OI\ line fits at $\lambda 971, 950, 948, 936, 929, 924$ are included during the \HI\ fitting 
but are held fixed (i.e., they are no longer allowed to vary while \HI\ is fit). 
We do this so that \emph{VoigtFit} knows about and fits \emph{around} the \OI, 
instead of masking out all these regions.

Sixth, the final \HI\ fits are obtained by iteratively adding, removing, shifting, 
and fixing \HI\ component guesses until the fit both converges and visually matches the data. 
Generally we strive for the minimum number of components to adequately fit the data, 
while keeping in mind which components are most likely present based on \OI, S\II, and \HI\ 21 cm data.
When possible, we also reference other metal transitions such as Si\II, Si\III, and C\II\ for
additional hints to component structure. At this point we also check for intervening deuterium absorption
associated with the strong, low-velocity components. Deuterium absorption can masquerade as weak IVC absorbers,
but we find little evidence of such contamination.
On occasion, an \OI, S\II, or \HI\ 21 cm feature appears to be two blended components that are 
very close in velocity. These features are very difficult to reproduce in the Lyman series fits, 
and often are instead fit with a single component. This can have the effect of biasing our
resulting \HI\ distribution toward higher columns, particularly in the log\,$N$(\HI) = $18 - 19$ column density range.
However, as this affects less than $20\%$ of our sightlines we expect this bias to be minor. 

Occasionally we found it necessary to fix a particular \HI\ component's velocity, linewidth, 
and/or column density for a fit to converge. In these cases we start by fixing only the 
component velocity and/or linewidth, which is informed by visually inspecting the spectra 
alongside \OI, S\II, and \HI\ 21 cm results. In only one case (\object{PG1211+143}) did we find it 
necessary to fix a component column density, using the value from 21 cm observations, and we 
do not include this component in any subsequent analyses. The components with fixed parameters 
are overwhelmingly in the low-velocity ISM regime, where high-$N$(\HI) components are 
preferentially found, and where the component velocity, linewidth, and column density are 
otherwise well constrained by \OI, S\II, and \HI\ 21 cm values. 

Table \ref{table:fuse_results} presents the complete list of fit 
components with errors and notes indicating which components were fixed.
We make a final quality cut for all subsequent analysis by including only components with 
total column density fit errors less than 3 dex, which removes three particularly uncertain components. 
Figure \ref{fig:overview} showcases example fits for four sightlines ranging in signal-to-noise along 
with an all-sky plot with the positions of all of our targets, which we discuss in more detail in 
Section \ref{section:components}.
Detailed figures showing our fits for each sightline are presented in Appendix \ref{section:fits}.

It is important to consider whether any bias is introduced by beginning with 
\OI\ and \ion{S}{2} component structure and applying this to the \HI. 
Since hydrogen is the dominant baryon in the Universe, there must be \HI\ co-existing with 
the observed \OI\ and \ion{S}{2}, and our fitting procedure will by design reveal the co-spatial 
portion of the \HI, but in principle there could be \emph{low-metallicity} \HI\ clouds 
\emph{without} corresponding \OI\ and \ion{S}{2}, and any such clouds would be missed 
by our metal-based fitting procedure.
However, we have addressed this issue by the iterative process of adding and removing 
\HI\ components until the data are matched with the minimum number of components as 
described above. The finding that there are no strong \HI\ components without 
\OI\ counterparts suggests that we are not missing a population of extremely metal-poor 
components, and therefore that this bias is small. 
In a future paper we will present the associated \OI\ and \ion{S}{2}
absorber population along with an analysis of the resulting metallicities.

\subsection{Determining Component Structure: Examples} \label{section:components}
In this section we examine the component structure and fitting choices in detail for the four example sightlines shown in Figure \ref{fig:overview}. The purpose of this exercise is to illustrate the effectiveness of our fitting methodology and to show how information can be extracted on the \HI\ properties \emph{even from datasets that suffer from saturation and low S/N}. The four sightlines cover a range of S/N and sky location. In each direction, Figure \ref{fig:overview} shows six Lyman series transitions and two metal line transitions, and the sightlines are ordered left-to-right from high-to-low S/N. The colored tick marks show the locations of individual components and correspond to the matching-color Voigt profile fits. The colors of the components from left-to-right in velocity additionally match between metal and \HI\ plots, further indicating how components match between metal and \HI\ data. The final combined fit profile is shown in red for the \HI\ data. The details of the component structure are now discussed for each of these four sightlines in turn.

Many of the individual \fuse\ spectra presented here have been published in earlier work on the Galactic halo \citep{sembach2002, collins2003, collins2004, collins2005, collins2007, fox2004, fox2005, fox2006, ganguly2005, zech2008, shull2011}, and several of these studies present \HI\ measurements in individual sightlines. However, no systematic \fuse\ \HI\ survey has been conducted before.

\subsubsection{MRK\,817}
The \fuse\ spectrum of MRK\,817, a high-quality dataset with $\rm (S/N)_{977}=18.5$, shows strong \HI\ absorption at both positive and negative velocities. This sightline passes through HVC Complex C \citep{collins2003}. All five of the \HI\ components we fit in the \fuse\ data are also seen in \OI\ and S\II, albeit somewhat shifted in velocity. In Figure \ref{fig:overview} we show \OI\ $\lambda1302$ and S\II\ $\lambda1253$ as metal-line examples. The weaker S\II\ $\lambda 1253$ transition clearly shows two components near zero velocity (fit in green and pink). The stronger \OI\ transition clearly shows a higher negative velocity component near $-$100 \kms\ (orange), and suggests additional weaker components on the highest negative and positive velocity edges of the absorption profile (blue and lime-green).

\subsubsection{HE\,0226-4110}
The \fuse\ spectrum of HE\,0226-4110, with $\rm (S/N)_{977}=11.2$, illustrates an example of low-metallicity gas detected in the Lyman series data \citep[see][]{fox2005}. The S\II\ $\lambda 1253$ transition clearly shows one strong, near-zero velocity component, and the \OI\ $\lambda 1302$ transition suggests two very weak additional components around $+80$ and $+150$ \kms\ (shown in orange and green). All three of these components are clearly visible in the \HI. However, two additional higher velocity components (pink and lime-green) are clearly seen in the \HI\ spectra. As these components both have column densities of log\,$N$(\HI)$\lesssim 17$, their \HI\ properties would be unknown if not for the \fuse\ data. We note that the orange component appears to be poorly fit in Ly$\mu$ (\HI\_12), but this is due to edge effects and model $\rm H_2$ over-correction artificially pulling up flux. This component is well fit throughout the rest of the Lyman series.

\subsubsection{PG\,1011-040}
The \fuse\ spectrum of PG\,1011-040, with a moderate $\rm (S/N)_{977}=8.9$, is an example of a highly saturated profile with a number of absorbers at positive velocities. Nevertheless, the constraints imposed by the combination of \HI, \OI, and S\II\ data illustrate how a saturated, moderate S/N \fuse\ spectrum is still useful and can yield reliable \HI\ fits. Here the \OI\ $\lambda 1302$ and S\II\ $\lambda 1253$ suggest two overlapping $\sim0$ \kms\ components, one broad (shown in blue) 
and one narrow (orange). The \HI\ is also best fit with both components as a single, higher column density component would lead to overly strong wings in Ly$\alpha$ and Ly$\beta$. Three additional components are seen in \OI\ and S\II\ around $+100$ \kms, which also clearly appear in the \HI. 
From the \HI\ spectra we see that there are additional, low-metallicity components needed. We start by adding a component at $\sim$325\kms\ (brown). However, there must be an additional component near $\sim$200\kms\ (yellow) as this region of the spectra remains saturated. Finally, the edges of the spectra between these yellow and brown components suggest the presence of an additional weak component (grey). 

Hence, while the \fuse\ data alone looks too saturated to achieve a unique solution, the constraints imposed by the metal lines leave little room for degeneracy even when there is metal-poor gas present. We also note that there is a flux-calibration issue with this COS sightline, such that the saturated core of the \OI\ $\lambda$1302 still shows substantial flux. The orange fit line, which dives below the data here, is well constrained by other \OI\ transitions.

\subsubsection{1H0707-495}
The \fuse\ spectrum of 1H0707-495 is our lowest S/N example (with $\rm (S/N)_{977}=6.4$), and does not have complementary COS or STIS data. However, \OI\ absorption seen in the \fuse\ data alone is still useful for determining the component structure. \OI\ $\lambda 1039$ shows a strong $\sim0$ \kms\ component, and good evidence for three weak, higher velocity components. Other \OI\ transitions, such as the \OI\ $\lambda 948$ shown, provide additional constraints on the strength and velocity of each component. These four components appear to nicely match the Lyman series profiles, even though the low S/N makes fitting difficult. 


\section{Results}\label{section:results}

\begin{figure}[!ht]
\centering
 \includegraphics[width=\linewidth]{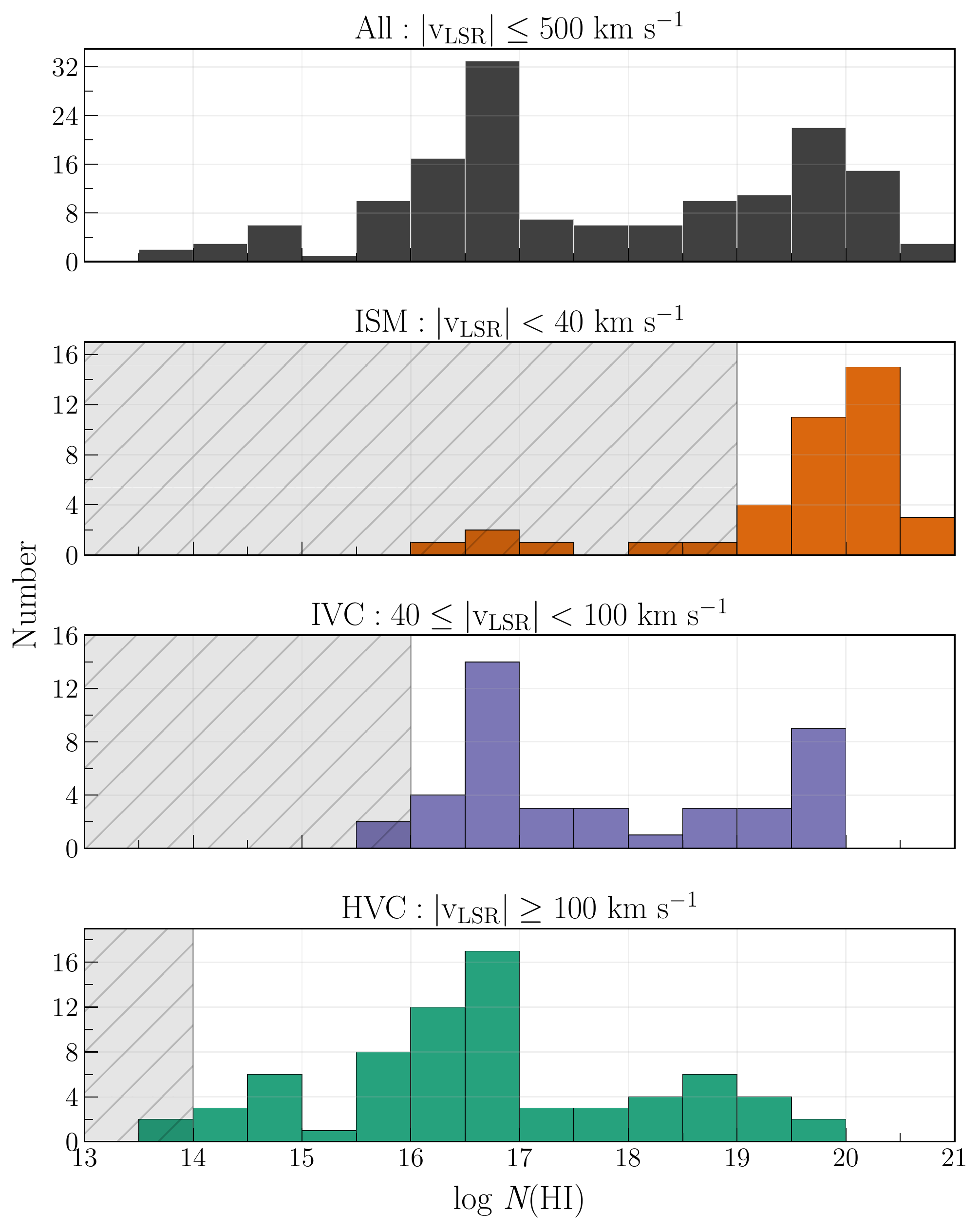}
  \caption{\small{The raw distribution of \HI\ component column densities observed by \fuse. The top panel shows all velocities (black). The next three panels show the distributions split by LSR velocity into ISM (orange), IVC (purple), and HVC (green) sub-samples. Hatched regions in the ISM, IVC, and HVC panels indicate the column density regions we are insensitive to due to saturation.}}
  \label{fig:N_hist}
  \vspace{5pt}
\end{figure}

\begin{figure}[h!]
\centering
  \includegraphics[width=1.0\linewidth]{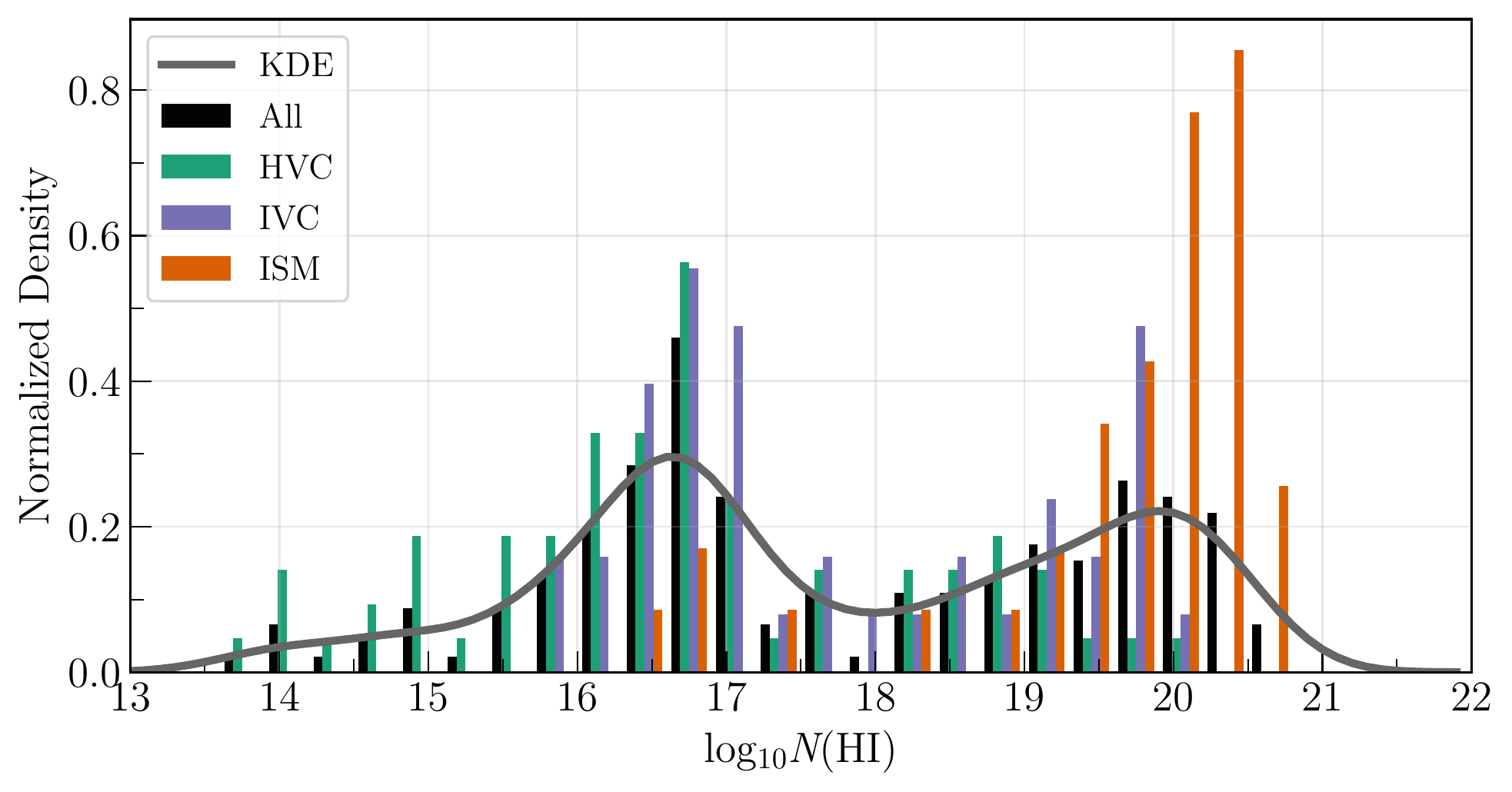}
  \caption{\small{
    Normalized histograms showing the column density distribution for the full sample (black). ISM, IVC, and HVC subsamples are shown similarly in orange, purple, and green histograms. The probability distribution function (PDF), approximated via a Kernel Density Estimation (KDE), is overplotted in grey and illustrates the smoothed column density distribution.}}
  \label{fig:KDE}
  \vspace{15pt}
\end{figure}

\begin{figure}[!ht]
\centering
 \includegraphics[width=\linewidth]{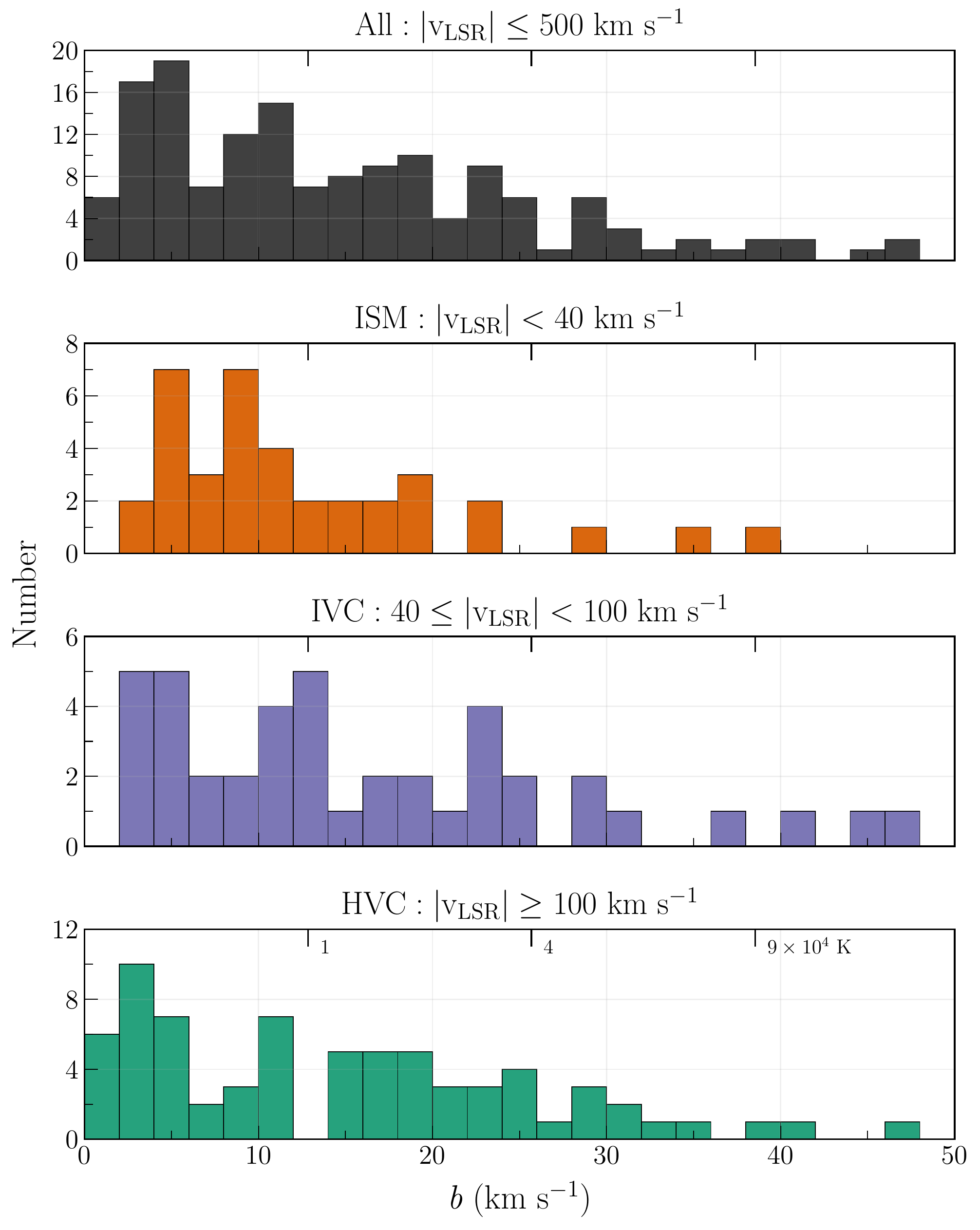} 
  \caption{\small{The distribution of \HI\ component Doppler $b$-parameters observed by \fuse. The top panel shows all velocities (black). The next three panels show the distributions split by LSR velocity into ISM (orange), IVC (purple), and HVC (green) subsamples. Two anomalously high values are not shown here: $b$=68\kms\ toward PG\,1116+215 and $b$=72\kms\ toward PKS\,2155-304 (both of which are ISM components that may be unresolved).}}
  \label{fig:b_hist}
   \vspace{5pt}
\end{figure}

The intention of this paper is to provide an empirical survey of the \HI\ absorption
in the Galactic halo.
We therefore focus our results on the following observational quantities:
the basic properties of the \HI\ sample (sample size, kinematics, and column densities;
Section~\ref{subsec:basicprop}),
the \HI\ errors (Section~\ref{subsec:nerrors}),
the dependence of \HI\ properties on sky location (Section~\ref{subsec:spatial}), 
the \HI\ sky covering fraction (Section~\ref{subsec:skycov}),
the \HI\ column density distribution function (Section~\ref{subsec:cddf}), 
and the relative prevalence of inflow and outflow (Section~\ref{subsec:flow}).

\subsection{Basic H I Properties} \label{subsec:basicprop}

\begin{figure*}[ht!]
\centering
  \includegraphics[width=.90\linewidth]{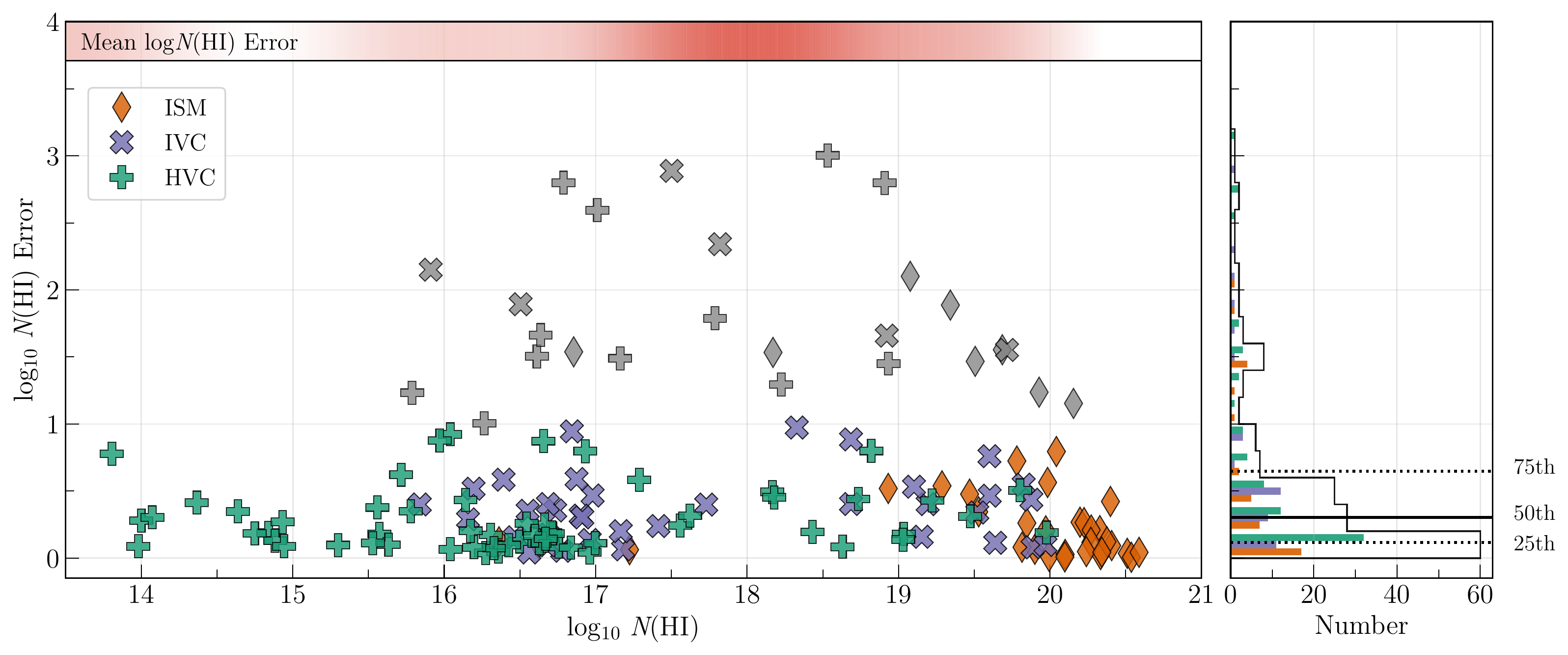}
  \caption{\small{
    The total \HI\ column density error (summed in quadrature) as a function of \HI\ column density for all components included in the final sample. The three components with error $>3$\,dex are not shown, but have log\,$N$(\HI) = 18.8, 18.4, and 18.4. The red-shaded horizontal bar illustrates the mean error as a function of column density, where darker red implies larger errors. On the right we include a marginal histogram showing the distribution of errors for the ISM (orange), IVC (purple), HVC (green), and combined sample (black outline). The quartiles of the distribution are indicated by the black solid and dashed lines and matching labels.}}
  \label{fig:e_logNHI}
  \vspace{5pt}
\end{figure*}

From our sample of 25 sightlines we arrive at a final sample of
152 individual \HI\ components, 
or an average of 6 components per target. The sample is composed of
39 ISM ($|v_{\rm LSR}|<40$\kms), 42 IVC ($40 \leq |v_{\rm LSR}|<100$\kms), 
and 71 HVC ($|v_{\rm LSR}|\ge100$\kms) components.
All sightlines have at least one HVC, and 24/25 have at least one IVC.
Taken together, there are 113 HVCs and IVCs in the sample. 64 of these 113 (58\%) 
have log\,$N$(\HI)$<$17.5 and are not detectable in 21\,cm emission, 
showing that the majority of the HVCs and IVCs in our sample are invisible 
to 21\,cm observations.

Figure~\ref{fig:N_hist} shows histograms of the component column densities for the full sample 
and for three bins of velocity: ISM, IVC, 
and HVC. These histograms show the basic distributions from 
which all subsequent \HI\ quantities are derived. 
The ISM components tend to show high column densities of log\,$N$(\HI)$\ga$19 
(as expected, since the strong \HI\ components in the Galactic disk are already well 
characterized by 21\,cm observations), whereas IVCs and HVCs show much lower values, 
each peaking near log\,$N$(\HI)=16.5. The HVC distribution also shows a low column density 
tail extending down to log\,$N$(\HI)$\approx$14, which is not seen in the IVC distribution.
We also visualize the combined, normalized dataset in Figure \ref{fig:KDE}. This figure includes
a Kernel Density Estimation (KDE) of the full sample of column densities (shown by the grey line).
The KDE is an estimate of the probability density function (PDF) of the sample, and in practice
produces a smoothed, non-parametric functional form for the column density distribution of the Galactic halo.

Figure \ref{fig:b_hist} shows histograms of the component $b$-values (Doppler line widths), 
with the same figure format as Figure~\ref{fig:N_hist}. The observed \HI\ $b$-values range 
from 0--50\kms\ for each of the ISM, IVC, and HVC populations. In each case, the distribution 
peaks at $b\approx10$\kms\ and then shows a long tail extending out to $\approx$50\kms. 
Note that the \fuse\ instrumental velocity resolution (FWHM=20\kms) corresponds to a 
limiting $b$-value of 12\kms, so a large fraction of the components are unresolved. 
We do not show components with $b>50$\kms\ (two such cases were found, both ISM components), 
since such broad lines may be unresolved blends of narrower components, especially given the 
low S/N of the data.

\subsection{The Saturation Problem} \label{subsec:nerrors}

The difficulty of accurately measuring saturated lines from the Lyman series 
results in large column density errors for many \HI\ components. To illustrate this, 
Figure \ref{fig:e_logNHI} shows the errors on log\,$N$(\HI) as a function of log\,$N$(\HI) 
for all components in the final sample. This error is a quadrature sum of the statistical and 
systematic errors listed in Table \ref{table:fuse_results}. The errors are largest in the
log\,$N$(\HI)$\sim$17--19 region where the lines are saturated but the wings are not yet 
strong enough to fit. The errors reach $\sim$1--3\,dex in this region. In Figure \ref{fig:e_logNHI}
we also include a red-shaded horizontal bar illustrating the mean error as a function of column
density. We include this bar in subsequent figures as a visual indicator of the 
regions most affected by saturation.

\begin{figure*}[!ht]
\centering
 \includegraphics[width=1.0\linewidth]{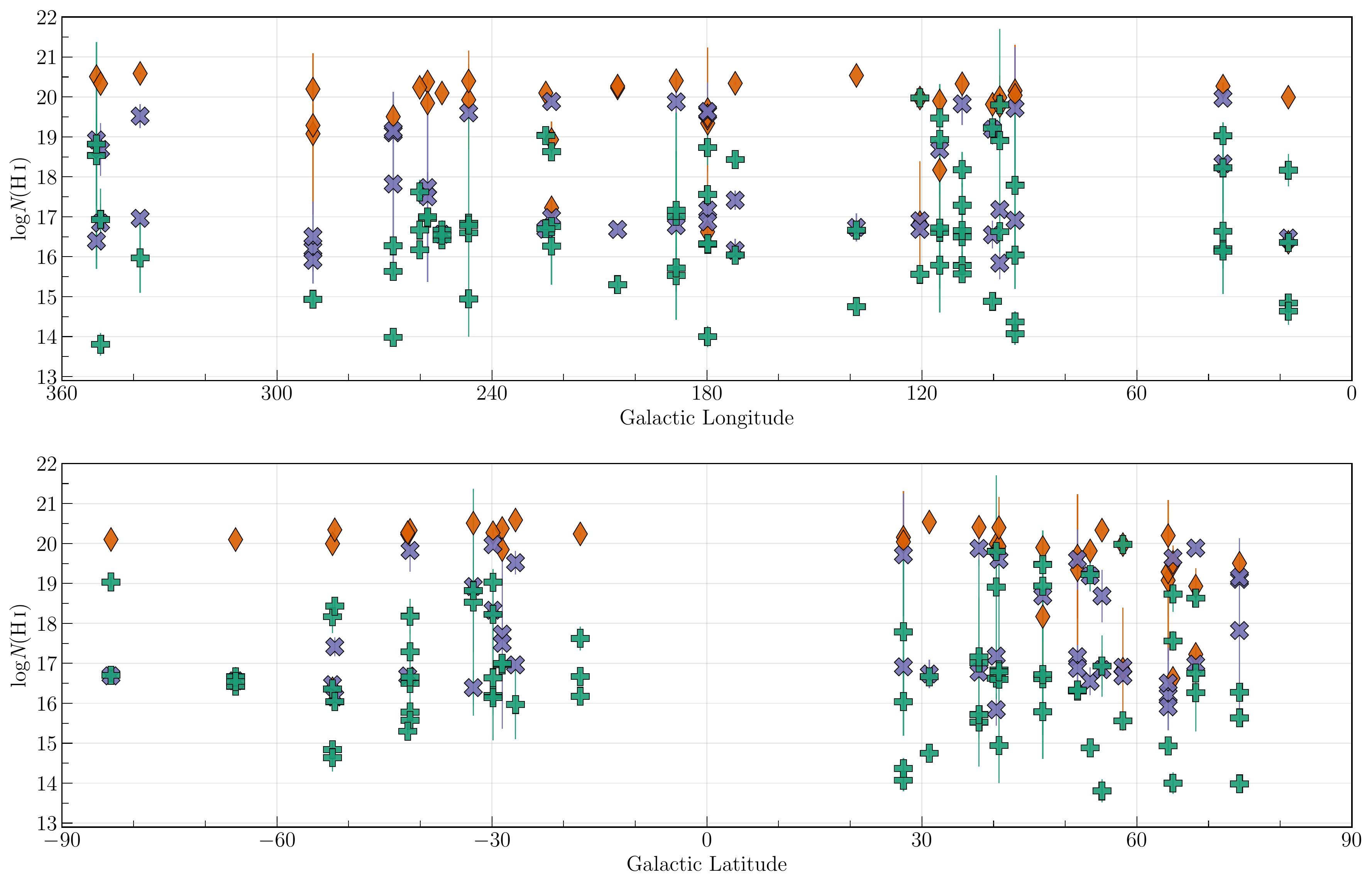}
  \caption{\small{\HI\ column densities as a function of Galactic longitude (top) and latitude (bottom)
  for HVCs (green pluses), 
  IVCs (purple crosses), 
  and ISM components (orange rhombuses). 
  Each column of points shows the components from a given sightline.
  No components are seen at $|b|<20\degr$ due to a lack of observable background AGN in this region.}}
  \label{fig:N_gal}
  \vspace{5pt}
\end{figure*}

\begin{figure*}[ht!]
\centering
  \includegraphics[width=1.0\linewidth]{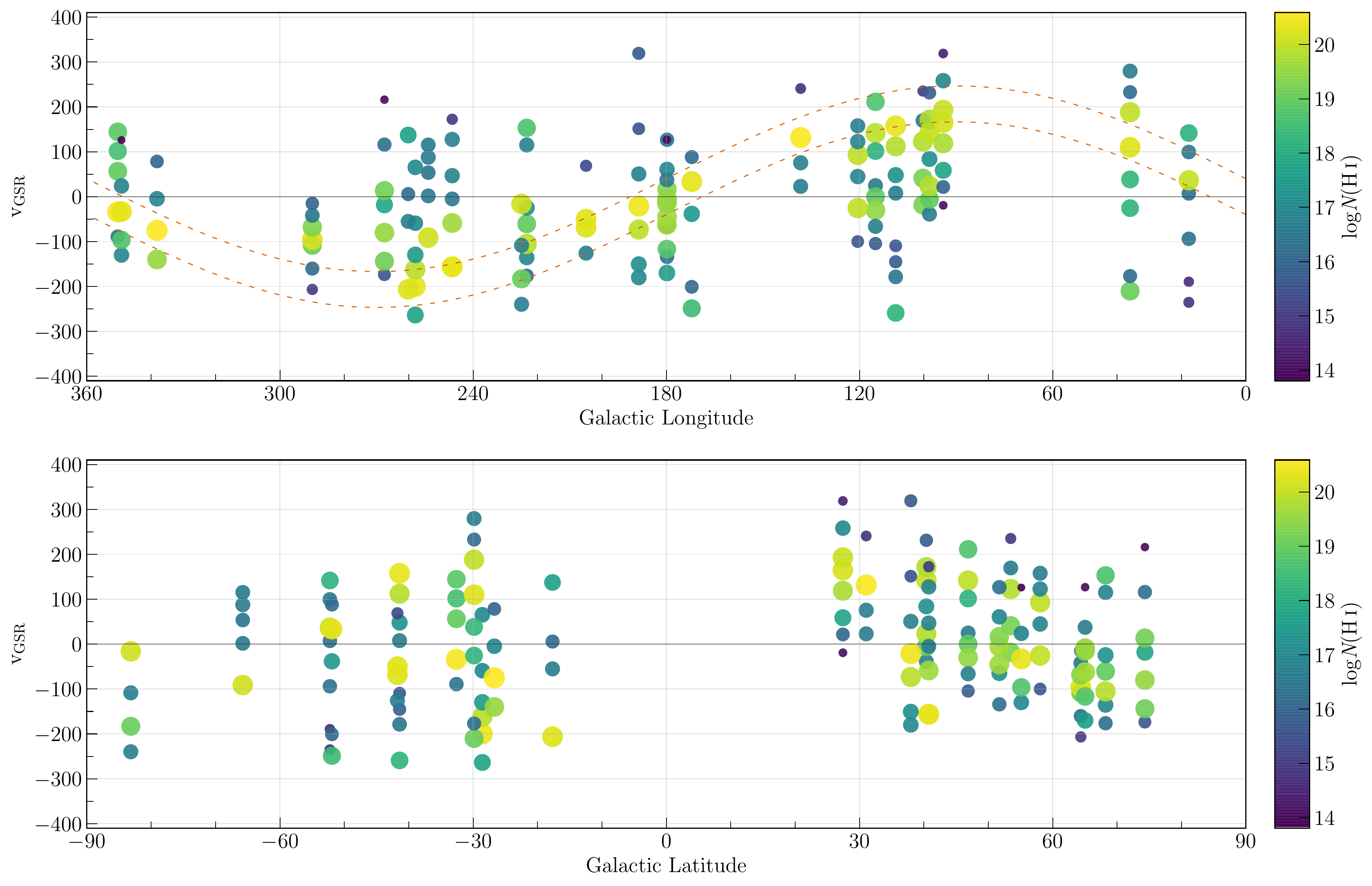}
  \caption{\small{
  The GSR velocities of \HI\ components as a function of Galactic longitude and latitude, with symbols color coded by $N$(\HI). The dashed orange lines (top) indicate the $|v_{\rm LSR}| \leq 40$ \kms\ boundary between ISM components and higher velocity gas, illustrating the sinusoidal motion of Galactic rotation. The height of each column of points indicates the interval over which components are detected in each spectrum and provides a measure of the kinematic complexity of the sightline.}}
  \label{fig:v_gal}
  \vspace{5pt}
\end{figure*}

For some of these
components the statistical fitting errors may be artificially inflated and/or unreliable due to
numerical errors in the fitting routine. This can happen because the code cannot resolve degeneracy
between components, even though we can be confident of component location and approximate strength 
from the combination of other constraints (e.g., \HI\ 21 cm and UV metal-line absorption). We highlight these highly uncertain 
components in Table \ref{table:fuse_results} and plot them in grey in Figure \ref{fig:e_logNHI}.

Below log\,$N$(\HI)$\sim$17, the errors become considerably smaller as the higher-order
Lyman lines become unsaturated. While a few uncertain
components remain in this region, the vast majority have errors $\lesssim 0.5$ dex. 
Above log\,$N$(\HI)$\sim$19, the errors become smaller as the damping wings of 
Ly$\alpha$ allow precise measurements.
We find that 45 (out of 152) components have an error $>$0.5\,dex and 24 have an error $>$1\,dex,
illustrating the challenging (but not insurmountable) nature of the measurements. These large errors
provide confidence that the \emph{VoigtFit} code is doing a reasonable job of identifying
uncertain components.

\subsection{Spatial Distribution} \label{subsec:spatial}

Next we consider the dependence of \HI\ component properties on Galactic longitude 
$l$ and latitude $b$, to look for spatial variation in the neutral gas.
Figure \ref{fig:N_gal} shows the \HI\ column densities against $l$ and $b$, with 
ISM, IVC, and HVC components colored separately. The \HI\ columns are found to be 
fairly evenly distributed in \emph{longitude}; this is true both in general and for 
each velocity category (ISM, IVC, and HVC), with the range of observed columns staying 
fairly constant across longitude. The range in columns observed is also fairly constant 
in \emph{latitude}; however, we observe more IVCs and HVCs in the north than the south. 
This may simply reflect that there are more northern sightlines in the survey than southern 
(our sample includes 14 N and 11 S), and many of those northern sightlines pierce large 
well-known \HI\ HVC complexes like Complex C and Complex A. To statistically check for
differences we performed Anderson-Darling 2-sample distribution tests for north versus south
latitudes and longitudes above and below $180^{\circ}$, which found no significant deviations
between these split samples. The gap at $|b|<20\degr$ is due to a lack of background AGN in this 
region, which is a consequence of dust extinction in the Galactic plane.

Figure \ref{fig:v_gal} shows the component velocity centroids in the Galactic Standard of 
Rest (GSR) reference frame against Galactic longitude (top) and latitude (bottom).
Since the Local Standard of Rest (LSR) frame in which the data are fit is a rotating frame, the
transformation to GSR attempts to remove the effect of Galactic rotation
and thus reveal a more valid distribution of cloud velocities in the Galactic rest frame. 
In this figure we have normalized the size and 
color of each point to reflect the \HI\ column density of each component, allowing a visual 
assessment of where the components are located in velocity-$N$(\HI) space.
The height of each column of points indicates the velocity interval 
over which components are detected, and therefore provides a measure of the 
kinematic complexity of each sightline.

In the top panel, we also plot dashed red lines to indicate where ISM gas at LSR velocities of
$|v_{\rm LSR}| \le 40$ \kms\ resides in GSR space. The highest column density components, as 
indicated by larger point size and yellow color, mostly fall within this velocity range.
We again performed Anderson-Darling 2-sample distribution tests to check for differences
between components at longitudes of $0 \le l < 180^{\circ}$ and $180 \le l < 360^{\circ}$.
The resulting $p$-value ($p=0.002$) suggests that there is a difference, with 
preferentially positive velocity components at $0 \le l < 180^{\circ}$ and 
negative at $180 \le l < 360^{\circ}$.
However, this is exactly what we would expect if the conversion of velocities from LSR-to-GSR
was imperfect (i.e., notice how the points appear to follow the dashed red lines). As we know
the LSR-to-GSR conversion is indeed imperfect, we do not assign much significance to this finding.

In the bottom panel, we also observe some structure with latitude. In particular,
we notice an absence of high positive velocity ($v>100$\kms) absorbers at high negative latitude 
(the top-left corner of the plot), whereas high positive velocity components \emph{are} found at 
high positive latitudes (the top-right corner). Both distributions show more negative velocity
components than positive, and this is particularly true in the south.
This may indicate either a deficit of outflowing gas in the southern sky, or a relative
excess of inflow. An Anderson-Darling test produces a $p$-value of $p=0.01$, 
suggesting a mildly significant difference between the high-and low-latitude distributions.

\subsection{Sky Covering Fraction} \label{subsec:skycov}

\begin{figure}[ht!]
\centering
 \includegraphics[width=\linewidth]{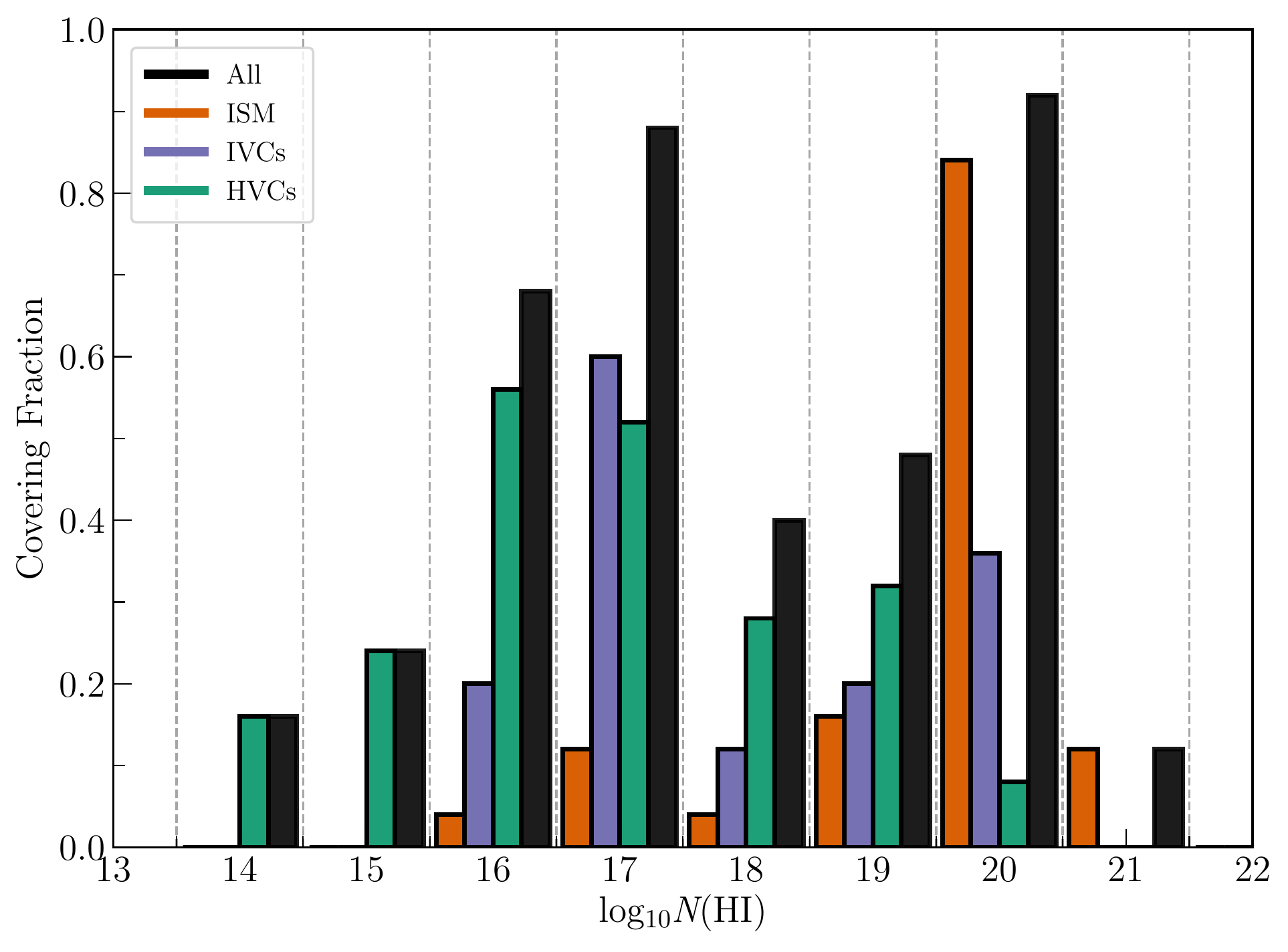} 
  \caption{\small{The discrete \HI\ sky covering fraction $f_{\rm sky}$ 
  as a function of column density interval log\,$N\pm\Delta$log\,$N$ where 
  $\Delta$log\,$N$=0.5. The distributions are given for all gas (black), 
  ISM (orange), IVC (purple), and HVC (green) populations. The vertical grey dotted lines
  denote the bin edges.}}
  \label{fig:f_c}
\end{figure}

\begin{figure}[ht!]
\centering
 \includegraphics[width=\linewidth]{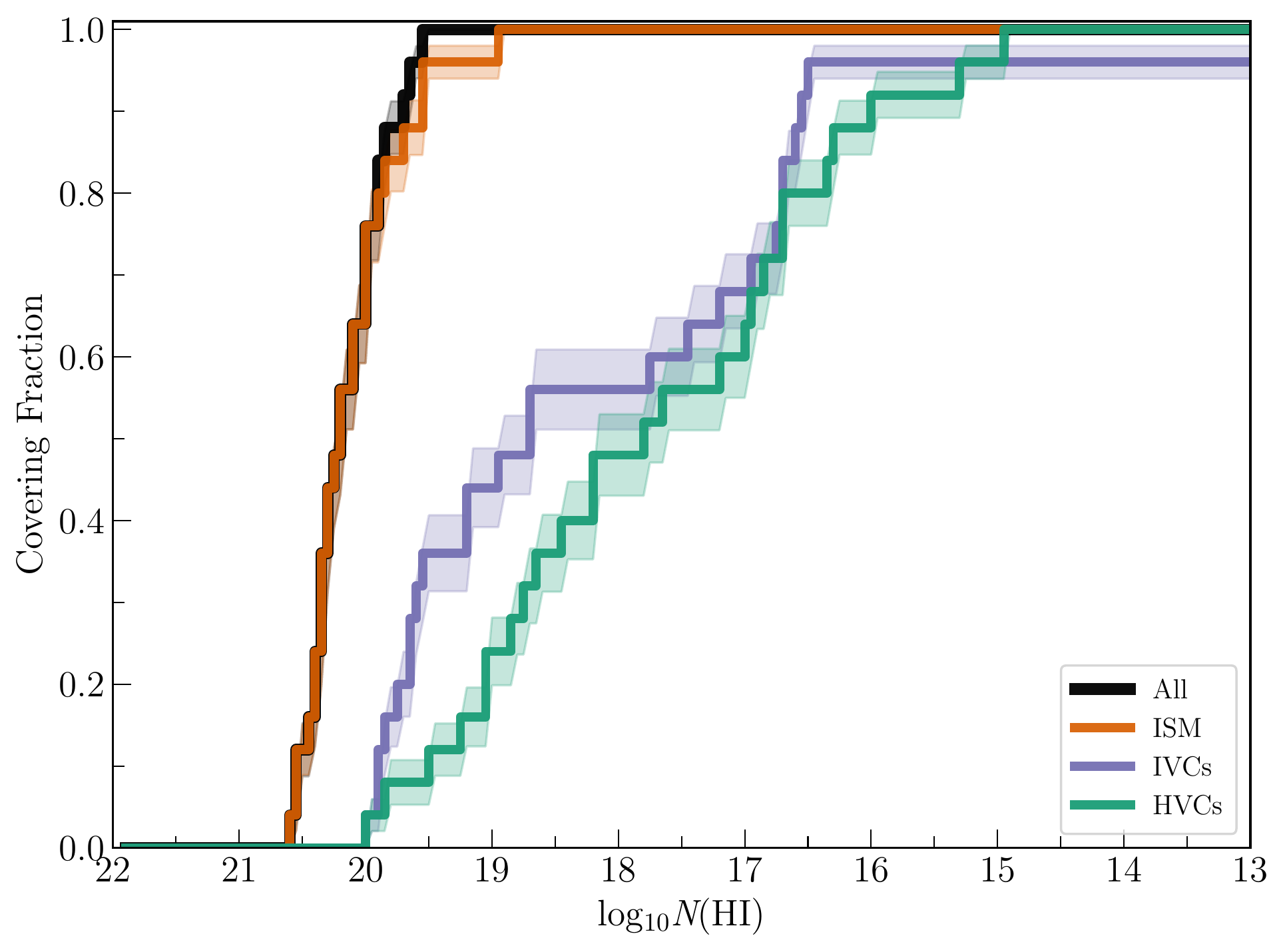}\\ 
 \includegraphics[width=\linewidth]{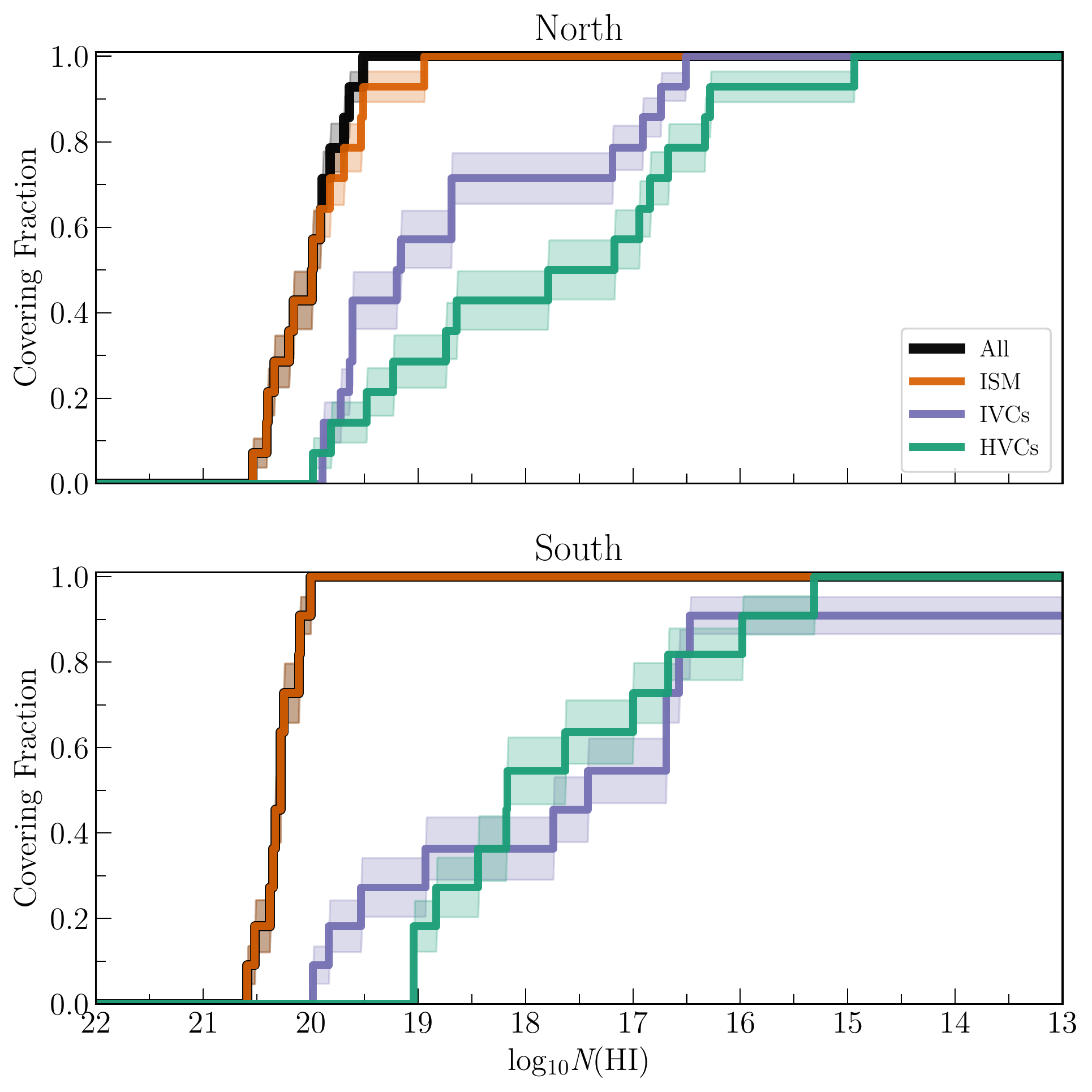} 
  \caption{\small{{\bf Top panel:} The cumulative \HI\ sky covering fraction $f_{\rm sky}(>N)$ as a function of \HI\ column density, for all absorbers (black) and for each velocity bin (colored lines). {\bf Lower two panels:} same as above but split into the northern and southern hemispheres.}}
  \label{fig:f_c_cum}
  \vspace{0pt}
\end{figure}

One fundamental \HI\ quantity we can calculate is the sky covering fraction $f_{\rm sky}(N)$, 
which is the fraction of sightlines containing at least one component of a given column density. 
We can calculate this in two primary ways.
First, we derive $f_{\rm sky}$ for discrete column density intervals log\,$N\pm\Delta$log\,$N$ in
Figure~\ref{fig:f_c}, where $\Delta$log\,$N$=0.5, chosen because this gives bin sizes of 1\,dex. 
This plot is similar to the raw distribution of 
components in Figure~\ref{fig:N_hist}, but with the key difference that a sightline may contain 
multiple components in any given log\,$N$ interval, whereas the sky covering fractions only count one 
component per bin even if multiple are present. In the full sample we observe a high value
$f_{\rm sky}>0.5$ down to 10$^{16}$\sqcm, with a peak in the number of components in the
log\,$N$(\HI)=16.5--17.5 column density bin. This peak is observed in both the IVC and HVC sub-samples, 
although we note that the measurement errors are generally high around these column densities.

Second, we calculate the \emph{cumulative} sky covering fraction as a function of column density, 
$f_{\rm sky}(>N)$, in Figure \ref{fig:f_c_cum}. This shows the percentage of sightlines in which 
we observe an absorber of column density $N$ or higher. We can repeat the exercise separately for 
the northern and southern absorbers, as shown in the lower two panels in the figure. 
This analysis shows that the sky is 100\% covered with ISM \HI\ above log\,$N$(\HI)=19, 
i.e. all sightlines in our survey have at least one ISM component at or above this column density. 
In contrast, 100\% coverage of the sky with HVCs not achieved until log\,$N$(\HI)=15, and 
maximal ($\approx95\%$) IVC sky coverage only at log\,$N$(\HI)=16.5.

We observe a difference between the northern and southern hemispheres, 
with higher sky covering fractions for IVCs in the north than in the south. 
To quantify this N-S asymmetry, we determine the \HI\ column at which
a certain $f_{\rm sky}$ is reached.
For IVCs, we find that $f_{\rm sky}$=0.8 (80\% sky coverage) is reached at log\,$N$(\HI)=17.0 in
the north and 16.5 in the south,
showing that one must go 0.5\,dex lower in $N$(\HI) in the south to 
reach this sky covering fraction. 
This excess of northern IVCs has already been reported in \HI\ and 
\ion{O}{6} studies \citep{savage2003}.
For HVCs, we find that $f_{\rm sky}$=0.8 is reached at log\,$N$(\HI)=16.5 
in both the north and in the south, so there is less asymmetry for HVCs than 
for IVCs in this low-column regime;
however, there is a tail of high-$N$(\HI) HVC gas extending to log\,$N$(\HI)=20 
in the north that is not present in the south. This is not surprising, as there
are several large, well-known HVC complexes in the North, such as Complex A and Complex C.

The fact that the cumulative $f_{\rm sky}$ of our HVC sample reaches unity indicates that 
we have probed the HVC population in far more depth than past studies using 21\,cm emission data. 
However, we can make a fairer comparison by placing a lower column-density limit to replicate 
the sensitivity of relevant 21\,cm results. \cite{wakker1991} find an HVC covering fraction of 
$18\%$ down to a detection limit of log\,$N$(\HI)=18.3, compared to our $40\%$ result at this 
column density. The more sensitive surveys of \cite{murphy1995} and \cite{lockman2002} probe 
down to log\,$N$(\HI)$\sim17.8$ and both find a $37\%$ covering fraction, compared to our 
$48\%$ result at this column density limit. Our slightly higher covering fractions may be due
to the locations of our sightlines not being entirely random, but with a slight 
bias toward known structures (e.g., five of our sightlines probe Complex C). Incompleteness
in the 21 cm results may also play a role \citep{wakker1997}.

\subsection{The Column Density Distribution Function} \label{subsec:cddf}

\subsubsection{Definition of CDDF} \label{subsubsec:def}

The \HI\ CDDF is defined \citep{tytler1987} as

\begin{equation}
    {\rm CDDF} = f(N) = \frac{d^2 n}{dN(\HI) dX},
\end{equation}

\noindent where $n$ is the number of \HI\ absorbers per column density interval
$d N(\rm \HI)$ per unit pathlength $dX$, where

\begin{equation}\label{eq:dX}
    dX = dv/c,
\end{equation}
with $dv$ equal to the velocity pathlength over which we searched for absorption. This is derived from the cosmological description of co-moving pathlength used by higher redshift surveys and defined as 
$dX=\frac{H_0}{H(z)}(1+z)^2\,dz$, which at $z$=0 reduces to Eq. \ref{eq:dX}.

The CDDF is a fundamental description of the distribution of a sample of QSO absorbers, 
just as the luminosity function is a fundamental description of a sample of galaxies. 
It encodes information on which absorbers host most of the \HI\ mass. 

The first two moments of the CDDF contain important physical information. 
The zeroth moment represents the incidence of absorbers per absorption distance 
$dn/dX$ [sometimes called $l(X)$ in the literature] between two column density 
limits $N_1$ and $N_2$:

\begin{equation}
dn/dX = \int_{N_1}^{N_2} f(N) dN.
\end{equation}

The first moment of the CDDF integrated over path length gives the \HI\ mass density between the two limits.

\begin{equation}
\rho(\HI) = m_{\rm H} (H_0/c) \int_{N_1}^{N_2} Nf(N) dN.
\end{equation}

For a power-law slope $\beta$ where $f(N)=C\,N^{\beta}$, a critical slope of $-$2 exists, above which 
the \HI\ mass budget will be dominated by high-column-density systems, 
so for slopes shallower than --2 the mass will be dominated by
the choice of $N_2$ in the integral. This is because 
$\rho$(\HI)$\propto\int N\,N^{\beta}dN$ so if $\beta$=--2, then 
$\rho$(\HI)$\propto\int N^{-1}dN$=ln\,$N$, which diverges \citep{prochaska2009, danforth2016}. 

\subsubsection{Construction of the CDDF} \label{subsubsec:construct}

We construct the Galactic CDDF by automatically binning the \HI\ column density 
distribution using the maximum of the Freedman Diaconis Estimator or Sturges methods, 
which takes into account both data variability and size.
The number of absorbers in each bin, $n$, is then divided by the column density 
size of each bin, $\Delta N(\HI)$, multiplied by the pathlength searched for absorption, $dX$.

The pathlength $dX$ effectively sets the vertical normalization of the CDDF, but also encodes 
key information concerning the incompleteness of our sample. Each absorption component obscures 
a section of the spectrum and thus reduces the pathlength over which additional components can 
be detected. To account for this incompleteness, we have simulated the observability of 
additional components in each sightline.

\begin{figure}[t!]
\centering
 \includegraphics[width=\linewidth]{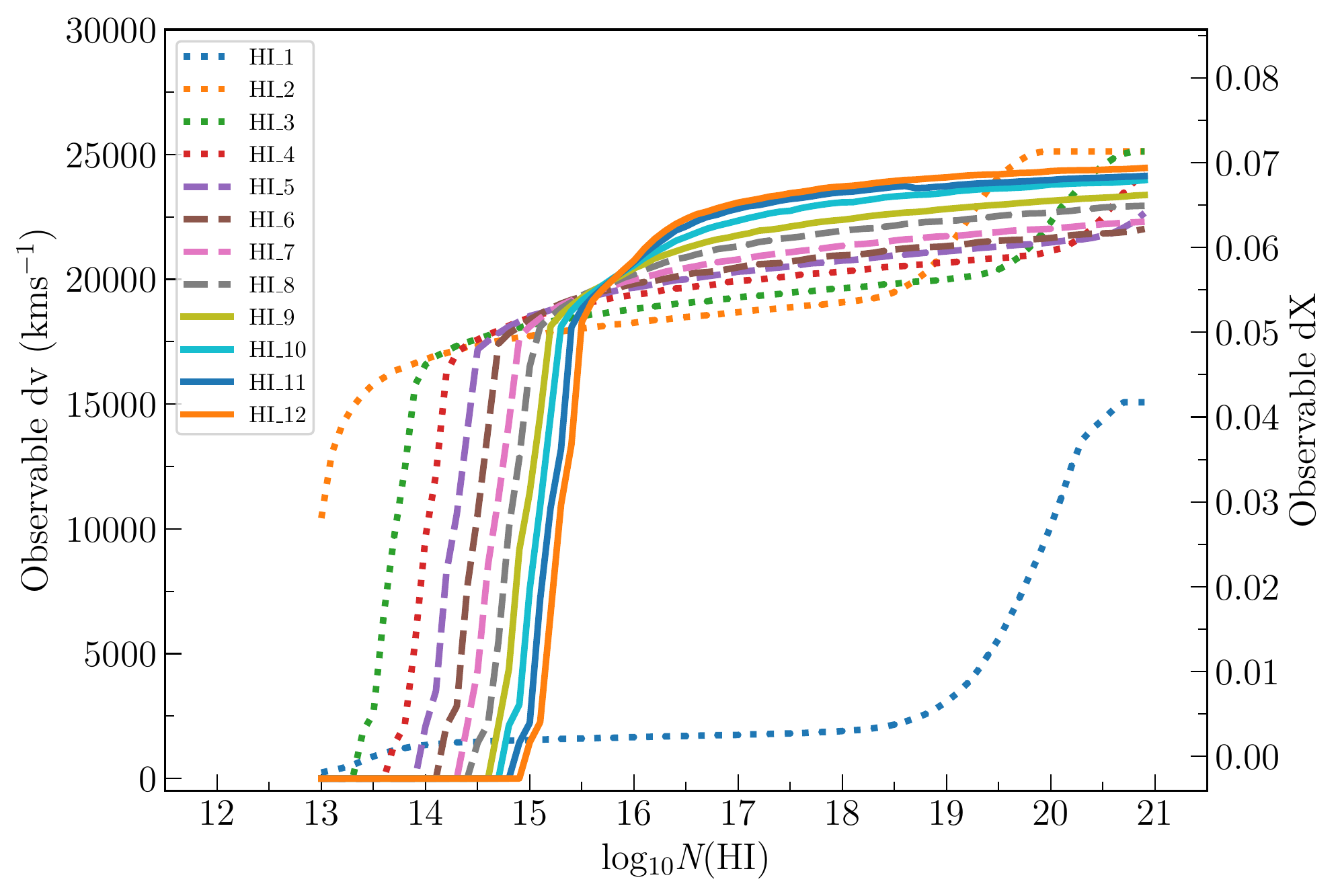} 
  \caption{\small{The observable pathlength ($dv$ on the left axis and $dX$ on the right axis) of 
  \HI\ components as a function of column density for each Lyman series transition. The upper envelope
  of this family of curves defines the total pathlength used for our CDDF calculations.}}
  \label{fig:dX_incompleteness}
\end{figure}

Our observable-pathlength simulation works by first recreating the 
absorption profile we fit for each target, then adding an additional component to the 
profile from a grid of column densities and velocities spanning the entire observable range,
i.e., column densities from 13$\leq$log\,$N$(\HI)$\leq$21, and velocities from 
$-500\leq v_{\rm LSR} \leq 500$ \kms. This modified profile is then subtracted from the 
original to yield a residual $R$, which represents the impact of the additional component 
on the line profile. 
If the maximum value of the residual is not significant at 3$\sigma$,
i.e., if $R_{\rm max}<3/(S/N)$, then we would not be able to detect 
that component if it were present, because it would be lost in the noise,
so we declare such a component unobservable. For example, 
in a spectrum with S/N=20, we can detect components that create residuals at the 15\% level, 
whereas if S/N=10, we can only detect residuals at the 30\% level, and if S/N=5, we are only sensitive
to 60\% residuals. The output of this 
simulation is a function for each Lyman series transition
that gives the observable pathlength $dX$ as a function of column density, 
taking into account the noise properties of each individual spectrum.
Figure \ref{fig:dX_incompleteness} shows the resulting pathlength $dX$ summed for the entire sample.
The pathlength used in our CDDF calculations is the upper envelope drawn 
from this curve, i.e. the Lyman line that gives the maximum pathlength at a given $N$(\HI).
This method ensures that each CDDF bin is individually normalized vertically 
to account for the incompleteness due to velocity shielding within each column density interval.

Finally, we calculate errors in each CDDF bin via a bootstrap resampling method. 
This method randomly resamples and rebins the entire distribution
10,000 times, where each component is itself drawn from the normal distribution 
defined by its VoigtFit measurement errors (a quadrature sum of the statistical 
and systematic fitting errors shown in Column 7 of Table \ref{table:fuse_results}). The final error for each bin is then the 16th and 84th percentile ($\pm 1\sigma$) of the
resampled distribution. This method thus takes into account both measurement and sampling uncertainties.

\subsubsection{Fitting the CDDF}\label{subsubsec:fitting}

We perform power-law fits to the CDDF using the affine-invariant 
Markov Chain Monte Carlo 
(MCMC) Ensemble sampler \emph{Emcee} \citep{Foreman_Mackey_2013}. \emph{Emcee} implements an 
MCMC sampling of likelihoods across the parameter space using an ensemble of walkers, 
each of which evolves for a number of steps to sample the posterior distribution. 
Our best fits are then reported as the median, 16th, and 84th percentiles ($1\sigma$) of 
the posterior distributions. Because the 17$\lesssim$log\,$N$(\HI)$\lesssim$19 region is
affected by saturation, we sum all components in this region into a single, larger bin prior to 
fitting the CDDF to help minimize the effect of highly uncertain components. 
All of the final fits are performed with this
larger bin for the saturated absorbers, although we also display the original, finer-binned data points in Figures 
\ref{fig:cddf} and \ref{fig:cddf_comp_fits} for completeness.

We also investigate the possibility of breaks in the full CDDF as a function of column 
density only. Figure \ref{fig:cddf_comp_fits} presents the CDDF fit with a 
three-component, piecewise, power-law function. The fitting process is again carried 
out using \emph{Emcee}, but with six free parameters representing the $y$-intercept, 
slopes for the three segments, and the locations of the two break points. We perform this fit
both for our complete sample of absorbers (``Galactic Halo + Disk") as well as a ``Galactic Halo" only version, which excludes the ISM components (i.e., $|v_{\rm LSR}| < 40$ \kms).
We find a best fit with breaks at log\,$N$(\HI) = \lowbreak\ (\lowbreakhalo) and \highbreak\ (\highbreakhalo) for the Galactic halo and disk (Galactic halo only). The results of these MCMC fits
can be found in Appendix \ref{section:fits} (Figure \ref{fig:corner}) displayed as corner plots.
The advantage of this piecewise method is that the break points are identified 
automatically in a statistically rigorous manner, rather than being inserted manually.

\subsubsection{Displaying the CDDF}\label{subsubsec:plotting}

\begin{figure*}[ht!]
\centering
 \includegraphics[width=0.495\linewidth]{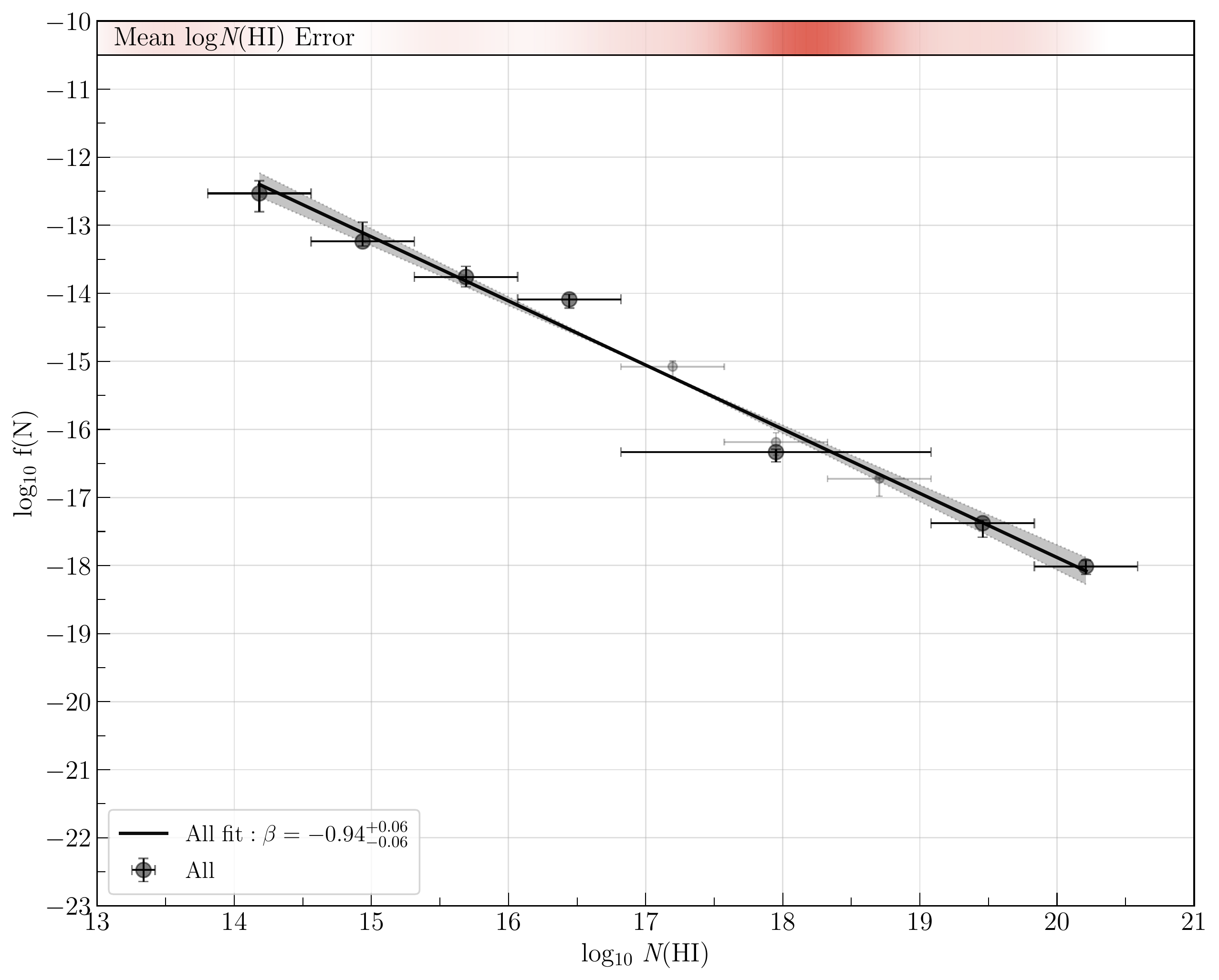} 
 \includegraphics[width=0.495\linewidth]{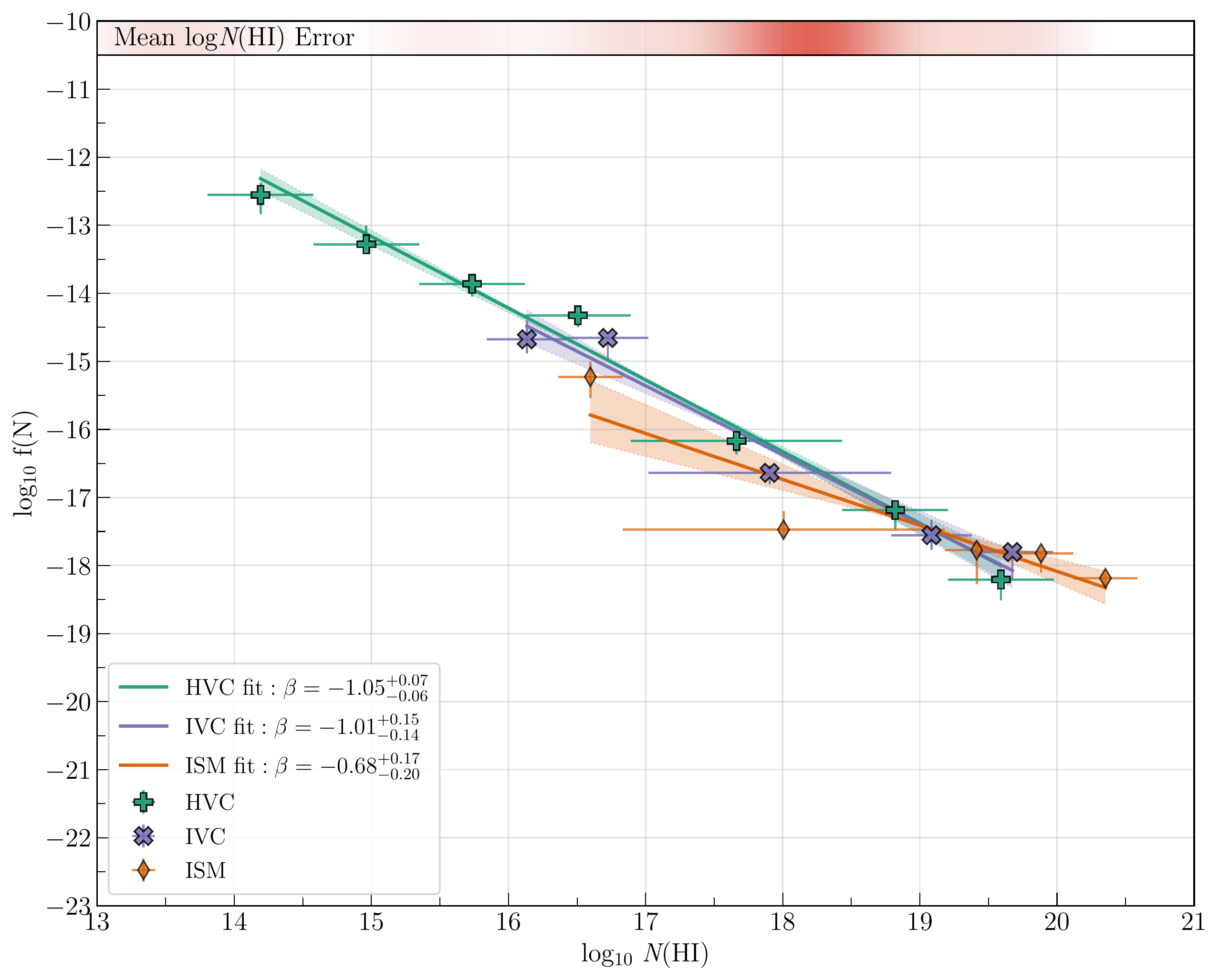}
  \caption{\small{{\bf Left:} The CDDF of \HI\ in the Galactic halo as observed by \fuse. 
  Components at all velocities are included. The best-fit power law is overplotted in solid-black, 
  with the $1\sigma$ fit errors shown in grey, and the slope indicated in the legend. 
  The three light-grey points indicate the bins that have been summed
  into a single, large bin across the high-error (saturated) region. 
  {\bf Right:} The \fuse\ \HI\ CDDF split into three velocity bins: 
  HVCs (green pluses), 
  IVCs (purple crosses), 
  and ISM components (orange rhombuses). 
  Each velocity bin is fit separately, with color coding matching the underlying symbols. 
  {\bf All: } The vertical error bars are 1-sigma quantiles calculated via bootstrap resampling, and the 
  horizontal error bars indicate the column-density width of each bin.
  The red-shaded horizontal bar illustrates the mean component fit-error as a function of column density as displayed in Figure \ref{fig:e_logNHI}.}}
  \label{fig:cddf}
\end{figure*}

\begin{figure*}[!ht]
\centering
 \includegraphics[width=0.495\linewidth]{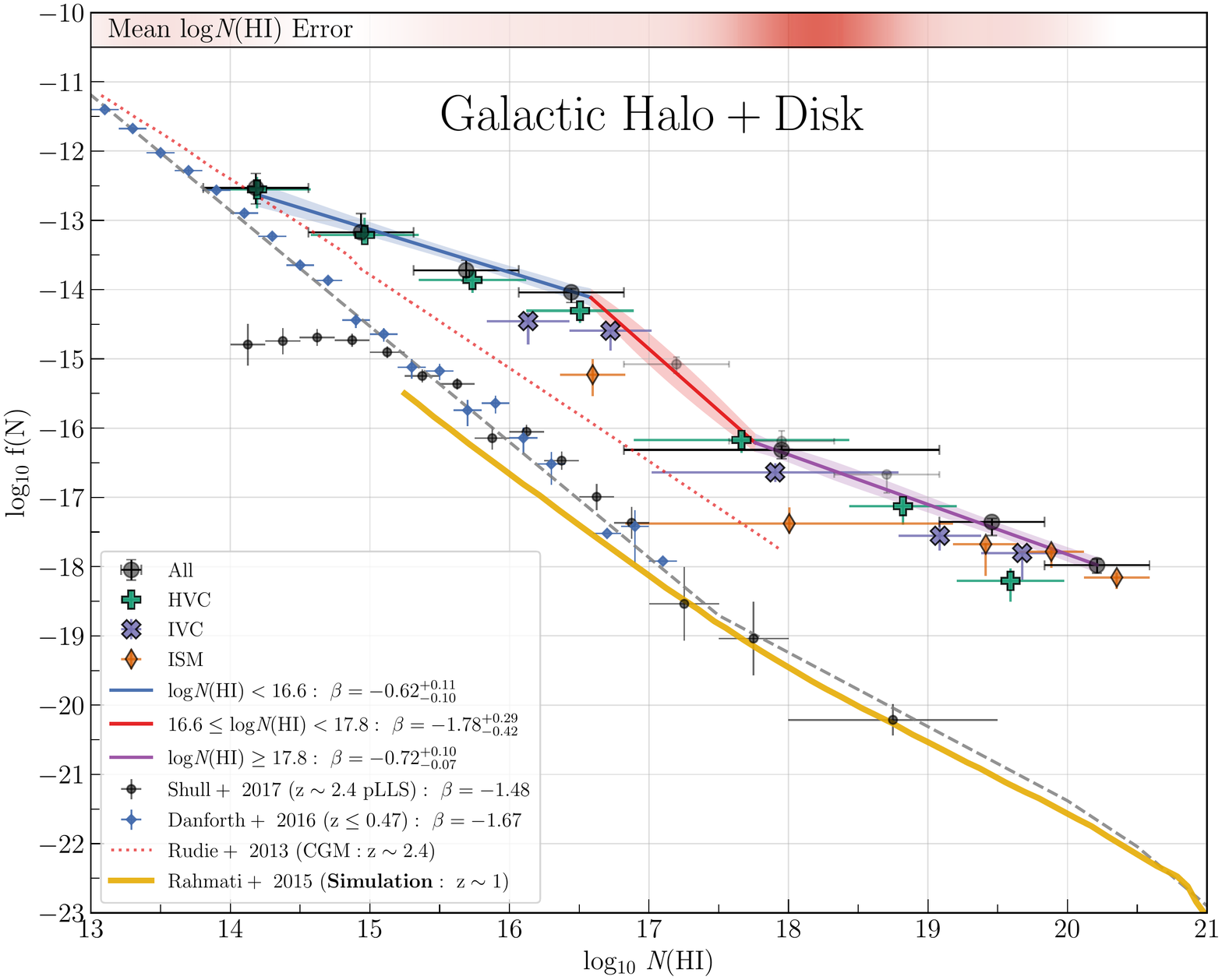}
 \includegraphics[width=0.495\linewidth]{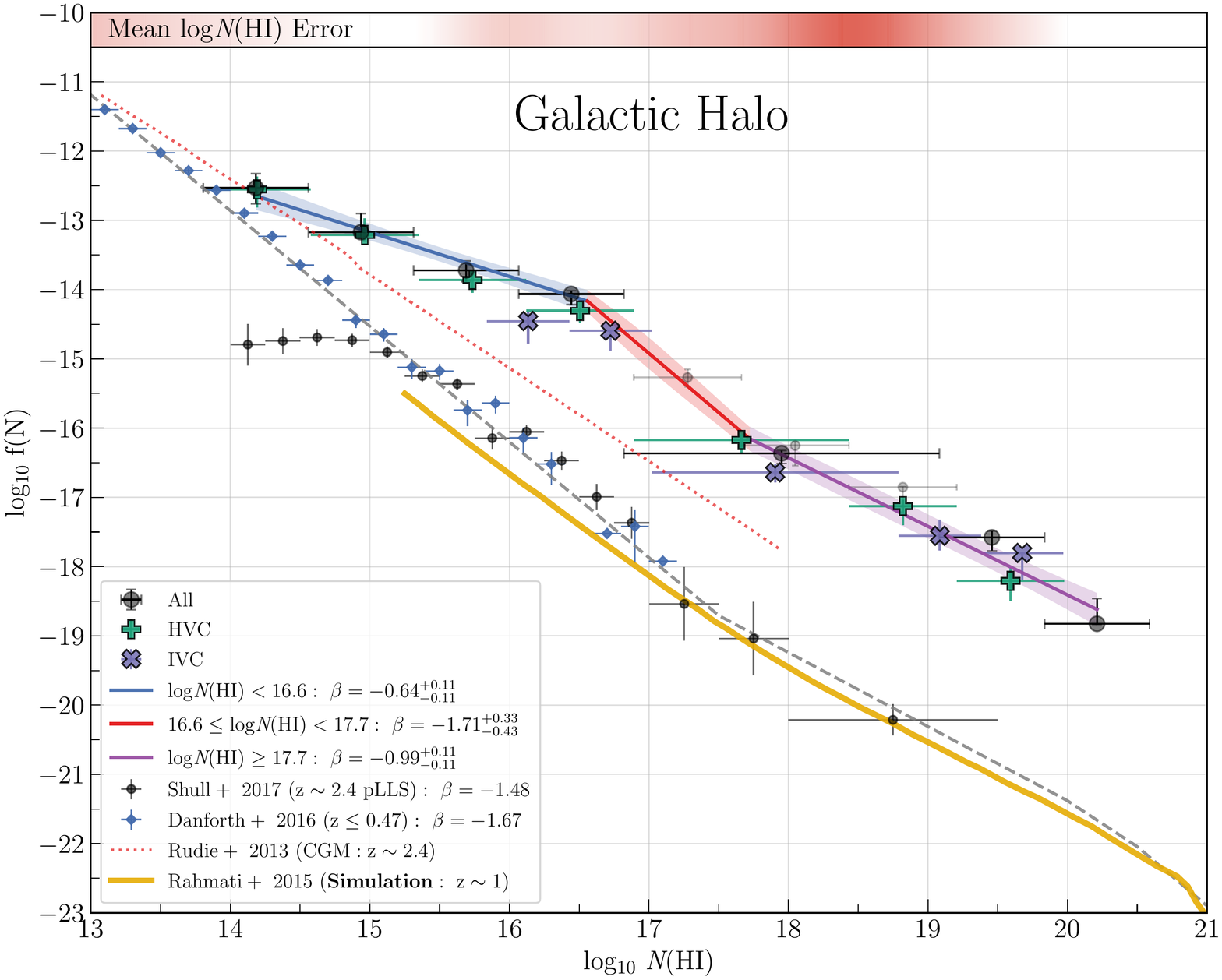}
  \caption{\small{\textbf{Left:} The complete \fuse\ \HI\ CDDF plotted together with a selection of extragalactic CDDFs from the literature including observations from \citet{omeara2013} (grey-dashed), 
  \citet{rudie2013} (red-dotted), \citet{danforth2016} (blue-diamonds), \citet{shull2017} (black-circles), and a simulation from \cite{rahmati2015} (solid-gold). Our 3-component 
  piecewise power-law fit to the \fuse\ data is shown for the three autonomously chosen $N$(\HI) regimes:
  log\,$N$(\HI)$\leq$16.6 (solid-blue), 16.6$<$log\,$N$(\HI)$\leq$17.8 (solid-red), and log\,$N$(\HI)$>$17.8 (solid-purple), all fitted simultaneously, and with slopes indicated in the legend. 
  The three small light-grey points show the bins that have been summed 
  into a single, large bin across the high-error (saturated) region.
  The vertical error bars are the 1-$\sigma$ quantiles calculated via bootstrap resampling, and the 
  horizontal error bars indicate the column density width of each bin.
  The red-shaded horizontal bar illustrates the mean component fit-error as a function of column density as displayed in Figure \ref{fig:e_logNHI}.
  \textbf{Right:} Same as on the left, but now only including FUSE \HI\ components with $\rm v_{LSR} \ge 40$ \kms\ (i.e., IVCs and HVCs only).}}
  \label{fig:cddf_comp_fits}
    \vspace{0.0pt}
\end{figure*}

We display the \HI\ CDDF in several formats to highlight different features of the distribution.
First, in the left panel of Figure \ref{fig:cddf} we show the global \HI\ CDDF 
(over all velocities) fit with a single power-law. 
Second, in the right panel of Figure \ref{fig:cddf}
we show a version split by velocity into 
HVC, IVC, and ISM bins, each fit with a separate power-law. 
Third, in Figure \ref{fig:cddf_comp_fits} we show the \HI\ CDDF together with the 
three-component piecewise power-law function described in Section~\ref{subsubsec:fitting}, 
split into Galactic halo and disk (all velocities) and Galactic halo only 
($\rm v_{LSR} \ge 40$ \kms) versions. The key difference of Figure \ref{fig:cddf_comp_fits} 
is that it breaks the sample down by \HI\ column density, rather than by velocity. In addition, 
Figure \ref{fig:cddf_comp_fits} includes a comparison of 
the Galactic CDDF to a selection of extragalactic CDDFs from the literature, including results from \citet{omeara2013}, \citet{rudie2013}, \citet{danforth2016}, and \citet{shull2017}, and simulation results from \citet{rahmati2015}.


We emphasize that the extragalactic CDDFs are derived from blind surveys, whereas the 
Galactic CDDF is derived from our vantage point inside the Milky Way, which is not a 
random position. That is why the Galactic CDDF has a much higher normalization than 
the extragalactic CDDF -- we are not in an unbiased location. Of more physical interest 
is the difference in slope. The slopes we measure from the various forms of the \HI\ 
CDDF are summarized in Table~\ref{tab:cddf-slopes}. A discussion of these slopes
is given in Section~\ref{subsubsec:cddf-slope}. Each figure also includes a 
red-shaded bar across the top edge which indicates the mean
component error as a function of column density (derived from Figure \ref{fig:e_logNHI}). 
This gives a visual indication of which regions are most affected by saturation and
blending issues.

\begin{deluxetable}{ll}[!t]
\tablewidth{0pt}
\tablecaption{Summary of CDDF Slopes}
\tablehead{Sample & $\beta$}
\startdata
All \HI\ & \allbeta    \\
ISM ($|v_{\rm LSR}|<40$\kms) & \ismbeta    \\
IVC ($40 \leq |v_{\rm LSR}|<100$\kms) & \ivcbeta   \\  
HVC ($|v_{\rm LSR}|\ge100$\kms) & \hvcbeta \\
\hline
\hline
Galactic Halo & \\
\hline
log\,$N$(\HI)$\gtrsim$\highbreakhalo & \highbetahalo\\ 
\lowbreakhalo$\lesssim$log\,$N$(\HI)$\lesssim$\highbreakhalo & \midbetahalo\\ 
log\,$N$(\HI)$\lesssim$\lowbreakhalo & \lowbetahalo\\ 
\hline
\hline
Galactic Halo + Disk & \\
\hline
log\,$N$(\HI)$\gtrsim$\highbreak & \highbeta\\ 
\lowbreak$\lesssim$log\,$N$(\HI)$\lesssim$\highbreak & \midbeta\\ 
log\,$N$(\HI)$\lesssim$\lowbreak & \lowbeta\\ 
\enddata 
\tablecomments{This table reports the best-fit power-law slopes of \HI\ CDDF as shown in Figures~\ref{fig:cddf} and \ref{fig:cddf_comp_fits}, where $f(N)=C\,N^{\beta}$.} 
\label{tab:cddf-slopes}
\end{deluxetable}

\subsubsection{Slope of the CDDF}\label{subsubsec:cddf-slope}

In Figure \ref{fig:cddf} (left), we find that a single power-law with a slope of $\beta$=\allbeta\ 
provides a reasonable fit to the full CDDF over six orders of magnitude from log\,$N$=14--20. The fit is not perfect however, as several data points lie at or beyond 1$\sigma$ from the 
line of best fit, particularly near log\,$N$(\HI)$\approx$16.5 where there is a marked excess 
of IVCs and HVCs.

Next, in Figure \ref{fig:cddf} (right), we individually fit the ISM, IVC, and
HVC subsamples each with a single powerlaw.
We find that IVCs and HVCs have nearly indistinguishable CDDF slopes, with 
$\beta_{\rm IVC}$=\ivcbeta\ and $\beta_{\rm HVC}$=\hvcbeta,
in line with the basic similarities we noted from their number counts seen in 
Figure \ref{fig:N_hist}. Conversely, the ISM components (which are heavily biased toward 
higher \HI\ column densities) show a shallower slope, $\beta_{\rm ISM}$=\ismbeta. This shallower
slope is largely due to a lack of ISM components in the 17$\lesssim$log\,$N$(\HI)$\lesssim$19
region, where saturation and velocity blending make it difficult to detect components.

These differing CDDF slopes may indicate that the ISM \HI\ shows a different behavior than 
ISM and HVC \HI, which perhaps reflects the different physical conditions in the disk and 
halo. However, the ISM fit in particular does not appear to capture the detailed structure
of the CDDF data points.
While the high-column end of the ISM is quite shallow, there appears to be a steepening
below log\,$N$(\HI)$\approx$18, evidenced by the last ISM point more closely matching the 
IVCs and HVCs. Hence, while a difference likely exists between the high- and low-velocity
regimes, the overall shape of the CDDF may be mostly a function of column density.

An alternative model for the full CDDF is a three-component piecewise power-law 
as shown in Figure \ref{fig:cddf_comp_fits}, in which the CDDF is fit over three separate
column density regimes.
We include two versions of this fit. First, for the complete survey including absorption
at all velocities (Figure \ref{fig:cddf_comp_fits} left: ``Galactic Halo + Disk"), and second for
just $v_{LSR} \ge 40$ \kms\ velocity absorption (Figure \ref{fig:cddf_comp_fits} right: ``Galactic Halo").
As mentioned earlier, our unique location within the Galaxy means that our sightlines include gas from
the Galactic disk as well as the halo. Extragalactic CDDF results, however, mostly consist of halo or IGM
material far from galaxies, so our ``Galactic halo" CDDF may provide a more fair comparison to these studies.

It is worth noting that the slope of the Galactic halo and disk CDDF is similar at the low end
[$\beta_{\rm low}$=\lowbeta\ for log\,$N$(\HI)$\lesssim$\lowbreak]
and high end 
[$\beta_{\rm high}$=\highbeta\ for log\,$N$(\HI))$\gtrsim$\highbreak], on either side of the saturation zone. So if we were to interpolate between the low and high ends, \emph{we would recover a slope similar to what is actually observed in the intermediate region where saturation is important}. This indicates that despite the large uncertainties on the individual saturated components, their effect on the CDDF is minor, because we would have derived a similar global CDDF slope even if we had excluded them from the sample entirely. 

To further test this point, we performed all of the CDDF fits again after excluding all components with errors
larger than 1 dex. Doing so results in almost the same CDDF slopes as before: 
$\beta_{\rm all\ \leq 1 dex} = -0.96^{+0.07}_{-0.07}$, 
$\beta_{\rm low\ \leq 1 dex} = -0.66^{+0.11}_{-0.11}$, 
$\beta_{\rm mid\ \leq 1 dex} = -1.86^{+0.29}_{-0.38}$, 
$\beta_{\rm high\ \leq 1 dex} = -0.66^{+0.11}_{-0.08}$. This occurs because the highly uncertain components
are much less common than the well-constrained components and are spread fairly evenly across a wide range of \HI\ column density. The finding that the highly uncertain data points in our sample do not substantially change the derived slopes of the CDDF indicates that our results are robust against the presence of a minority of uncertain components. When excluding the low-velocity ISM components 
for our Galactic halo only version, the high column end rises to $\beta_{\rm high} \sim -1$, 
as expected from the individual IVC and HVC powerlaw fit results.

\subsubsection{Discussion of the CDDF}\label{subsubsec:cddf-disc}

Our survey has extended our knowledge of the Galactic \HI\ distribution by four orders of magnitude,
pushing down from the limits of 21\,cm-based \HI\  
\citep[which reach log\,$N$(\HI)$\approx$18;][]{wakker1991, murphy1995, lockman2002, moss2013, westmeier2018}
down to our UV limit of log\,$N$(\HI)$\approx$14.

Our measured near-unity slope for the full-sample CDDF ($\beta$=\allbeta)
is much shallower than the critical slope of $-$2, telling us that the \HI\ mass 
is dominated by high $N_{\rm H\,I}$ systems. 
This measured slope is also significantly shallower than extragalactic results
from blind surveys, which tend to find CDDF slopes with $\beta =$ -1.4 to -1.7
 \citep{kim2002b, lehner2007, prochaska2010, prochaska2014, ribaudo2011, tilton2012, kim2013, omeara2013, rudie2013, danforth2016, shull2017}. Hence, the distribution of \HI\ in the Galactic halo appears to be tipped 
toward higher \HI\ columns. This result may not be unique to our own Galaxy: 
although most extragalactic CDDF measurements are composed of absorbers in the 
IGM and extended CGM, far from galaxies, \cite{rudie2013} constructed CDDFs 
from \HI\ both ``near" and ``far" from galaxies, and found shallower power-law 
slopes for absorbers within 300 proper kpc and 300\kms\ of galaxies. 
This supports the idea that galaxy environments lessen the slope of the CDDF
and host different absorber populations than the IGM, a result
also supported by metal-line analyses \citep{richter2009}.

Moving to our alternative piecewise power-law model,
the high-column-density end (log\,$N$(\HI)$\gtrsim$\highbreak)
has a slope $\beta=$\highbeta\ (or $\beta=$\highbetahalo\ for the Galactic halo only),
which aligns well
with extragalactic results from \cite{ribaudo2011}, who report 
$\beta = -0.8_{-0.1}^{+0.3}$ for their SLLS regime of
$19.1 \leq {\rm log}\,N(\HI) \leq 20.2$. However, this value is significantly
shallower than results from Galactic 21 cm studies, such as those found by 
\cite{putman2002}, and \cite{moss2013}, who find $\beta\sim-2$.
This could be the result of several competing effects, including beam-smearing,
velocity resolution, and differences in definition between absorption lines and
emission peaks. In a future paper we will compare in detail our \HI\ results
to those from a variety of 21 cm observatories in an attempt to untangle these
effects.

Our intermediate column-density region, 
\lowbreak $\leq {\rm log}\,N(\HI) \leq$\highbreak\ 
shows a 
significantly steepened slope of $\beta$=\midbeta\ 
($\beta=$\midbetahalo\ for the Galactic halo only). 
A steepening below log\,$N(\HI) \lesssim 17.5$ has also been reported in blind surveys at 
$z>1$ by \cite{prochaska2010}, \cite{ribaudo2011}, \cite{omeara2013}, and \cite{rudie2013}.
At lower redshift \citet{shull2017} found a slightly shallower slope of $\beta = -1.48 \pm 0.05$
for partial Lyman Limit Systems (pLLS), 
although \citet{danforth2016} also found evidence of $\beta$ steepening with redshift between $0.01 < z<0.47.$

The steepening in the Galactic CDDF may be related to ionization:
the optical depth to ionizing radiation reaches $\sim$1 at log\,$N(\HI)\approx17.2$,
so absorbers below this threshold are expected to be significantly ionized. However,  
this effect would be expected to lead to a \emph{flattening} of the \HI\ slope, 
not a steepening, so this does not explain the observed excess of clouds in this range. 
This excess is visible in the raw component 
distributions shown in Figure~\ref{fig:N_hist} as well,
so it is not related to the way the CDDF is constructed. 
Recent theoretical work on the size scale of fragmenting, pressure-confined
cool gas favors characteristic columns of
log\,$N$(\HI)=17 \citep{mccourt2018} to log\,$N$(\HI)=18 \citep{gronke2018},
so this is one potential explanation for the observed abundance of these absorbers and
resulting uptick in the CDDF slope. 
However, we caution that this intermediate column density region also corresponds 
to the regime where 
saturation effects and fitting errors are the highest (as illustrated by the red shading
in the mean log\,$N(\HI)$ error bar along the top edge of each CDDF figure). 
The transition to unsaturated absorption at the low edge of this region could also
cause some fitting artifacts, which could be contributing to the distribution peak.
Thus, it is possible the break at log\,$N(\HI) \sim 18$ is exaggerated by fitting errors. 
That being said, the 
fact that a majority of other 
CDDF studies \emph{also} see a slope change near this region (as seen in 
Figure \ref{fig:cddf_comp_fits}) does provide reason to believe that a slope change is real.
Indeed, we note that the overall shape of the CDDF from
\lowbreakhalo\ $\lesssim$ log\,$N$(\HI) $\lesssim$ \highbreakhalo\ for the Galactic halo-only model matches well
with the extragalactic CDDFs, as seen in Figure \ref{fig:cddf_comp_fits}.

In the low-column-density regime (log\,$N$(\HI)$\lesssim$17),
we find a significantly shallower slope 
($\beta$=\lowbeta) than the rest of the Galactic CDDF and extragalactic CDDF results
(e.g., \citealt{kim2002b}, \citealt{lehner2007}, \citealt{tilton2012, danforth2016}). 
This flattened slope is likely due to a combination of incompleteness and 
physical differences between the Galactic halo and the Ly$\alpha$ forest environments
probed by other UV surveys.

In this lower column density range, low-$z$ extragalactic absorbers are generally far from 
galaxies and reside in dark-matter filaments. They have lower densities and higher
ionization parameters than Galactic halo gas (e.g., \citealt{collins2005, shull2011}).
Conversely, the low column density absorbers in our sample are all IVCs and HVCs,
which are likely relatively close to the Galaxy: \citealt{richter2017} find
$d\la2$\,kpc for IVCs, and \citealt{lehner2011} find $d\approx12$\,kpc for HVCs, although the
lowest column density clouds are less well constrained. 
This difference in environment and location also results in an ionization effect, in which our 
Galactic \HI\ sample is exposed to a significantly enhanced radiation field compared to 
the extragalactic sample, because the Galactic
ionizing radiation field dominates the extragalactic UV background within $\approx$50\,kpc 
\citep{giroux1997, blandhawthorn1999, fox2005}.
This ionization effect means that hydrogen exists preferentially in the form of \ion{H}{2}
toward lower $N$(\HI), which would lead to a depression in the relative amount of 
low column density \HI. Hence, it is not altogether unexpected for the low-column density 
end of the Galactic CDDF to differ
from that found in extragalactic surveys.

The other potential effect is incompleteness: the inability to detect some components 
because they are hidden in the data beneath our ability to detect them.
This effect increases toward lower $N$(\HI), since weaker components are easier to hide.
We have carefully accounted for the effect of reduced pathlength $dX$ for lower $N$(\HI)
absorbers using our observable-pathlength simulation (Section~\ref{subsubsec:def}), 
so we have dealt with incompleteness as much as is possible,
but we are likely still incomplete in the low $N$(\HI) regime because the 
\emph{velocity density} of absorbers (number per unit velocity)
is not uniform across $dX$.
We most readily detect weak 
(log\,N(\HI) $\lesssim$ \lowbreak) absorbers at high velocities ($|v|\gtrsim100$\kms), 
where the contamination from ISM gas is lower, but conversely find that the density of 
absorption across all $N$(\HI) is highest at low velocities ($|v|<100$\kms).

If there 
was a high velocity-density of weak absorbers at low velocities, we would be insensitive 
to them, and thus they would be missing from our statistics. This bias has been
estimated in recent simulations of the Galactic CGM \citep{zheng2020}, but 
observational results from the QuaStar project on \emph{HST} have found surprisingly 
little missed low-velocity gas \citep{bish2021}. 
This discrepancy could be explained by invoking a flared disk or flattened halo model,
in which the majority of gas is located at low latitudes along an extended disk 
\citep[e.g.][]{Qu_2020}. 
In this case, high latitudes sightlines, such as in our survey, would miss the most 
significant reservoir of Galactic halo gas. There is indeed some observational 
evidence supporting such flattened halos in extragalactic \HI\ studies (\citealt{french2017})
and several studies of rotating cold gas in the CGM \citep{Ho_2017, Martin_2019, French_2020}

An alternative approach to quantifying the incompleteness is to use upper limits 
on the CDDF slope to back out an absorber distribution. If we extend the 
intermediate $N_{\rm H\,I}$ region of our fit out to ${\rm log}\,N(\HI) = 14$, 
closely approximating the slopes found by \cite{omeara2013}, \cite{prochaska2014}, 
\cite{rahmati2015}, and others, we calculate we would have to have missed over 800 
low-column density components, or $\sim$33 absorbers per sightline,
with a distribution peaking around ${\rm log}\,N(\HI) = 14.5$).
While we cannot completely rule out such a large missed distribution, we emphasize that they would be forced to live entirely in the $|v_{\rm LSR}| \leq 100$ \kms\ region. Beyond this velocity in the HVC range, only a handful of additional components could be hidden. Thus, while the \emph{overall} low-column slope could be steeper than our best fit, the HVC-only CDDF results are relatively robust. Furthermore, the similarity between our HVC and IVC CDDFs indicates that the $40 \leq |v_{\rm LSR}| \leq 100$ \kms\ velocity range is not strongly affected by incompleteness either.


\subsection{Inflow vs Outflow} \label{subsec:flow}

Our dataset allows us to assess the relative prevalence of inflowing and outflowing gas.
Following \citet{fox2019}, we approximate clouds with $v_{\rm GSR}<0$ as inflowing and those 
with $v_{\rm GSR}>0$ as outflowing, thereby splitting the observed cloud population into two.
This may artificially identify some clouds as inflowing or outflowing depending on their distance and Galactic position, but we expect these imperfections to average out across the aggregate distribution (see, e.g., \citealt{wakker1991, kalberla2006, peek2008}). This assumption is only valid for distant clouds, so is more appropriate for HVCs than IVCs.

\begin{figure}[!ht]
\centering
 \includegraphics[width=\linewidth]{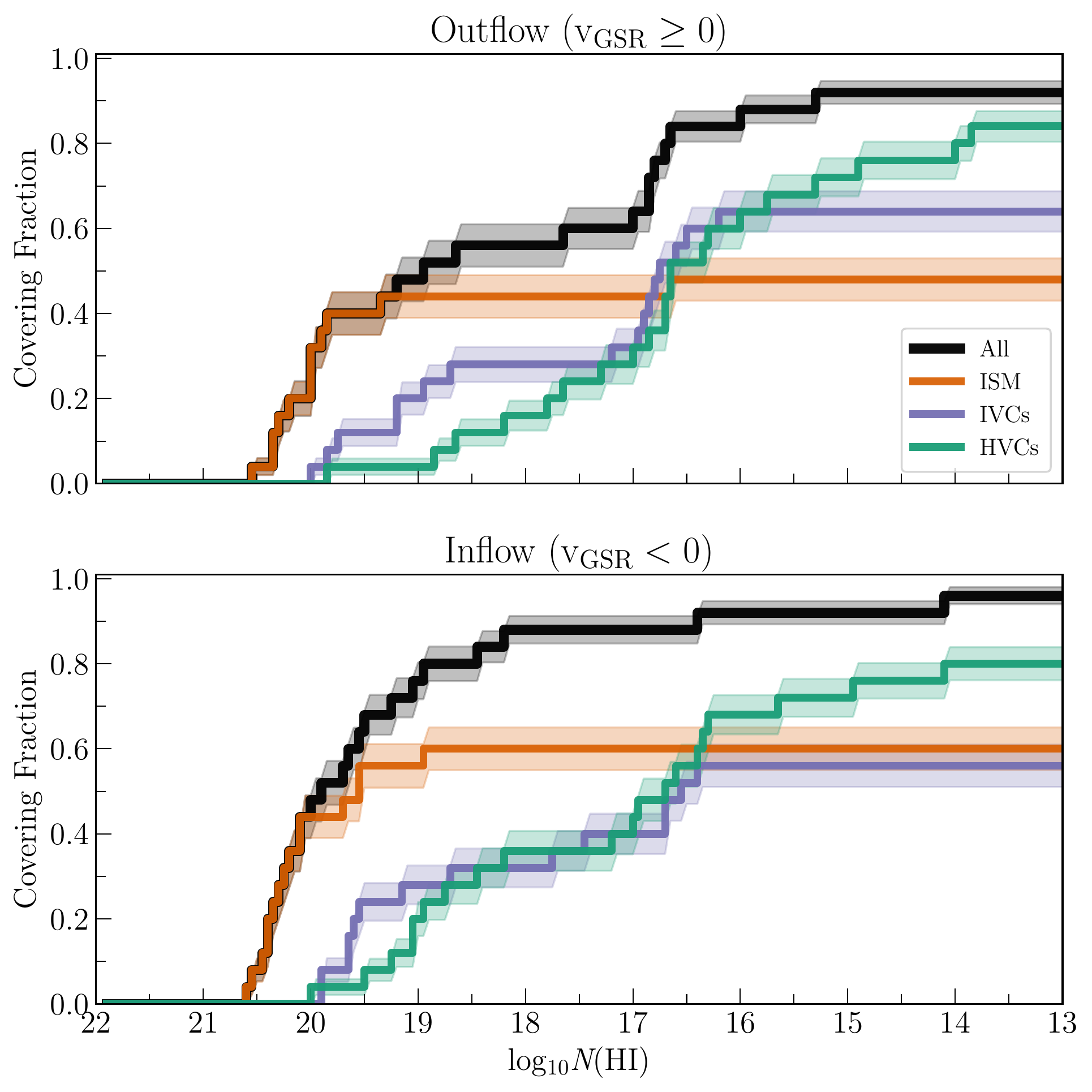} 
  \caption{\small{The \HI\ sky covering fraction split into inflowing and outflowing subsets, for all absorbers (black) and for each velocity bin (colored lines). The shaded regions around each line show the 1$\sigma$ error regions derived from a bootstrap re-sampling routine.}}
  \label{fig:f_c_vgsr}
  \vspace{1pt}
\end{figure}

Compared with \citet{fox2019} we have a significant advantage,
because whereas that metal-based survey relied on a metallicity correction, 
the \HI\ survey results presented here rely on no such correction, 
since we observe the neutral baryons directly.
We calculate the sky covering fraction separately for inflowing and outflowing samples, 
as shown in Figure \ref{fig:f_c_vgsr}, finding that the outflowing gas covers less of the 
sky than inflowing gas, with a large jump in covering fraction below log\,$N$(\HI)=17. 
This excess of inflowing gas is seen independently in the IVC and HVC sub-samples, as 
well as in the full sample.

\begin{figure}[!ht]
\centering
 \includegraphics[width=\linewidth]{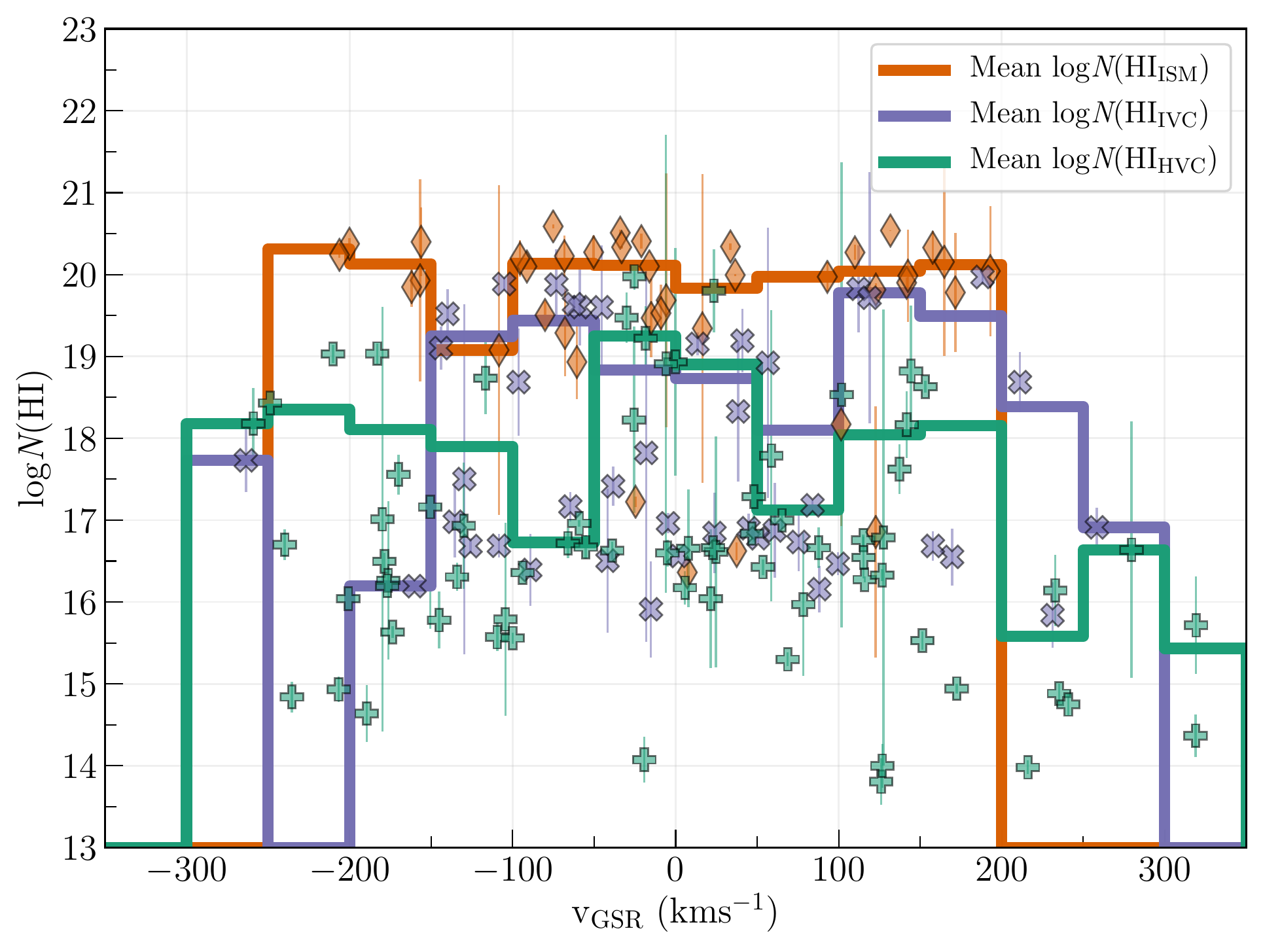} 
  \caption{\small{The \HI\ column densities as a function of GSR velocity for all absorbers in our sample. Histograms showing the mean column density within each 50\kms\ bin of $v_{\rm GSR}$ are overplotted for the ISM, IVC, and HVC velocity ranges, allowing the inflowing ($v_{\rm GSR}<0$) and outflowing ($v_{\rm GSR}>0$) populations to be identified.}}
  \label{fig:vel_vs_N}
  \vspace{1pt}
\end{figure}

We can then compare the mean 
\HI\ column density of the inflowing and outflowing clouds,
and look for an excess of one versus the other.
We perform this exercise in Figure~\ref{fig:vel_vs_N}, which shows the \HI\ column density 
of each component as a function of $v_{\rm GSR}$, with histograms overplotted showing 
the mean log\,$N$(\HI) in 50\kms\ bins for each of the ISM, IVC, and HVC categories.

In order to calculate the \HI\ mass and mass flow rate represented by our
absorbers, we adopt a partially-covered spherical shell model following \citet{fox2019}.
This calculation should be viewed as a back-of-an-envelope estimation rather than a precision calculation; nonetheless, the results still offer useful physical insight into the relative strength of inflow and outflow. 
In this model, the halo gas clouds are located in a shell at distance $d$ with
a sky covering fraction $f_{\rm sky}$ and mean \HI\ column $\langle N$(\HI)$\rangle$.
Then the total \HI\ mass in inflowing and outflowing clouds can be written as:

\begin{eqnarray}
    M_{\rm in} & = & 1.4m_{\rm H}f_{\rm sky, in} \langle N_{\rm H\,I, in} \rangle\,4\pi d^2 \hspace{0.25cm}{\rm and} \label{eqn:mass}\\
    M_{\rm out}& = & 1.4m_{\rm H}f_{\rm sky, out}\langle N_{\rm H\,I, out}\rangle\,4\pi d^2,
    \label{eqn:mass2}
\end{eqnarray}

\noindent where $m_{\rm H}$ is the mass of the hydrogen atom, 
the factor 1.4 accounts for the mass in helium and metals, 
and the covering fractions are assumed to be independent of distance. 
We then combine $M_{\rm in}$ and $M_{\rm out}$ with the mean flow velocities 
$\langle v_{\rm in}\rangle$ and $\langle v_{\rm out}\rangle$
to derive the mass flow rates $dM/dt$ for the inflowing and 
outflowing cloud populations:

\begin{eqnarray}
    dM_{\rm in}/dt & = & M_{\rm in}\langle v_{\rm in} \rangle/d \hspace{0.25cm}{\rm and}
    \label{eqn:massflow}\\
    dM_{\rm out}/dt& = & M_{\rm out}\langle v_{\rm out} \rangle/d.
    \label{eqn:massflow2}
\end{eqnarray}

Each of the terms on the right hand side of Equations~\ref{eqn:mass} to
\ref{eqn:massflow2} can be directly constrained by our observations, except the distance, which
we take from literature values:
We adopt a mean distance to HVCs of 12$\pm$4\,kpc from \citet{lehner2011} for both 
the inflowing and outflowing populations. We caution that this distance is highly uncertain, 
particularly for clouds with low \HI\ column densities.

\begin{deluxetable*}{l c c c r l r}[!ht]
\tablewidth{0pt}
\tablecaption{Inflow vs Outflow: \HI\ Masses and Mass Flow Rates}
\tablehead{Sample & 
log\,$N$(\HI) \tm{a} &
$\langle$ log\,$N$(\HI) $\rangle$\tm{b} & 
$f_{\rm sky}$ &
$\langle v_{\rm GSR}\rangle$\tm{c} & 
$M$(\HI) & $dM$(\HI)/$dt$ \\
    & 
($N$ in \sqcm) & 
($N$ in \sqcm) &  
    & 
(\kms) & 
($10^7$\,\msun) & 
(\msun\,yr$^{-1}$)}
\startdata
Inflowing HVCs & 18--19 (7) & 18.54 $\pm$ 0.12 & 0.28 $\pm$ 0.06 & $-$99 $\pm$ 43 &  1.9 $\pm$ 0.4 & $-$0.16 $\pm$ 0.08 \\
               & 19--20 (5) & 19.35 $\pm$ 0.18 & 0.20 $\pm$ 0.04 & $-$93 $\pm$ 42 & 9.0 $\pm$ 1.7 & $-$0.72 $\pm$ 0.35\\ 
\hline
Outflowing HVCs & 18--19 (4) & 18.54 $\pm$ 0.14 & 0.12 $\pm$ 0.06 & 135 $\pm$ 11 & 0.8 $\pm$ 0.4 & 0.10 $\pm$ 0.05 \\
                & 19--20 (1) & 19.80 $\pm$ 0.51 & 0.04 $\pm$ 0.04 & 23 $\pm$ 4\phn & 5.1 $\pm$ 4.8 & 0.10 $\pm$ 0.09 \\ 
\enddata 
\tablecomments{The \HI\ masses and mass flow rates are calculated using Equations~\ref{eqn:mass} to \ref{eqn:massflow2} for HVCs in two $N$(\HI) intervals.
Inflowing and outflowing clouds are defined as those with $v_{\rm GSR}<0$ and $v_{\rm GSR}>0$, respectively. A mean distance of 12\,kpc is assumed for HVCs. Negative mass flow rates denote inflow.} 
\tn{a}{The column density range of this bin (number of absorbers).}
\tn{b}{Mean log\,$N$(\HI) with its standard error.}
\tn{c}{Mean GSR velocity with its standard error.}
\label{tab:flow-rates}
\end{deluxetable*}

The calculated values for the masses and mass flow rates are presented in Table~\ref{tab:flow-rates}. 
They are calculated separately for inflowing and outflowing HVCs.
Furthermore, for each we give the results for two bins of $N$(\HI), [18-19] and [19-20], revealing how the mass and mass flow rate is distributed over $N$(\HI).
These two bins span the column density range containing the vast majority of
the mass and mass flow rate, because they are both proportional to the mean $N$(\HI) in the bin. 
Additionally, the [18-19] bin contains many poorly constrained saturated
absorbers, while the [19-20] bin contains mostly well-measured components 
(as shown in Figure \ref{fig:e_logNHI}).
By summing over these bins, we can calculate the \emph{total} \HI\ mass and 
total \HI\ mass flow rate in each of the four sub-samples.

\newcommand{\hvcinflow}{--0.88$\pm$0.40 ($d$/12 kpc)}
\newcommand{\hvcoutflow}{0.20$\pm$0.10 ($d$/12 kpc)}
\newcommand{\ivcinflow}{--0.15$\pm$0.04 ($d$/1 kpc)}
\newcommand{\ivcoutflow}{0.11$\pm$0.05 ($d$/1 kpc)}

We find that inflowing HVC components carry $\approx$2 times as much \HI\ mass as 
outflowing components, which results in a more than four times larger mass 
flow rate, with $\approx$\hvcinflow \msy\ of HVC inflow versus $\approx$\hvcoutflow \msy\ of HVC outflow).
A similar excess of HVC inflow over outflow was also found by \citet{fox2019} based on the 
\citet{richteretal2017} sample of UV metal-line HVCs, so we have independently confirmed
the inflow excess using \fuse\ \HI\ data, with the key advantage of not having to 
correct for dust or metallicity, because in \HI\ we are directly seeing the neutral baryons.
An excess of inflow (in the northern hemisphere) was also reported in the \ion{Si}{4} analysis
of \citet{Qu_2020}.

Additionally, we note that most of the mass flow occurs within the highest 
column density bin, log\,$N$(\HI) = 19--20, where our measurement and fitting
errors are the smallest.
This is because the larger column densities in this bin dominate the mass and flow rate calculations, with distance, covering fraction, and mean velocity being second-order effects.
At even lower column densities (log\,$N$(\HI)$<$18), increasing ionization 
will result in the total hydrogen mass being dominated by H\II.
Because the ionization fraction $N$(H\II)/$N\rm(H_{total}) \propto N$(\HI)$^{-1}$, the total (neutral+ionized)
inflow and outflow masses will be considerably higher than those based on \HI\ alone \citep{lehner2011}. Photoionization modeling is needed to quantify this.


\section{Summary} \label{sec:summary}

We have presented a survey of neutral gas in the Milky Way halo down to limiting column 
densities of $N$(\HI)$\sim10^{14}$\sqcm\ using measurements of \HI\ Lyman-series absorption 
from the \emph{Far Ultraviolet Spectroscopic Explorer} toward 25 AGN sightlines spread evenly 
across the sky above $|b|\sim$20$\degr$. By simultaneously fitting multi-component Voigt 
profiles to the 11 Lyman-series absorption transitions covered by \fuse\ (Ly$\beta$--Ly$\mu$) 
as well as additional data from \hst/COS and STIS covering Ly$\alpha$, 
we have derived complete \HI\ component parameters in each sightline. 
Our sample consists of 152 \HI\ components, including 39 ISM, 42 IVC, and 71 HVC components
and extends down to \HI\ columns four orders of magnitude lower than can be probed by radio 
21\,cm observations. Our survey has led to the following results.

\begin{enumerate}
    
    \item The \HI\ components cover approximately 7\,dex of column density, though the range 
    depends on velocity: ISM components cover a range of log\,$N$(\HI) from $\approx$18--21 
    whereas HVCs cover $\approx$14--19 and IVCs cover $\approx$16--20. The differences in column
    density ranges appear to be mostly due to incompleteness from saturation and velocity 
    crowding of components. These ranges of column density are observed at longitudes and
    latitudes across the sky with no strong spatial dependence.
    
    \item Taken together, there are 113 HVCs and IVCs in the sample. 64 of these 113 (58\%) 
    have log\,$N$(\HI)$<$17.5 and are not detectable in 21\,cm emission, 
    showing that the majority of the HVCs and IVCs in our sample are invisible 
    to 21\,cm observations.
    
    \item Saturation prevents accurate measurements of many components with column densities 
    17$\lesssim$log\,$N$(\HI)$\lesssim$19, but we derive robust measurements of components with
    log\,$N$(\HI)$\lesssim$17 and log\,$N$(\HI)$\gtrsim$19. Saturated components are kept in the 
    sample (with their associated large uncertainties) since excluding them does not significantly 
    change our results on the \HI\ distribution function.
    
    \item The \HI\ components have $b$-values $<$50\kms. For all three velocity categories 
    (ISM, IVC, and HVC), the distribution of $b$-values is peaked near 10\kms\ 
    (close to the \fuse\ instrumental resolution) and shows an extended tail out to 50\kms. 
    
    \item We have computed the sky covering fractions 
    for all, ISM, IVC, and HVC \HI\ components, 
    both in the discrete [$f_{\rm sky}(N+ \Delta N)$]
    and cumulative [$f_{\rm sky}(>N)$] forms. 
    We find an excess of IVCs in the northern hemisphere compared to the south. 
    To quantify this excess, we show that $f_{\rm sky}$=0.7 is reached at log\,$N$(\HI)=18.7 
    in the north but 16.7 in the south.
    For HVCs, the sky covering fractions are similar in the north and the south, though there
    is more high column-density \HI\ in the north than the south.

    \item We have computed the first UV column density distribution function (CDDF) for 
    Galactic \HI. This extends the range of the Galactic CDDF by over four orders of 
    magnitude deeper than existing 21\,cm surveys 
    \citep{wakker1991, murphy1995, lockman2002, moss2013, westmeier2018}.
    Our CDDF includes an incompleteness correction based on simulating the observability
    of mock components inserted into the data as a function of velocity and $N$(\HI), and we 
    assess the impact of saturation as a function of $N$(\HI). 
    
    \item We find that IVCs and HVCs show statistically indistinguishable CDDF slopes, with 
    $\beta_{\rm IVC}$=\ivcbeta\ and $\beta_{\rm HVC}$=\hvcbeta.
    In contrast, ISM clouds (at low velocity) have a shallower slope 
    $\beta_{\rm ISM}$=\ismbeta. These ISM clouds tend to have larger $N$(\HI) column densities.
    This slope difference from the HVC and IVC distributions suggests a significantly 
    different behavior and reflecting the different physical conditions in the disk and halo. 
    Although incompleteness affects the ISM and IVC samples at low column densities, 
    and saturation affects all components at intermediate column densities, the high
    column density results are relatively robust. Furthermore, the HVC slope is not 
    strongly affected by incompleteness effects since that preferentially affects the 
    low-velocity gas.
    
    \item We fit the global \HI\ CDDF as a function of column density with a 
    piecewise three-component fit, in which the breakpoints are automatically identified. 
    We have presented separate ``Galactic halo" and combined ``Galactic halo + disk"
    versions of our CDDF to aid in the comparison of Galactic and extragalactic studies.
    We found evidence that three components are needed for an adequate fit, with a marked
    steepening of the slope ($\beta$=\midbeta) in the intermediate column-density range 
    \lowbreak$\leq$log\,$N$(\HI)$\leq$\highbreak, driven by an excess of absorbers observed near 
    log\,$N$=17. This excess is present for both IVCs and HVCs, though we note it appears
    at the column densities where saturation becomes significant and so the values of log\,$N$ 
    are most uncertain. The slopes at the low end
    (log\,$N$(\HI)$<$\lowbreak; $\beta$=\lowbeta) and high end (log\,$N$(\HI)$>$\highbreak; 
    $\beta$=\highbeta) are shallower.

    \item By comparing the Galactic CDDF to a number of extragalactic CDDFs drawn from 
    the literature, we have 
    considered the relationship between the Galactic halo and extragalactic
    quasar absorption-line 
    systems. We find that the slope of the Galactic CDDF is most similar to the extragalactic CDDFs
    \citep{ribaudo2011, omeara2013, rudie2013, shull2017} in the intermediate column-density range (Lyman Limit Systems), but is flatter than them below log\,$N$(\HI)$\approx$\lowbreak. 
    While this flattening could be related to incompleteness, 
    it is more likely due to the difference
    in physical environment between Galactic halo and Ly$\alpha$ forest absorbers. 
    The shallower Galactic CDDF slope may reflect the enhanced ionizing radiation field in the
    Galactic halo compared to the IGM, as well as stripping and thermal instabilities resulting from
    infall.
    Overall, the shape of the high- and intermediate-column
    density range of the Galatic halo CDDF appears similar to extragalactic findings.
    
    \item We find an excess of inflowing \HI\ over outflowing \HI\ in HVCs, as revealed by 
    higher gas masses and mass flow rates for components with negative $v_{\rm GSR}$ than for 
    those with positive $v_{\rm GSR}$. Using a partially-covered spherical shell model, we show that 
    inflowing HVCs carry \hvcinflow \msy\ whereas outflowing HVCs represent only 
    \hvcoutflow \msy.
    The excess of inflow agrees with 
    results from UV metal-line HVCs \citep{fox2019}, but our new measurements are based on 
    \HI\ so they have the advantage of directly measuring the (neutral) baryons, without the 
    need for metallicity or dust corrections. These mass flow rates show that the halo overall 
    experiences a net accretion of cool gas, which can contribute to future star formation.

\end{enumerate}

By using the diagnostic power of the Lyman series absorption lines combined with the 
information content of the \fuse\ archives, this survey has 
provided a new, metal-independent view of the Galactic halo down to 
very sensitive limits. The measurements presented here constitute a rich database of the 
cool gas properties of the Milky Way that will be valuable for constraining future 
simulations of the formation and evolution of the Galactic halo.

\acknowledgments
We thank the dedicated team that operated the \fuse\ satellite. More than a decade after 
the mission ended, the data are still leading to new scientific insights. 
We gratefully acknowledge support from the NASA Astrophysics Data Analysis Program (ADAP) 
under grant 80NSSC18K0421, {\it Surveying the H I Content of the Galactic Halo via Lyman 
Series Absorption.} We thank Vanessa Moss, J. X. Prochaska, and Max Gronke for useful discussion 
on the CDDF, and we are grateful to the referee for a thorough report.

\facilities{\fuse, \hst/STIS, \hst/COS}

\software{\emph{Astropy} \citep{astropy2018}, \emph{CalFUSE} \citep{dixon2007}, 
\emph{Corner} \citep{corner}, \emph{Emcee} \citep{Foreman_Mackey_2013},
\emph{Matplotlib} \citep{matplotlib}, \emph{Numpy} \citep{harris2020}, 
\emph{pandas} \citep{pandas}, \emph{SciPy} \citep{scipy},
\emph{VoigtFit} \citep{voigtfit}}

\clearpage
\bibliography{bib}
\bibliographystyle{aasjournal}

\clearpage
\startlongtable
\begin{deluxetable*}{l D D D R R R l}
\tabletypesize{\footnotesize}
\tablecaption{\HI\ Fit Results \label{table:fuse_results}}
\tablehead{
\colhead{Sightline} & \multicolumn2c{$l$}   & \multicolumn2c{$b$}  	& \multicolumn2c{$\rm (S/N)_{977}$} & \colhead{$v \pm \sigma_{st} \pm \sigma_{sys}$} & \colhead{$b \pm \sigma_{st} \pm \sigma_{sys}$} & \colhead{$\rm log_{10} \emph{N} (\HI) \pm \sigma_{st} \pm \sigma_{sys}$} & \colhead{Note} \\
			   & \multicolumn2c{(deg)} & \multicolumn2c{(deg)}	&     \multicolumn2c{}   & \colhead{(km\,s$^{-1}$)}  & \colhead{(km\,s$^{-1}$)}  &   & \\
\colhead{(1)}  & \multicolumn2c{(2)} & \multicolumn2c{(3)} & \multicolumn2c{(4)} & \colhead{(5)} & \colhead{(6)} & \colhead{(7)} & \colhead{(8)}}
\decimals
\startdata
1H0707-495 & 260.17 & -17.67 & 6.41  &     0$\pm 3\pm 1$   &  36$\pm 1\pm 1$  &       20.24$\pm 0.038\rm \substack{-0.030 \\ +0.029}$   &  .    \\
&            .   &     .   &     .   &   151$\pm 6\pm 1$   &  32$\pm 11\pm 1$ &       16.67$\pm 0.143\rm \substack{-0.018 \\ +0.017}$   &  .    \\
&            .   &     .   &     .   &   212$\pm 12\pm 2$  &  29$\pm 5\pm 1$  &       16.174$\pm 0.203\rm \substack{-0.009 \\ +0.006}$  &  .    \\
&            .   &     .   &     .   &   344$\pm 1\pm 0$   &  18$\pm 1\pm 1$  &       17.623$\pm 0.303\rm \substack{-0.090 \\ +0.081}$  &  .    \\
3C273  & 289.95  &  64.36  &  19.59  &  -117$\pm 2\pm 4$.  &   3$\pm 1\pm 2$  &       14.933$\pm 0.155\rm \substack{+0.135 \\ +0.221}$  &  .    \\
&            .   &     .   &     .   &   -71$\pm 3\pm 17$  &  18$\pm 3\pm 14$ &       16.194$\pm 0.075\rm \substack{+0.318 \\ +0.512}$  &  .    \\
&            .   &     .   &     .   &   -19 ...  ...       &  13$\pm 4\pm 5$ &  \sim 19.077   & $\dagger\ \rm f_v$  \\
&            .   &     .   &     .   &    -6 ...  ...       &   5 ...  ...    &       20.198$\pm 0.226\rm \substack{-0.143 \\ -0.085}$  & $\rm f_v\rm ~f_b$    \\
&            .   &     .   &     .   &    22 ...  ...       &  5$\pm 9\pm 8$  &       19.287$\pm 0.533\rm \substack{+0.029 \\ -0.079}$  & $\rm f_v$     \\
&            .   &     .   &     .   &    48$\pm 24\pm 20$ & 14$\pm 31\pm 11$ &       16.505$\pm 0.882\rm \substack{+1.662 \\ +1.675}$  &  .    \\
&            .   &     .   &     .   &    74$\pm 8\pm 12$  & 10$\pm 4\pm 7$   &       15.911$\pm 0.587\rm \substack{+2.069 \\ +2.031}$  &  .    \\
ESO141-G55 & 338.18 & -26.71 &  8.26 &    -67$\pm 2\pm 1$  &  9$\pm 1\pm 1$   &       19.52$\pm 0.297\rm \substack{+0.123 \\ +0.165}$   &  .    \\
&            .   &     .   &     .   &     -2$\pm 3\pm 4$  & 10$\pm 1\pm 1$   &       20.588$\pm 0.026\rm \substack{-0.034 \\ -0.027}$  &  .    \\
&            .   &     .   &     .   &     68$\pm 3\pm 1$  & 23$\pm 2\pm 0$   &       16.961$\pm 0.125\rm \substack{-0.051 \\ +0.027}$  &  .    \\
&            .   &     .   &     .   &    151$\pm 2\pm 0$  &  2$\pm 1\pm 1$   &       15.97$\pm 0.871\rm \substack{+0.074 \\ +0.102}$   &  .    \\
H1821+643 &  94.00 &  27.42  &  7.64 &   -214$\pm 13\pm 1$ & 29$\pm 20\pm 0$  &       14.073$\pm 0.280\rm \substack{-0.029 \\ +0.117}$  &  .    \\
&            .   &     .   &     .   &   -173$\pm 2\pm 1$  &   3$\pm 1\pm 0$  &       16.04$\pm 0.849\rm \substack{-0.024 \\ +0.362}$   &  .    \\
&            .   &     .   &     .   &   -136$\pm 2\pm 0$  &   8$\pm 4\pm 1$  &  \sim 17.788  &  $\dagger$  \\
&            .   &     .   &     .   &    -76$\pm 72\pm 5$ &  12$\pm 24\pm 3$ &  \sim 19.716  &  $\dagger$  \\
&            .   &     .   &     .   &    -30 ...  ...       &  19$\pm 3\pm 1$  &  \sim 20.155  &  $\dagger\ \rm f_v$ \\
&            .   &     .   &     .   &     -2 ...  ...       &   6 ...  ...       &       20.042$\pm 0.794\rm \substack{-0.004 \\ +0.031}$  & $\rm f_v \rm ~f_b$ \\
&            .   &     .   &     .   &     64$\pm 2\pm 1$  &  16$\pm 3\pm 3$   &       16.912$\pm 0.243\rm \substack{-0.025 \\ +0.182}$  &  .    \\
&            .   &     .   &     .   &    124$\pm 13\pm 2$ &  18$\pm 13\pm 10$ &       14.367$\pm 0.262\rm \substack{-0.085 \\ +0.320}$  &  .    \\
HE0226-4110 & 253.94 & -65.77 & 11.2 &     -4$\pm 1\pm 0$  &  15$\pm 0\pm 1$   &       20.098$\pm 0.008\rm \substack{-0.004 \\ +0.004}$  &  .    \\
&            .   &     .   &     .   &     88$\pm 1\pm 0$  &  22$\pm 2\pm 1$   &       16.562$\pm 0.042\rm \substack{-0.003 \\ +0.004}$  &  .    \\
&            .   &     .   &     .   &    140$\pm 4\pm 0$  &  11$\pm 4\pm 0$   &       16.427$\pm 0.097\rm \substack{-0.004 \\ +0.005}$  &  .    \\
&            .   &     .   &     .   &    174$\pm 6\pm 0$  &  12$\pm 8\pm 1$   &       16.663$\pm 0.249\rm \substack{-0.003 \\ +0.002}$  &  .    \\
&            .   &     .   &     .   &    202$\pm 7\pm 0$  &  18$\pm 3\pm 1$   &       16.544$\pm 0.259\rm \substack{-0.004 \\ +0.002}$  &  .    \\
MRK279 &  115.04  & 46.86  &  20.62  &     75$\pm 1\pm 2$  &   2$\pm 0\pm 0$   &       18.686$\pm 0.369\rm \substack{-0.064 \\ -0.163}$  &  .    \\
&            .   &     .   &     .   &         5 ...  ...    &  19$\pm 1\pm 1$   &       19.90$\pm 0.063\rm \substack{+0.009 \\ +0.019}$   &  $\rm f_v$  \\
&            .   &     .   &     .   &       -35 ...  ...    &  39$\pm 13\pm 4$  &  \sim 18.171  &  $\dagger\ \rm f_v$ \\
&            .   &     .   &     .   &   -112$\pm 5\pm 13$ &  10$\pm 50\pm 12$ &  \sim 16.612  &  $\dagger$   \\
&            .   &     .   &     .   &   -136$\pm 15\pm 1$ &   4$\pm 5\pm 0$   &  \sim 18.933  &  $\dagger$   \\
&            .   &     .   &     .   &   -166$\pm 6\pm 2$  &   6$\pm 2\pm 1$   &       19.473$\pm 0.307\rm \substack{+0.049 \\ -0.004}$  &  .    \\
&            .   &     .   &     .   &   -202$\pm 5\pm 1$  &  15$\pm 3\pm 1$   &       16.718$\pm 0.182\rm \substack{-0.002 \\ +0.016}$  &  .    \\
&            .   &     .   &     .   &   -241$\pm 2\pm 0$  &   2$\pm 1\pm 1$   &  \sim 15.788  &  $\dagger$  \\
MRK335  & 108.76 & -41.42  & 13.22   &   -415$\pm 2\pm 1$  &  15$\pm 1\pm 2$   &       18.179$\pm 0.439\rm \substack{+0.039 \\ -0.111}$  &  .    \\
&            .   &     .   &     .   &   -335$\pm 6\pm 1$  &  29$\pm 8\pm 1$   &       16.497$\pm 0.105\rm \substack{+0.005 \\ -0.065}$  &  .    \\
&            .   &     .   &     .   &   -301$\pm 6\pm 3$  &   6$\pm 11\pm 3$  &       15.779$\pm 0.348\rm \substack{+0.021 \\ -0.030}$  &  .    \\
&            .   &     .   &     .   &   -266$\pm 13\pm 5$ &  39$\pm 9\pm 4$   &       15.572$\pm 0.170\rm \substack{-0.061 \\ +0.019}$  &  .    \\
&            .   &     .   &     .   &   -148$\pm 2\pm 3$  &   3$\pm 1\pm 3$   &       16.656$\pm 0.717\rm \substack{+0.198 \\ -0.497}$  &  .    \\
&            .   &     .   &     .   &   -108$\pm 4\pm 3$  &   11$\pm 5\pm 2$  &       17.288$\pm 0.558\rm \substack{+0.119 \\ -0.172}$  &  .    \\
&            .   &     .   &     .   &    -44 ...  ...     &   13$\pm 3\pm 3$  &       19.826$\pm 0.529\rm \substack{-0.117 \\ +0.142}$  & $\rm f_v$     \\
&            .   &     .   &     .   &     1$\pm 17\pm 1$  &   24$\pm 5\pm 0$  &       20.329$\pm 0.204\rm \substack{+0.009 \\ -0.020}$  &  .    \\
MRK421 &  179.83  & 65.03  &  14.62  &  -170$\pm 1\pm 1$   &    3$\pm 0\pm 1$  &       17.558$\pm 0.243\rm \substack{+0.012 \\ -0.011}$  &  .    \\
&            .   &     .   &     .   &  -117$\pm 4\pm 0$   &   16$\pm 2\pm 1$  &       18.734$\pm 0.441\rm \substack{+0.003 \\ -0.001}$  &  .    \\
&            .   &     .   &     .   &   -63$\pm 4\pm 0$   &    4$\pm 1\pm 1$  &       19.638$\pm 0.114\rm \substack{-0.002 \\ +0.002}$  &  .    \\
&            .   &     .   &     .   &   -15 ...  ...      &   12 ...  ...     &       19.469$\pm 0.477\rm \substack{+0.000 \\ +0.001}$  & $\rm f_v\rm ~f_b$    \\
&            .   &     .   &     .   &    -9 ...  ...      &    4 ...  ...     &       19.529$\pm 0.345\rm \substack{+0.000 \\ +0.000}$  & $\rm f_v\rm ~f_b$    \\
&            .   &     .   &     .   &   37$\pm 1\pm 0$    &   10$\pm 1\pm 1$  &       16.623$\pm 0.107\rm \substack{-0.011 \\ +0.011}$  &  .    \\
&            .   &     .   &     .   &  127$\pm 2\pm 0$    &    1$\pm 3\pm 1$  &       14.00$\pm 0.268\rm \substack{-0.078 \\ +0.077}$  &   .    \\
MRK509  &  35.97 &  -29.86  & 7.02   & -322$\pm 4\pm 2$    &   10$\pm 1\pm 1$  &       19.03$\pm 0.112\rm \substack{+0.080 \\ +0.070}$  &   .    \\
&            .   &     .   &     .   & -289$\pm 2\pm 1$    &   48$\pm 2\pm 1$  &       16.197$\pm 0.069\rm \substack{-0.041 \\ -0.010}$  &  .    \\
&            .   &     .   &     .   & -138$\pm 1\pm 0$    &    7$\pm 2\pm 2$  &  \sim 18.225   &  $\dagger$   \\
&            .   &     .   &     .   &  -74$\pm 2\pm 3$    &   13$\pm 2\pm 1$  &       18.328$\pm 0.860\rm \substack{+0.456 \\ +0.384}$  &  .    \\
&            .   &     .   &     .   &   -2$\pm 3\pm 3$    &   12$\pm 1\pm 0$  &       20.27$\pm 0.093\rm \substack{-0.072 \\ -0.027}$  &   .    \\
&            .   &     .   &     .   &   76$\pm 5\pm 1$    &   11$\pm 2\pm 0$  &       19.97$\pm 0.093\rm \substack{+0.031 \\ +0.026}$  &   .    \\
&            .   &     .   &     .   &  121$\pm 14\pm 1$   &   17$\pm 7\pm 1$  &       16.141$\pm 0.433\rm \substack{-0.017 \\ +0.008}$  &  .    \\
&            .   &     .   &     .   &  168$\pm 2\pm 2$    &    2$\pm 1\pm 1$  &  \sim 16.637  &  $\dagger$  \\
MRK817 &  100.30 &  53.48  &  18.54  & -147$\pm 13\pm 1$   &   18$\pm 4\pm 1$  &       19.222$\pm 0.394\rm \substack{+0.015 \\ -0.172}$  &  .    \\
&            .   &     .   &     .   &  -88$\pm 103\pm 2$  &   13$\pm 42\pm 1$ &       19.192$\pm 0.390\rm \substack{+0.020 \\ +0.099}$  &  .    \\
&            .   &     .   &     .   &   -6 ...  ...       &   15$\pm 3\pm 2$  &       19.815$\pm 0.077\rm \substack{-0.003 \\ +0.025}$  & $\rm f_v$  \\
&            .   &     .   &     .   &   41$\pm 13\pm 0$   &   22$\pm 6\pm 1$  &       16.551$\pm 0.348\rm \substack{-0.005 \\ +0.055}$  &  .    \\
&            .   &     .   &     .   &  106$\pm 2\pm 0$    &    8$\pm 2\pm 1$  &       14.883$\pm 0.120\rm \substack{-0.043 \\ +0.050}$  &  .    \\
MRK876 &  98.27  &  40.38  &  10.9   & -205$\pm 4\pm 1$    &   22$\pm 2\pm 1$  &       16.628$\pm 0.110\rm \substack{-0.005 \\ +0.004}$  &  .    \\
&            .   &     .   &     .   & -172$\pm 4\pm 0$    &    5$\pm 3\pm 1$  &  \sim 18.908   &  $\dagger$  \\
&            .   &     .   &     .   & -142$\pm 4\pm 1$    &    4$\pm 1\pm 0$  &       19.802$\pm 0.505\rm \substack{-0.007 \\ +0.007}$  &  .    \\
&            .   &     .   &     .   &  -82$\pm 5\pm 0$    &   36$\pm 14\pm 1$ &       17.181$\pm 0.065\rm \substack{-0.005 \\ +0.005}$  &  .    \\
&            .   &     .   &     .   &  -23$\pm 6\pm 1$    &    4$\pm 2\pm 1$  &       19.984$\pm 0.564\rm \substack{-0.006 \\ +0.007}$  &  .    \\
&            .   &     .   &     .   &    6$\pm 4\pm 0$    &    4$\pm 1\pm 1$  &       19.781$\pm 0.724\rm \substack{-0.003 \\ +0.003}$  &  .    \\
&            .   &     .   &     .   &   41$\pm 2\pm 0$    &    7$\pm 6\pm 1$  &  \sim 18.755   &  $\dagger$  \\
&            .   &     .   &     .   &   65$\pm 18\pm 1$   &   24$\pm 9\pm 1$  &       15.839$\pm 0.400\rm \substack{-0.005 \\ +0.005}$  &  .    \\
MRK1383 & 349.22 & 55.13   &  12.11  & -106$\pm 1\pm 4$    &    4$\pm 1\pm 1$  &       16.932$\pm 0.771\rm \substack{+0.082 \\ +0.204}$  &  .    \\
&            .   &     .   &     .   &  -73$\pm 1\pm 11$   &   6$\pm 1\pm 1$   &       18.686$\pm 0.659\rm \substack{-0.591 \\ -0.543}$  &  .    \\
&            .   &     .   &     .   &  -10$\pm 2\pm 2$    &   13$\pm 1\pm 5$  &       20.334$\pm 0.018\rm \substack{+0.002 \\ +0.019}$  &  .    \\
&            .   &     .   &     .   &   47$\pm 2\pm 19$   &    6$\pm 2\pm 4$  &       16.843$\pm 0.492\rm \substack{+0.807 \\ +0.673}$  &  .    \\
&            .   &     .   &     .   &  150$\pm 10\pm 32$  &   24$\pm 14\pm 10$ &      13.806$\pm 0.286\rm \substack{+0.571 \\ +0.723}$  &  .    \\
NGC1068 & 172.10  &  -51.93  & 9.8   & -267$\pm 1\pm 0$    &   10$\pm 1\pm 1$  &       18.432$\pm 0.193\rm \substack{-0.032 \\ -0.020}$  &  .    \\
&            .   &     .   &     .   & -219$\pm 3\pm 1$    &   18$\pm 2\pm 0$  &       16.041$\pm 0.065\rm \substack{-0.009 \\ +0.008}$  &  .    \\
&            .   &     .   &     .   &  -57$\pm 13\pm 1$   &   44$\pm 4\pm 0$  &       17.414$\pm 0.239\rm \substack{-0.007 \\ -0.003}$  &  .    \\
&            .   &     .   &     .   &   15$\pm 7\pm 1$    &   10$\pm 2\pm 0$  &       20.342$\pm 0.040\rm \substack{-0.017 \\ +0.018}$  &  .    \\
&            .   &     .   &     .   &   70$\pm 1\pm 0$    &    6$\pm 1\pm 1$  &       16.157$\pm 0.287\rm \substack{-0.028 \\ +0.031}$  &  .    \\
PG0804+761 & 138.28 & 31.03 &  13.97 & -103$\pm 7\pm 1$    &   30$\pm 4\pm 0$  &       16.664$\pm 0.125\rm \substack{-0.028 \\ -0.003}$  &  .    \\
&            .   &     .   &     .   &  -50$\pm 5\pm 0$    &    14$\pm 7\pm 1$ &       16.733$\pm 0.355\rm \substack{-0.011 \\ +0.011}$  &  .    \\
&            .   &     .   &     .   &    6$\pm 1\pm 0$    &     9$\pm 1\pm 1$ &       20.537$\pm 0.001\rm \substack{-0.006 \\ +0.006}$  &  .    \\
&            .   &     .   &     .   &   115$\pm 4\pm 2$   &    4$\pm 4\pm 1$  &       14.749$\pm 0.127\rm \substack{-0.134 \\ +0.128}$  &  .    \\
PG0844+349 & 188.56 & 37.97 &  6.44  &   -154$\pm 2\pm 0$  &    3$\pm 2\pm 1$  &  \sim 17.011   &  $\dagger$  \\
&            .   &     .   &     .   &   -125$\pm 2\pm 0$  &    5$\pm 2\pm 1$  &  \sim 17.162   &  $\dagger$  \\
&            .   &     .   &     .   &    -90$\pm 2\pm 0$  &    5$\pm 4\pm 1$  &  \sim 18.351   &  $\dagger$  \\
&            .   &     .   &     .   &    -47$\pm 3\pm 0$  &    7$\pm 1\pm 1$  &       19.877$\pm 0.435\rm \substack{-0.013 \\ +0.005}$  &  .    \\
&            .   &     .   &     .   &      5$\pm 3\pm 1$  &    7$\pm 1\pm 1$  &       20.407$\pm 0.093\rm \substack{-0.003 \\ +0.007}$  &  .    \\
&            .   &     .   &     .   &     76$\pm 5\pm 1$  &   46$\pm 7\pm 1$  &       16.779$\pm 0.054\rm \substack{-0.008 \\ +0.008}$  &  .    \\
&            .   &     .   &     .   &    177$\pm 6\pm 0$  &   24$\pm 4\pm 0$  &       15.529$\pm 0.111\rm \substack{-0.014 \\ +0.015}$  &  .    \\
&            .   &     .   &     .   &    345$\pm 1\pm 1$  &    3$\pm 1\pm 0$  &       15.716$\pm 0.593\rm \substack{-0.188 \\ -0.009}$  &  .    \\
PG0953+414 & 179.79 & 51.71 &  10.52 &   -135$\pm 6\pm 1$  &   22$\pm 4\pm 0$  &       16.307$\pm 0.170\rm \substack{-0.035 \\ +0.012}$  &  .    \\
&            .   &     .   &     .   &    -65 ...  ...     &   41$\pm 5\pm 3$  &       17.166$\pm 0.181\rm \substack{+0.032 \\ +0.087}$  & $\rm f_v$  \\
&            .   &     .   &     .   &    -46 ...  ...     &   15 ...  ...     &       19.60$\pm 0.760\rm \substack{-0.009 \\ -0.001}$  &  $\rm f_v\rm ~f_b$  \\
&            .   &     .   &     .   &     -6 ...  ...     &    8 ...  ...     &  \sim 19.685   & $\dagger\ \rm f_v ~f_b$\  \\
&            .   &     .   &     .   &     16 ...  ...     &    7$\pm 15\pm 3$ &  \sim 19.342   & $\dagger\ \rm f_v$\  \\
&            .   &     .   &     .   &    60$\pm 22\pm 3$  &   29$\pm 12\pm 3$ &       16.874$\pm 0.581\rm \substack{-0.101 \\ +0.088}$  &  .    \\
&            .   &     .   &     .   &   126$\pm 2\pm 1$   &   15$\pm 1\pm 1$  &       16.329$\pm 0.071\rm \substack{-0.023 \\ -0.012}$  &  .    \\
PG1011-040 & 246.50 & 40.75 &  8.89  &    -4 ...  ...      &   29$\pm 5\pm 1$  &  \sim 19.928   & $\dagger\ \rm f_v$\ \\
&            .   &     .   &     .   &    -3 ...  ...      &    7 ...  ...     &       20.399$\pm 0.421\rm \substack{+0.018 \\ +0.030}$  & $\rm f_v\rm ~f_b$  \\
&            .   &     .   &     .   &    94$\pm 3\pm 2$   &    3$\pm 1\pm 1$  &       19.602$\pm 0.465\rm \substack{+0.006 \\ -0.012}$  &  .    \\
&            .   &     .   &     .   &   118$\pm 5\pm 2$   &    4$\pm 6\pm 0$  &  \sim 18.327   &  $\dagger$  \\
&            .   &     .   &     .   &   148$\pm 6\pm 1$   &   15$\pm 11\pm 3$ &       16.599$\pm 0.163\rm \substack{-0.039 \\ +0.034}$  &  .    \\
&            .   &     .   &     .   &   200$\pm 6\pm 2$   &   36$\pm 4\pm 3$  &       16.836$\pm 0.078\rm \substack{+0.008 \\ -0.005}$  &  .    \\
&            .   &     .   &     .   &   280$\pm 2\pm 2$   &    2$\pm 1\pm 1$  &  \sim 16.788   &  $\dagger$  \\
&            .   &     .   &     .   &   325$\pm 2\pm 1$   &   24$\pm 3\pm 1$  &       14.943$\pm 0.058\rm \substack{-0.043 \\ +0.066}$  &  .    \\
PG1116+215 & 223.36 & 68.21 &  12.19 &  -120$\pm 1\pm 1$   &    3$\pm 1\pm 1$  &       16.265$\pm 0.969\rm \substack{-0.016 \\ +0.267}$  &  .    \\
&            .   &     .   &     .   &   -79$\pm 3\pm 1$   &   10$\pm 3\pm 1$  &       16.98$\pm 0.439\rm \substack{+0.036 \\ -0.159}$  &   .    \\
&            .   &     .   &     .   &   -49$\pm 4\pm 2$   &    3$\pm 1\pm 1$  &       19.885$\pm 0.078\rm \substack{+0.006 \\ -0.019}$  &  .    \\
&            .   &     .   &     .   &    -4 ...  ...      &   10$\pm 2\pm 1$  &       18.932$\pm 0.450\rm \substack{-0.126 \\ +0.259}$  & $\rm f_v$  \\
&            .   &     .   &     .   &    31$\pm 4\pm 1$   &   68$\pm 3\pm 3$  &       17.224$\pm 0.061\rm \substack{-0.003 \\ -0.014}$  &  .    \\
&            .   &     .   &     .   &   171$\pm 1\pm 1$   &   21$\pm 2\pm 0$  &       16.756$\pm 0.070\rm \substack{+0.006 \\ +0.047}$  &  .    \\
&            .   &     .   &     .   &   209$\pm 1\pm 1$   &    2$\pm 0\pm 0$  &       18.63$\pm 0.075\rm \substack{+0.005 \\ -0.038}$  &   .    \\
PG1211+143 & 267.55 & 74.31 &  11.58 &  -114$\pm 1\pm 4$   &   11$\pm 1\pm 2$  &       15.633$\pm 0.073\rm \substack{+0.068 \\ +0.038}$  &  .    \\
&            .   &     .   &     .   &   -85$\pm 1\pm 23$  &    2$\pm 0\pm 5$  &       19.102$\pm 0.266\rm \substack{+0.461 \\ +0.104}$  &  .    \\
&            .   &     .   &     .   &   -21 ...  ...      &   17$\pm 1\pm 2$  &       19.506$\pm 0.163\rm \substack{-1.458 \\ -0.018}$  & $\rm f_v$  \\
&            .   &     .   &     .   &    -9 ...  ...      &    9 ...  ...     &  \sim 20.186   & $\dagger\ \rm f_v ~f_b ~f_N$\ \\
&            .   &     .   &     .   &    41$\pm 1\pm 3$   &    6$\pm 3\pm 1$  &  \sim 17.821   & $\dagger$ \\
&            .   &     .   &     .   &    73$\pm 1\pm 0$   &    3$\pm 0\pm 1$  &       19.153$\pm 0.137\rm \substack{+0.081 \\ +0.009}$  &  .    \\
&            .   &     .   &     .   &   175$\pm 1\pm 0$   &   20$\pm 1\pm 1$  &       16.274$\pm 0.036\rm \substack{-0.014 \\ +0.014}$  &  .    \\
&            .   &     .   &     .   &   276 ...  ...      &   17 ...  ...     &       13.981$\pm 0.083\rm \substack{-0.027 \\ +0.028}$  & $\rm f_v\rm ~f_b$  \\
PG1259+593 & 120.56  & 58.05 &  13.7 &  -200$\pm 3\pm 16$  &   24$\pm 4\pm 13$ &       15.559$\pm 0.066\rm \substack{-0.372 \\ +0.273}$  &  .    \\
&            .   &     .   &     .   &  -126$\pm 7\pm 15$  &   15$\pm 2\pm 8$  &       19.977$\pm 0.168\rm \substack{-0.089 \\ -0.031}$  &  .    \\
&            .   &     .   &     .   &   -55$\pm 12\pm 24$ &  30$\pm 20\pm 33$ &      16.901$\pm 0.175\rm \substack{+0.264 \\ +0.137}$  &  .    \\
&            .   &     .   &     .   &    -7$\pm 5\pm 12$  &    4$\pm 1\pm 1$  &       19.972$\pm 0.173\rm \substack{+0.072 \\ +0.107}$  &  .    \\
&            .   &     .   &     .   &    22$\pm 6\pm 8$   &   10$\pm 13\pm 25$ & \sim 16.855   &  $\dagger$  \\
&            .   &     .   &     .   &    58$\pm 7\pm 14$  &   20$\pm 4\pm 14$ &       16.685$\pm 0.181\rm \substack{-0.115 \\ -0.360}$  &  .    \\
PKS0405-12 & 204.93 & -41.76 & 6.88  &   -57$\pm 6\pm 1$   &   31$\pm 3\pm 2$  &       16.683$\pm 0.092\rm \substack{-0.025 \\ +0.020}$  &  .    \\
&            .   &     .   &     .   &     1 ...  ...      &    8$\pm 2\pm 1$  &       20.225$\pm 0.255\rm \substack{-0.049 \\ +0.070}$  & $\rm f_v$  \\
&            .   &     .   &     .   &    19 ...  ...      &    9$\pm 1\pm 1$  &       20.272$\pm 0.202\rm \substack{+0.024 \\ -0.056}$  & $\rm f_v$  \\
&            .   &     .   &     .   &   138$\pm 3\pm 1$   &   19$\pm 3\pm 2$  &       15.30$\pm 0.085\rm \substack{-0.042 \\ +0.047}$  &   .    \\
PKS0558-504 & 257.96 & -28.57 & 8.6  &   -75$\pm 4\pm 2$   &   22$\pm 3\pm 1$  &       17.73$\pm 0.389\rm \substack{+0.034 \\ +0.088}$  &   .    \\
&            .   &     .   &     .   &   -11$\pm 4\pm 2$   &    5$\pm 1\pm 1$  &       20.376$\pm 0.068\rm \substack{-0.101 \\ -0.058}$  &  .    \\
&            .   &     .   &     .   &    27$\pm 1\pm 3$   &    5$\pm 1\pm 2$  &       19.847$\pm 0.238\rm \substack{+0.108 \\ +0.097}$  &  .    \\
&            .   &     .   &     .   &    59$\pm 2\pm 2$   &    5$\pm 3\pm 1$  &  \sim 17.502   &  $\dagger$  \\
&            .   &     .   &     .   &   130$\pm 2\pm 1$   &   40$\pm 2\pm 1$  &       16.961$\pm 0.040\rm \substack{-0.015 \\ +0.009}$  &  .    \\
&            .   &     .   &     .   &   254$\pm 1\pm 0$   &   24$\pm 1\pm 1$  &       16.997$\pm 0.105\rm \substack{-0.035 \\ +0.028}$  &  .    \\
PKS2005-489 & 350.37 & -32.60 & 6.62 &   -58$\pm 16\pm 18$ &  25$\pm 7\pm 9$  &       16.392$\pm 0.440\rm \substack{+0.385 \\ +0.316}$  &  .    \\
&            .   &     .   &     .   &    -3$\pm 6\pm 1$   &   19$\pm 3\pm 1$  &       20.51$\pm 0.035\rm \substack{+0.006 \\ +0.004}$  &   .    \\
&            .   &     .   &     .   &    88$\pm 4\pm 1$   &   10$\pm 3\pm 1$  &  \sim 18.922   &  $\dagger$  \\
&            .   &     .   &     .   &   133$\pm 2\pm 4$   &    3$\pm 1\pm 1$  &  \sim 18.532   &  $\dagger$  \\
&            .   &     .   &     .   &   175$\pm 1\pm 1$   &    9$\pm 2\pm 0$  &       18.82$\pm 0.799\rm \substack{-0.018 \\ -0.007}$  &   .    \\
PKS2155-304 & 17.73 & -52.25 & 12.5  &  -276$\pm 7\pm 1$   &   26$\pm 4\pm 1$  &       14.84$\pm 0.188\rm \substack{-0.017 \\ +0.012}$  &   .    \\
&            .   &     .   &     .   &  -231$\pm 15\pm 2$  &  33$\pm 21\pm 1$  &       14.638$\pm 0.348\rm \substack{+0.004 \\ -0.010}$  &  .    \\
&            .   &     .   &     .   &  -135$\pm 1\pm 0$   &   20$\pm 2\pm 1$  &       16.358$\pm 0.041\rm \substack{-0.024 \\ +0.014}$  &  .    \\
&            .   &     .   &     .   &   -34$\pm 16\pm 3$  &   72$\pm 8\pm 2$  &       16.362$\pm 0.117\rm \substack{-0.015 \\ +0.033}$  &  .    \\
&            .   &     .   &     .   &    -4$\pm 3\pm 2$   &   16$\pm 1\pm 1$  &       19.994$\pm 0.012\rm \substack{+0.003 \\ +0.004}$  &  .    \\
&            .   &     .   &     .   &    58$\pm 5\pm 4$   &   17$\pm 3\pm 2$  &       16.467$\pm 0.141\rm \substack{-0.065 \\ -0.001}$  &  .    \\
&            .   &     .   &     .   &   101$\pm 1\pm 1$   &    1$\pm 0\pm 1$  &       18.167$\pm 0.407\rm \substack{-0.279 \\ -0.197}$  &  .    \\
TON\_S210 & 224.97  & -83.16 & 7.69  &  -221$\pm 8\pm 0$   &   28$\pm 3\pm 1$  &       16.70$\pm 0.191\rm \substack{-0.018 \\ +0.011}$  &   .    \\
&            .   &     .   &     .   &  -165$\pm 7\pm 1$   &   12$\pm 3\pm 1$  &       19.034$\pm 0.125\rm \substack{+0.099 \\ -0.130}$  &  .    \\
&            .   &     .   &     .   &   -90$\pm 8\pm 8$   &   25$\pm 10\pm 8$ &       16.687$\pm 0.141\rm \substack{+0.103 \\ -0.094}$  &  .    \\
&            .   &     .   &     .   &     2$\pm 4\pm 5$   &   22$\pm 1\pm 2$  &       20.10$\pm 0.013\rm \substack{-0.024 \\ +0.021}$  &   .    \\
\hline 
\enddata
\tablecomments{The Note column indicates if the velocity ($\rm f_v$), $b$-parameter ($\rm f_b$), or $\rm log\emph{N}$ (\HI) ($\rm f_N$) was manually fixed for this component. Fixed component parameters consequently do not have error estimates. A dagger ($\dagger$) and approximate symbol ($\sim$) indicate column density measurements that are very uncertain or possibly suffer from unreliable error estimates.}
\end{deluxetable*}
\vspace{5pt}

\clearpage
\appendix

\section{Ly$\alpha$ Inclusion}
\label{section:lya}

\begin{figure*}[ht!]
\centering
 \includegraphics[width=0.7\linewidth]{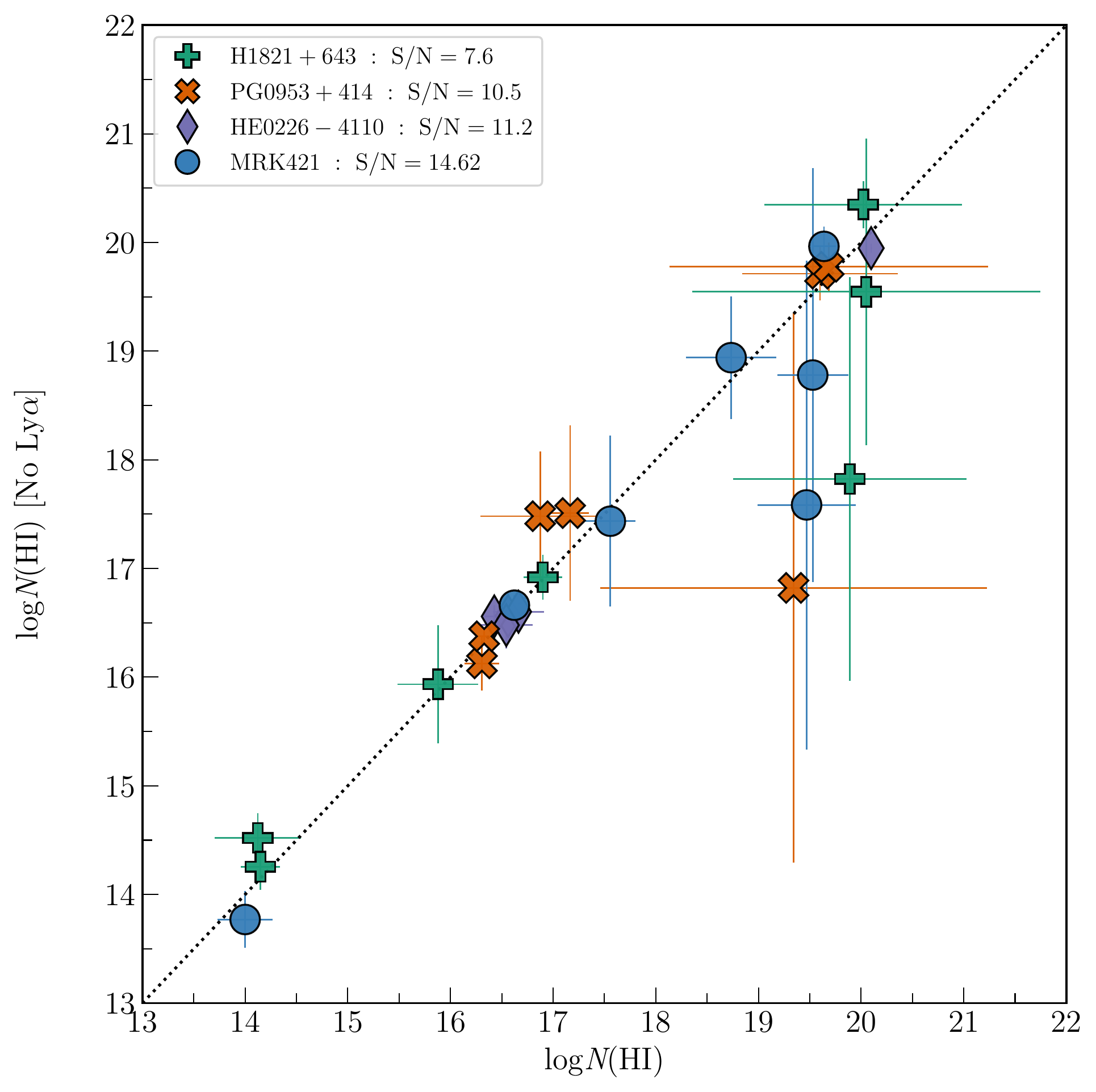}
  \caption{\small{A comparison of component column densities resulting from including or excluding Ly$\alpha$ data when fitting four sightlines. At high values of $N$(\HI), excluding Ly$\alpha$ often results in a substantial underestimate of the column density.}} 
  \label{fig:lya_comparison}
  \vspace{10pt}
\end{figure*}

We have tested the effect of excluding the Ly$\alpha$ transition from our fitting analysis. Ly$\alpha$ 1215.6700 lies outside the wavelength coverage of \fuse, thus including it requires obtaining spectra from \hst/COS or STIS. This adds significant overhead to the analysis, as well as 
complications due to combining spectra with different S/N, resolution, and line spread functions. To test if this is necessary we chose a sample of four sightlines of varying signal-to-noise, and fit them while both including and excluding Ly$\alpha$. The result is shown in Figure \ref{fig:lya_comparison}.

We find that below column densities of $\sim$10$^{19}$\sqcm, there is little difference between the fit results. Above this point, the no-Ly$\alpha$ fits often under-predict the column densities by substantial amounts (up to 3\,dex). All four sightlines show at least one component where the no-Ly$\alpha$ fit produced a column density 1$\sigma$ or more away from the Ly$\alpha$-included fits. For this reason we have included Ly$\alpha$ in the fitting procedure whenever data is available.

\section{Spectra and Fits}
\label{section:fits}

\begin{figure*}[h!]
\centering
 \includegraphics[width=0.595\linewidth]{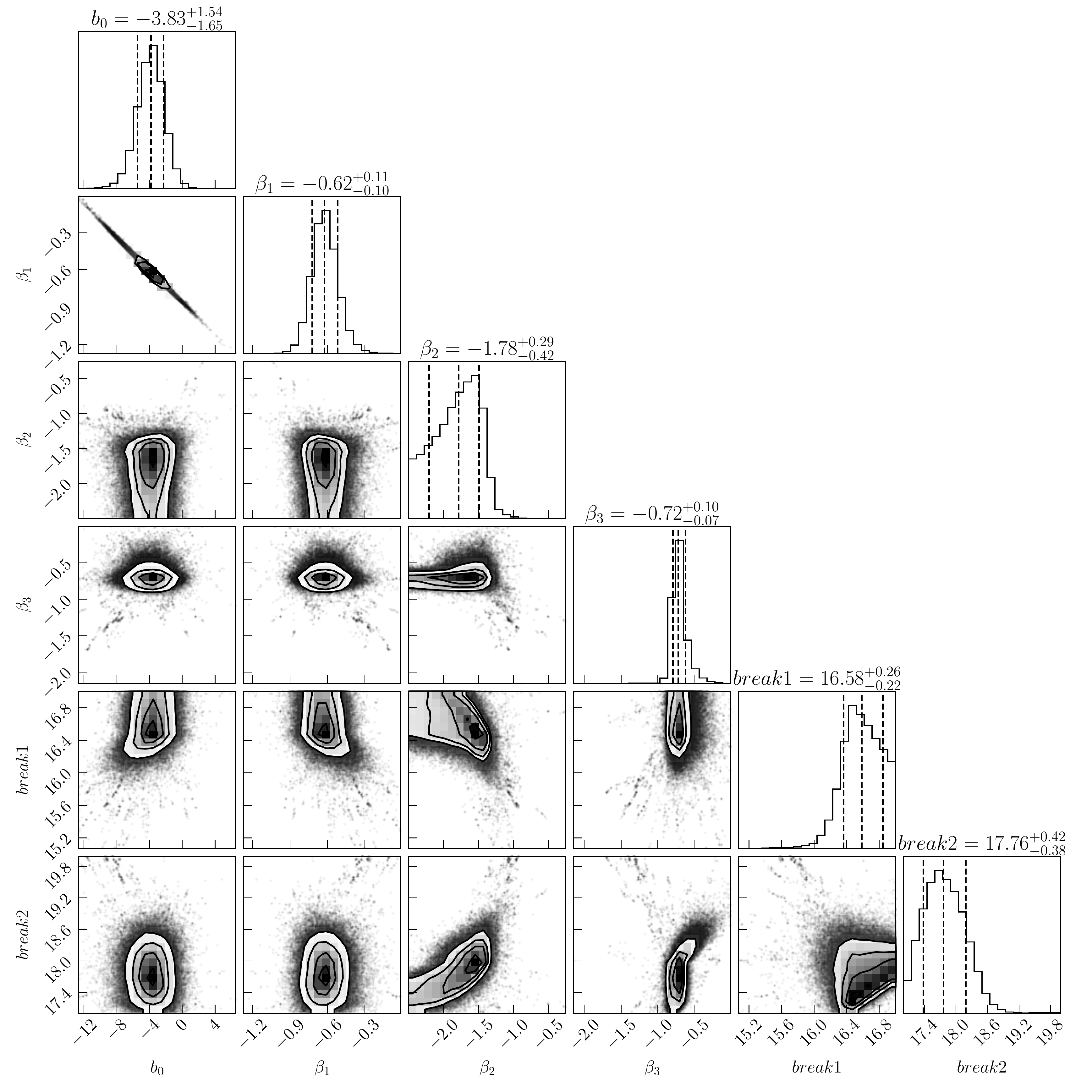}
 
 \includegraphics[width=0.595\linewidth]{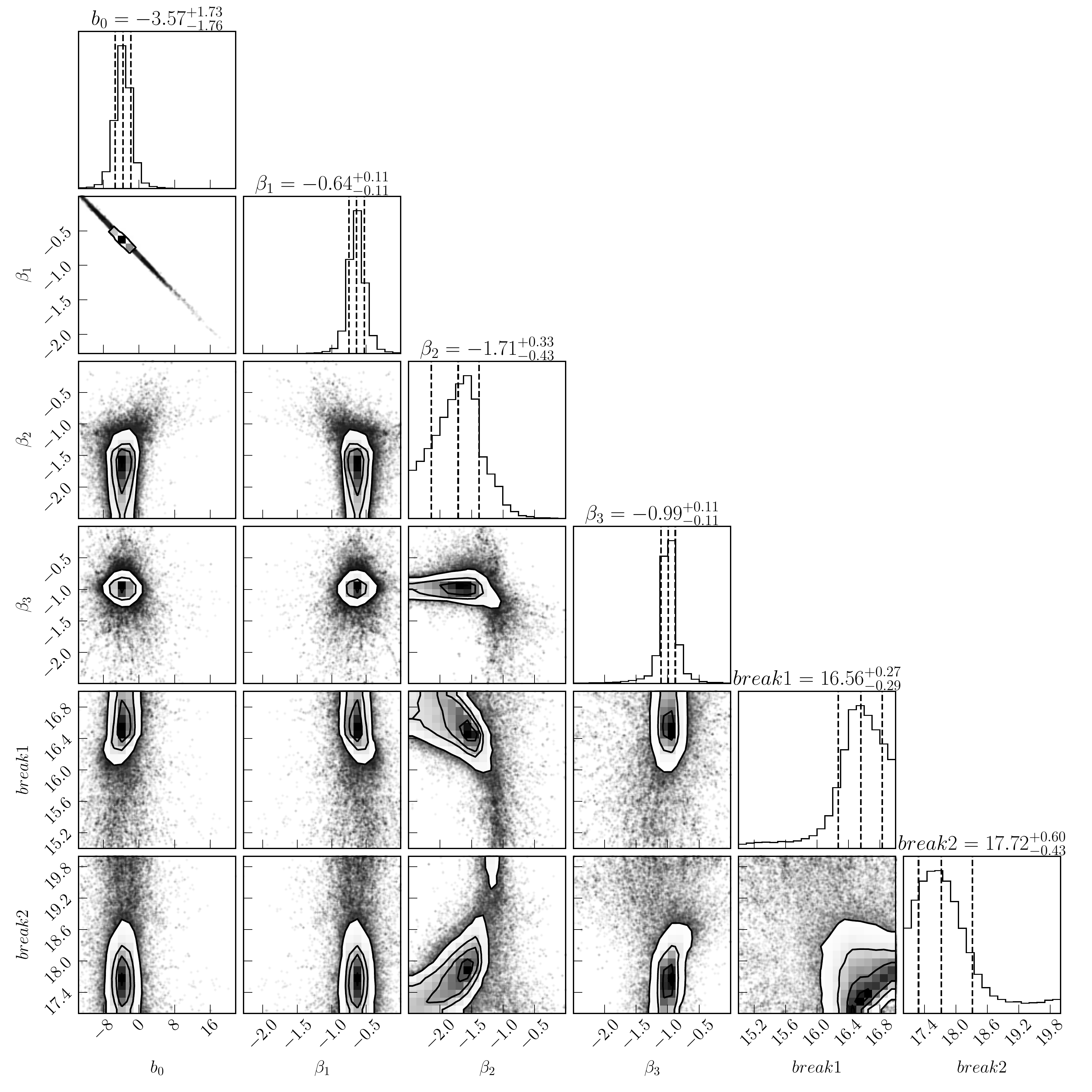}
  \caption{\small{Corner plots from our MCMC piecewise CDDF fits. \textbf{Top: } Full dataset fit including ISM, IVC, and HVC components (Galactic halo and disk). \textbf{Bottom: } Galactic halo only fit ($\rm v_{LSR} \ge 40$ \kms). \textbf{Both: } The histograms along the outer diagonals show the probability density functions (PDFs) for the normalization (i.e., y-intercept; $b_0$), low-, mid-, and high-column density slopes ($\beta_1$, $\beta_2$, $\beta_3$), and low- and high-column density break locations. The contour plots below each diagonal PDF show the joint posterior PDFs for the row and column.}} 
  \label{fig:corner}
\end{figure*}

Figure \ref{fig:corner} presents corner plots from our MCMC piecewise CDDF fits. The histograms along the outer diagonal show the probability density functions (PDFs) for the normalization (i.e., y-intercept; $b_0$), low-, mid-, and high-column density slopes ($\beta_1$, $\beta_2$, $\beta_3$), and low- and high-column density break locations. The contour plots below each diagonal PDF show the joint posterior PDFs for the row and column.

\begin{figure*}[ht!]
\centering
 \includegraphics[width=0.85\linewidth]{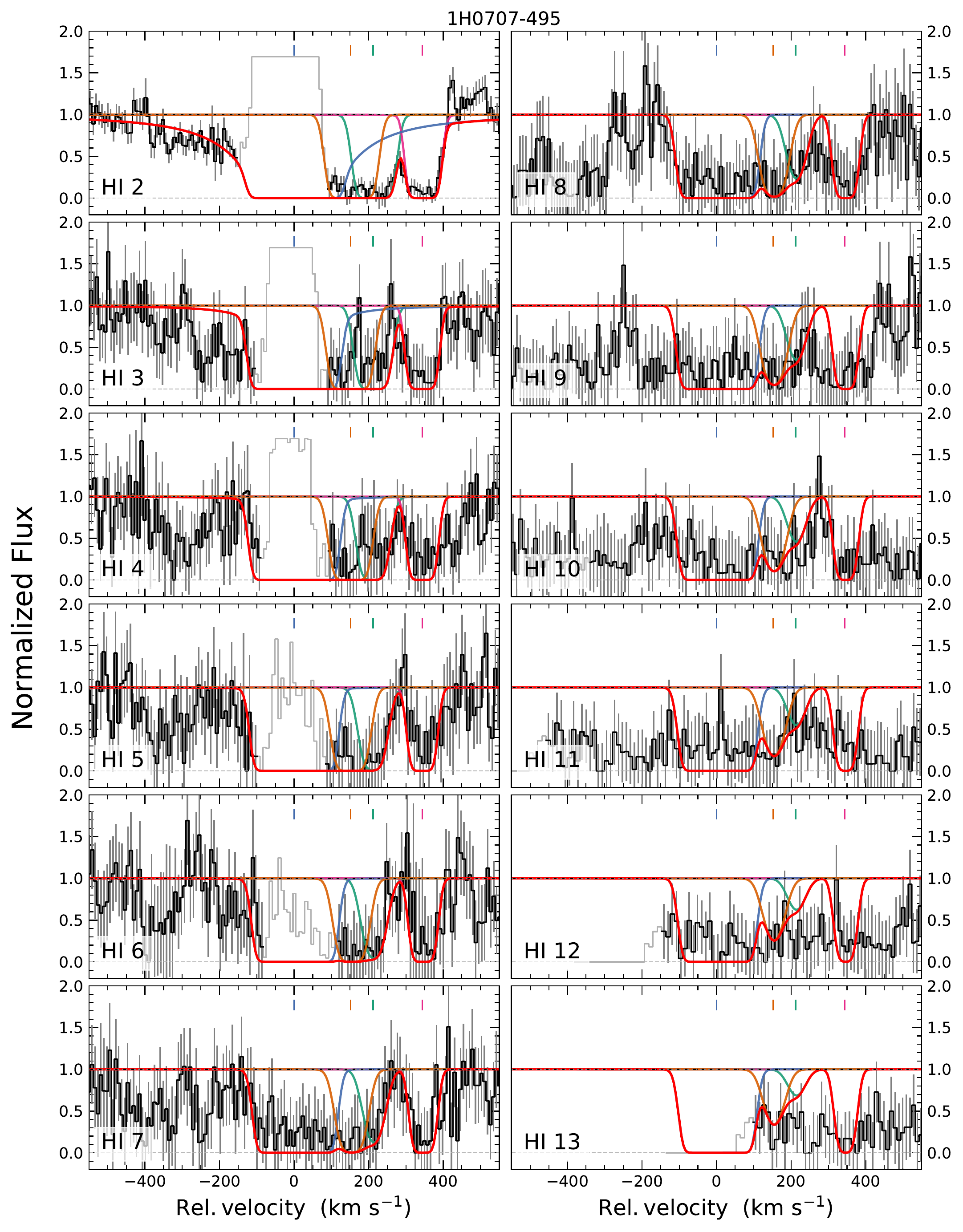}
  \caption{\small{Fits for 1H0707-495. The data are shown in black, errors and/or masked regions in light grey, and the composite fit in red. Each contributing component is plotted with a unique color, and the matching tick marks in the top of each panel show the centroid velocity. Note: No \hst\/COS or STIS data is available for this sightline covering the Ly$\alpha$ transition.}}
  \label{fig:1H0707-495N}
  \vspace{10pt}
\end{figure*}

\begin{figure*}[ht!]
\centering
 \includegraphics[width=0.9\linewidth]{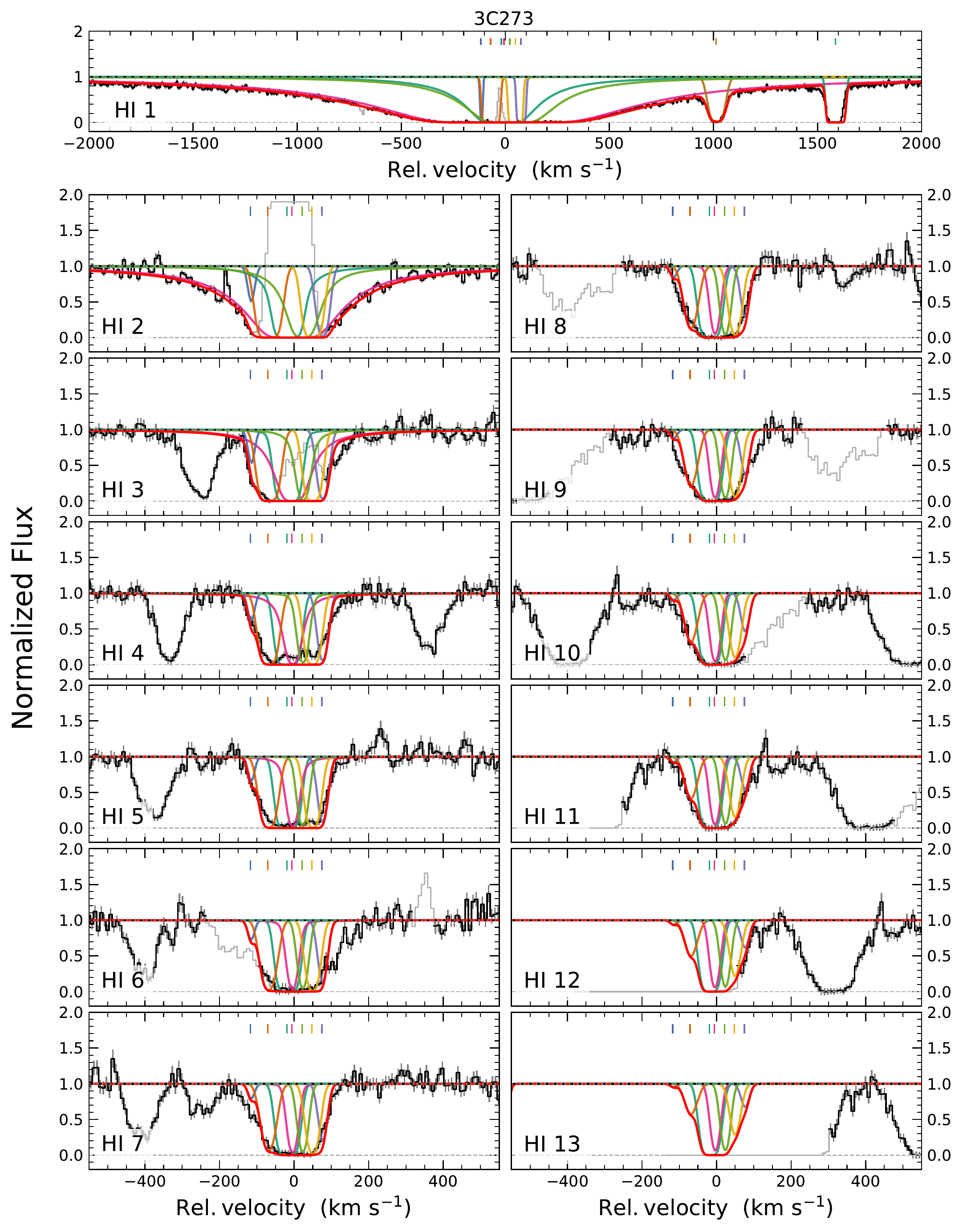}
  \caption{\small{Fits for 3C273. The data are shown in black, errors and/or masked regions in light grey, and the composite fit in red. Each contributing component is plotted with a unique color, and the matching tick marks in the top of each panel show the centroid velocity.}}
  \label{fig:3C273}
  \vspace{10pt}
\end{figure*}

\begin{figure*}[ht!]
\centering
 \includegraphics[width=0.9\linewidth]{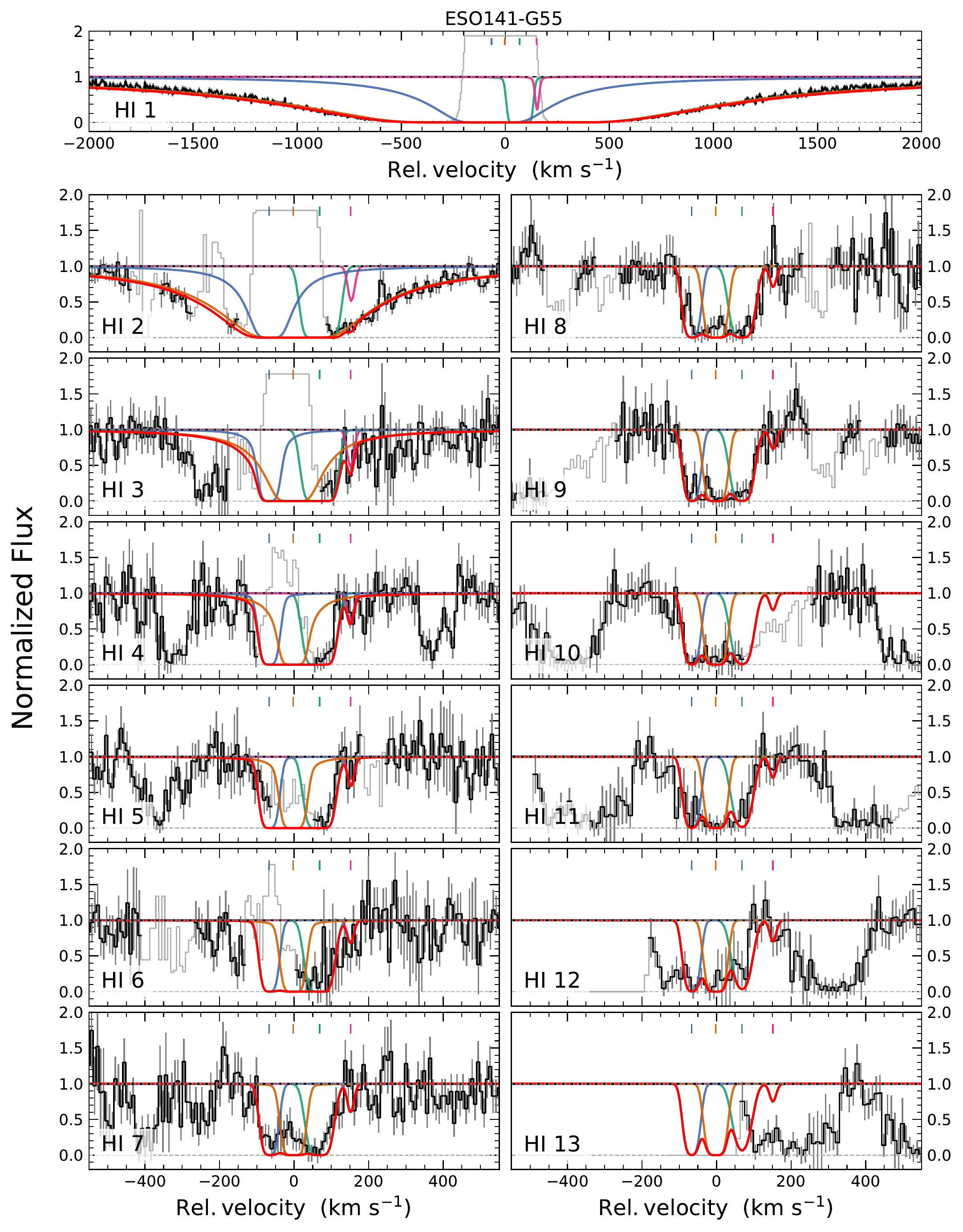}
  \caption{\small{Fits for ESO141-G55. The data are shown in black, errors and/or masked regions in light grey, and the composite fit in red. Each contributing component is plotted with a unique color, and the matching tick marks in the top of each panel show the centroid velocity. Note: Ly$\alpha$ lies on a non-linear region of the spectra due to the low redshift of ESO141-G55, resulting in an uncertain continuum placement.}}
  \label{fig:ESO141-G55}
  \vspace{10pt}
\end{figure*}

\begin{figure*}[ht!]
\centering
 \includegraphics[width=0.9\linewidth]{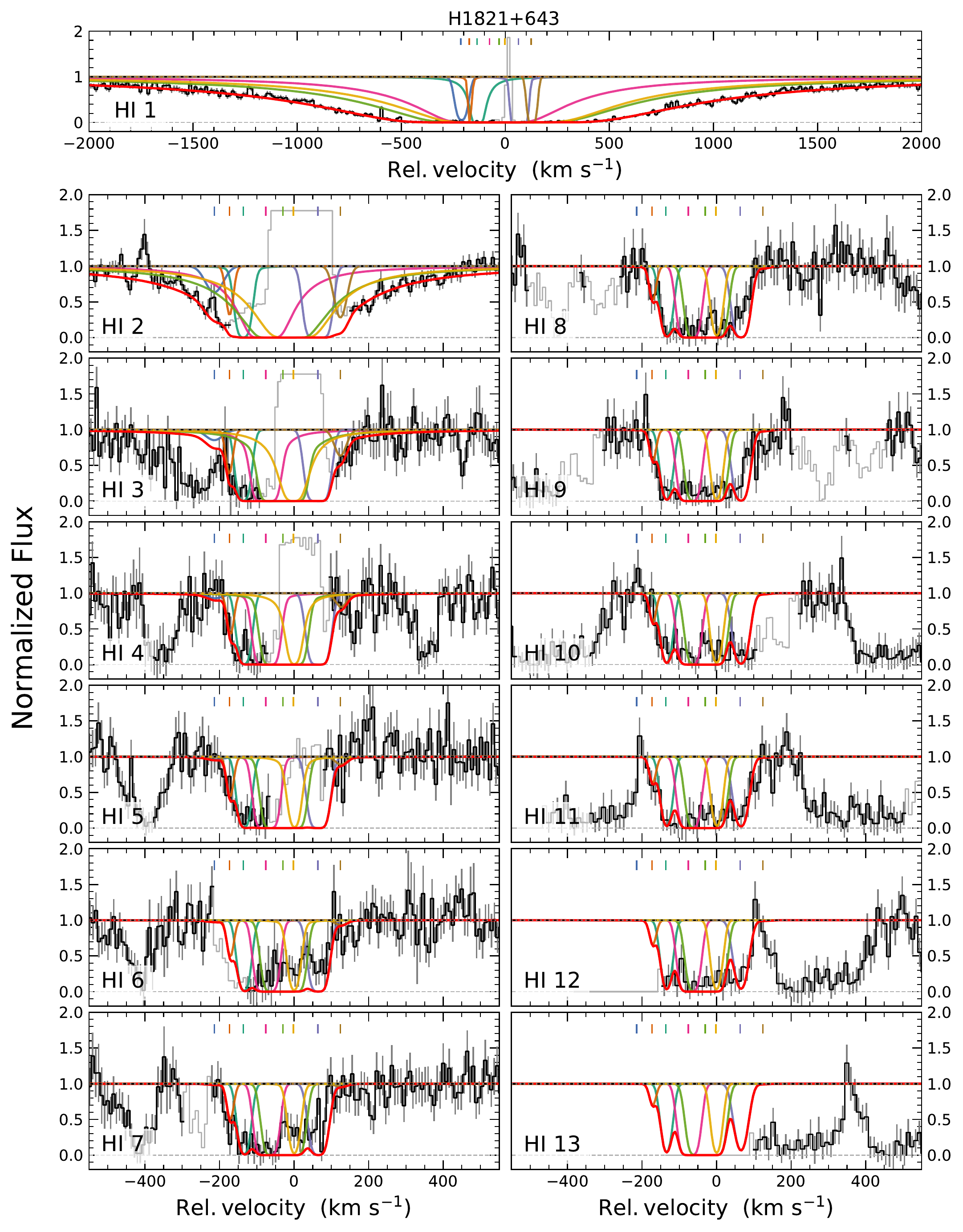}
  \caption{\small{Fits for H1821+643. The data are shown in black, errors and/or masked regions in light grey, and the composite fit in red. Each contributing component is plotted with a unique color, and the matching tick marks in the top of each panel show the centroid velocity.}}
  \label{fig:H1821+643}
  \vspace{10pt}
\end{figure*}

\begin{figure*}[ht!]
\centering
 \includegraphics[width=0.9\linewidth]{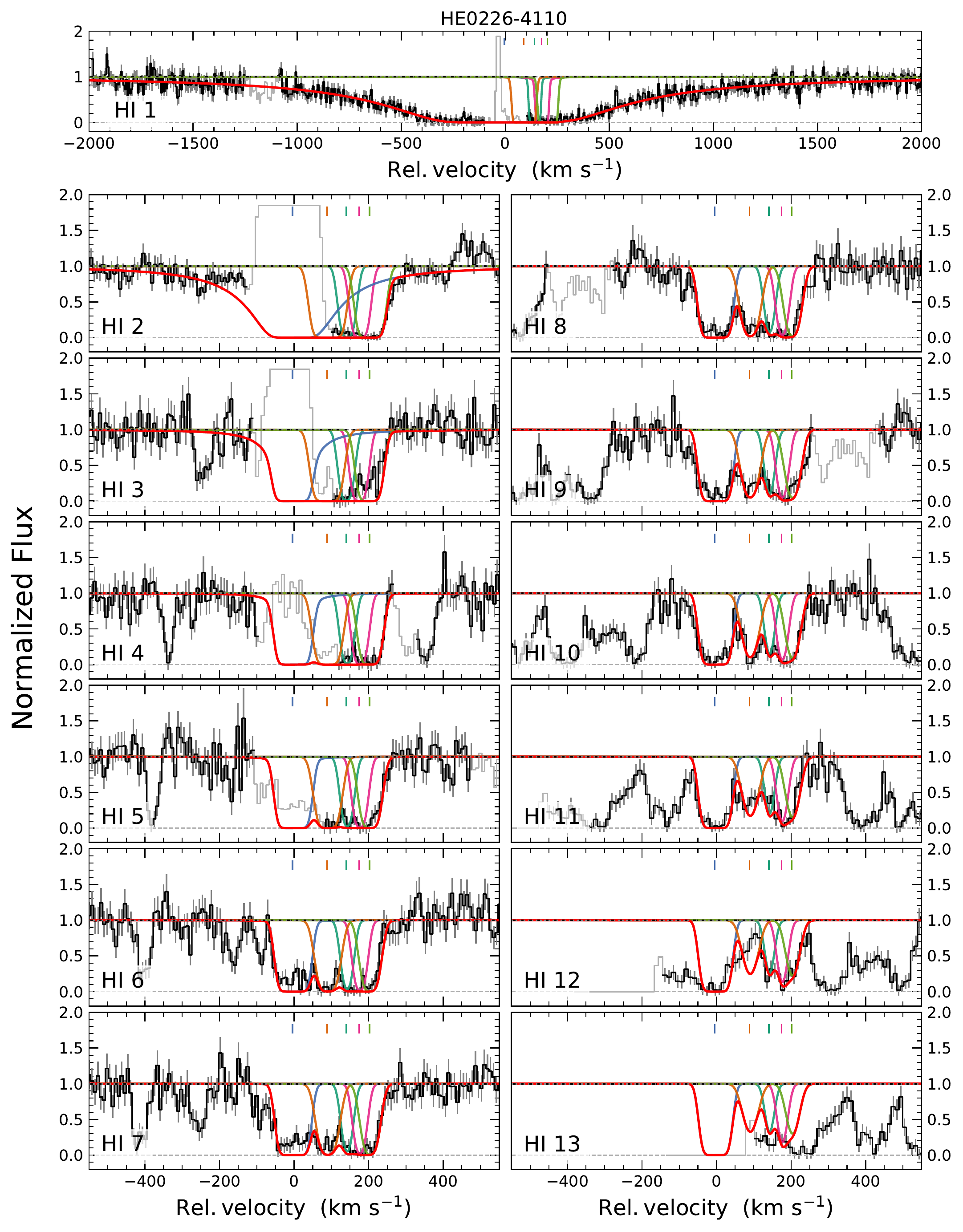}
  \caption{\small{Fits for HE0226-4110. The data are shown in black, errors and/or masked regions in light grey, and the composite fit in red. Each contributing component is plotted with a unique color, and the matching tick marks in the top of each panel show the centroid velocity.}}
  \label{fig:HE0226-4110}
  \vspace{10pt}
\end{figure*}

\begin{figure*}[ht!]
\centering
 \includegraphics[width=0.9\linewidth]{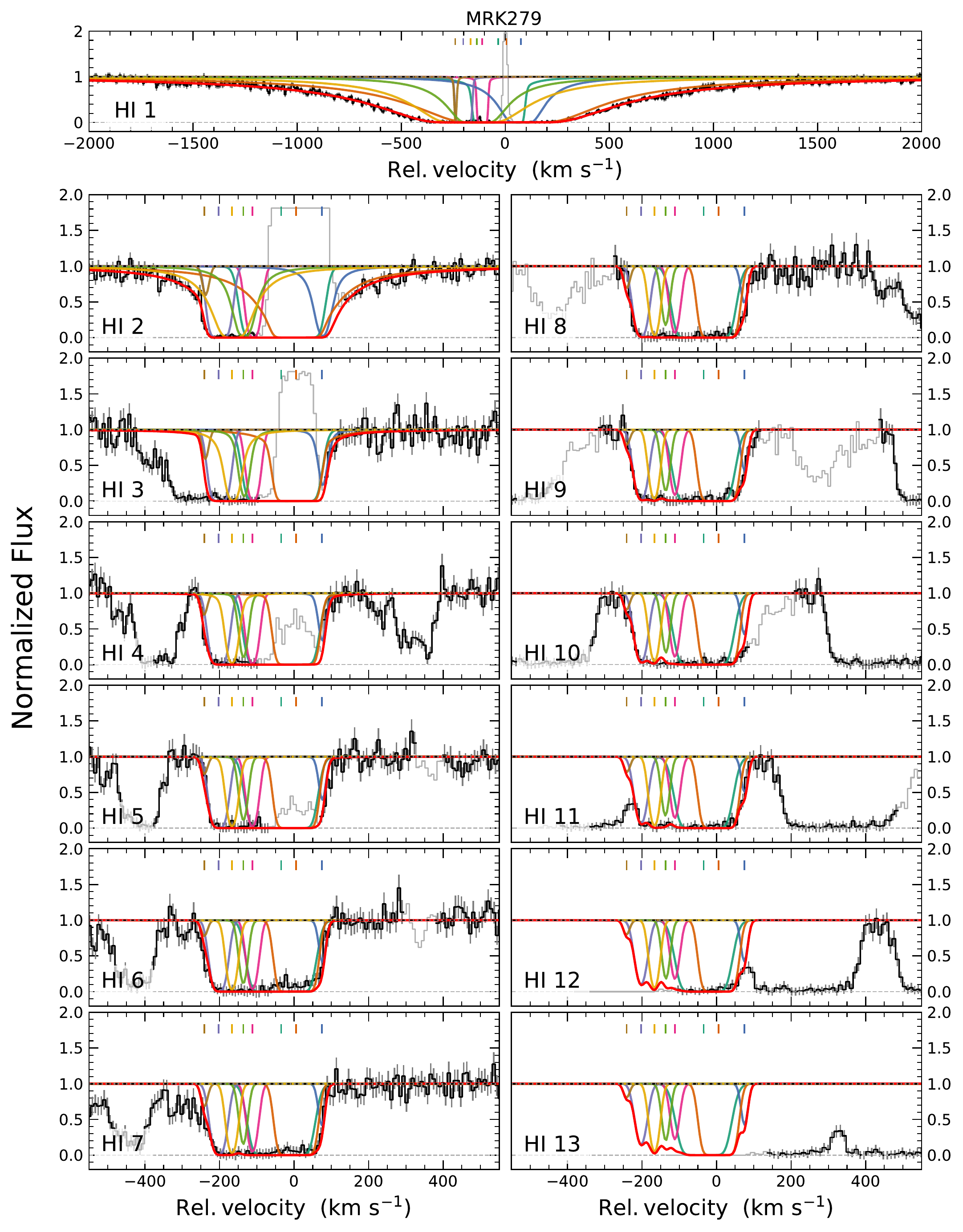}
  \caption{\small{Fits for MRK279. The data are shown in black, errors and/or masked regions in light grey, and the composite fit in red. Each contributing component is plotted with a unique color, and the matching tick marks in the top of each panel show the centroid velocity.}}
  \label{fig:MRK279}
  \vspace{10pt}
\end{figure*}

\begin{figure*}[ht!]
\centering
 \includegraphics[width=0.9\linewidth]{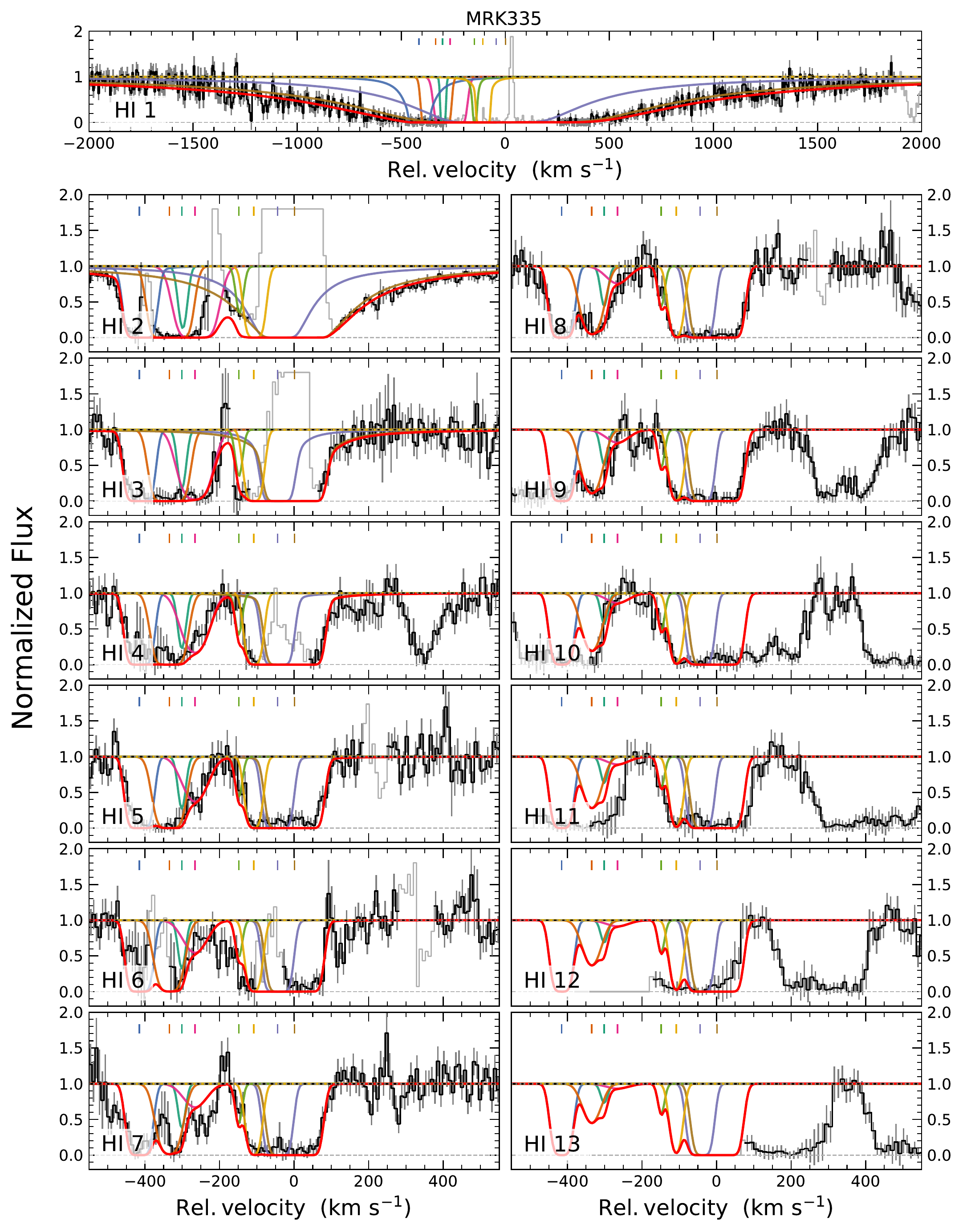}
  \caption{\small{Fits for MRK335. The data are shown in black, errors and/or masked regions in light grey, and the composite fit in red. Each contributing component is plotted with a unique color, and the matching tick marks in the top of each panel show the centroid velocity.}}
  \label{fig:MRK335}
  \vspace{10pt}
\end{figure*}

\begin{figure*}[ht!]
\centering
 \includegraphics[width=0.9\linewidth]{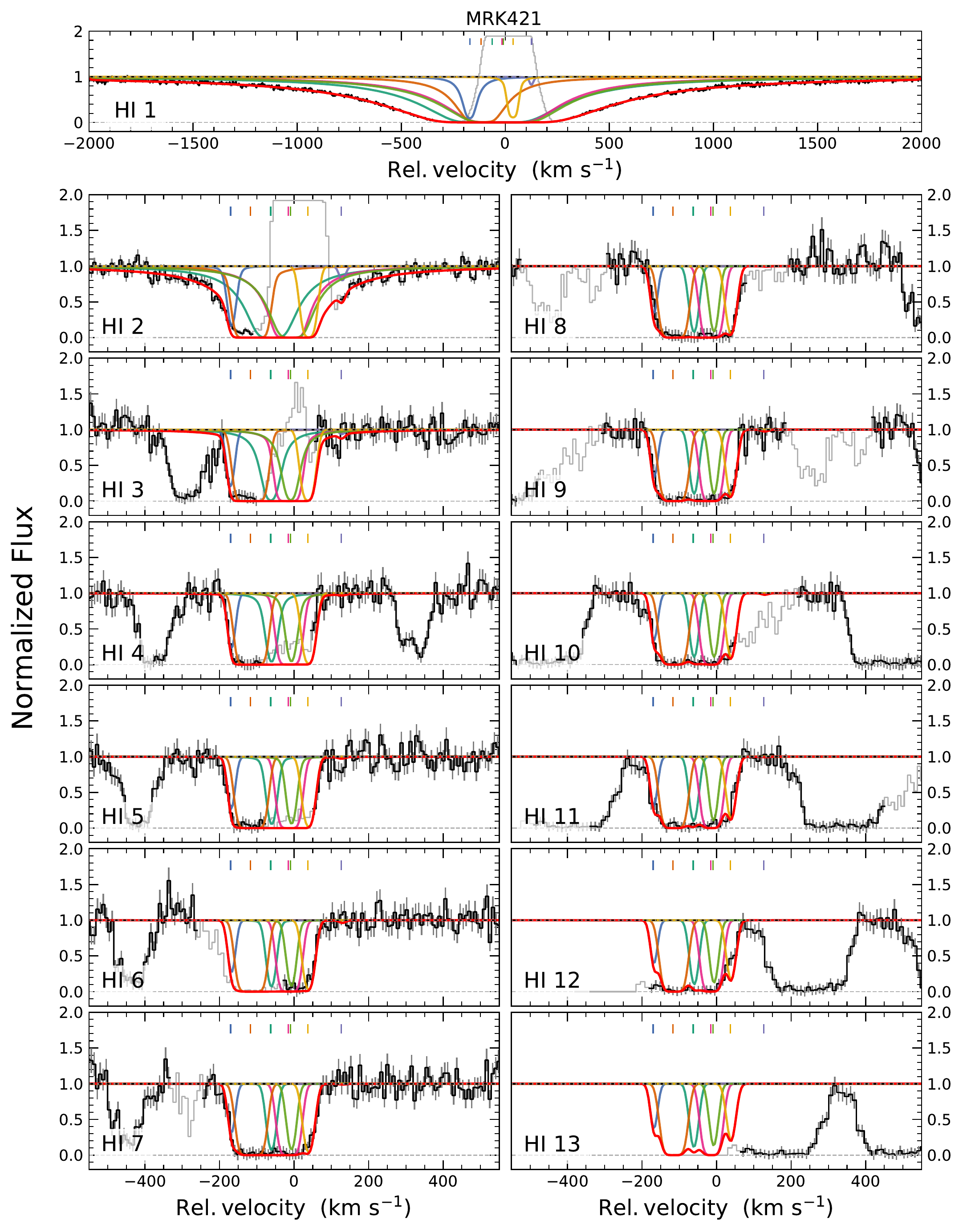}
  \caption{\small{Fits for MRK421. The data are shown in black, errors and/or masked regions in light grey, and the composite fit in red. Each contributing component is plotted with a unique color, and the matching tick marks in the top of each panel show the centroid velocity.}}
  \label{fig:MRK421}
  \vspace{10pt}
\end{figure*}

\begin{figure*}[ht!]
\centering
 \includegraphics[width=0.9\linewidth]{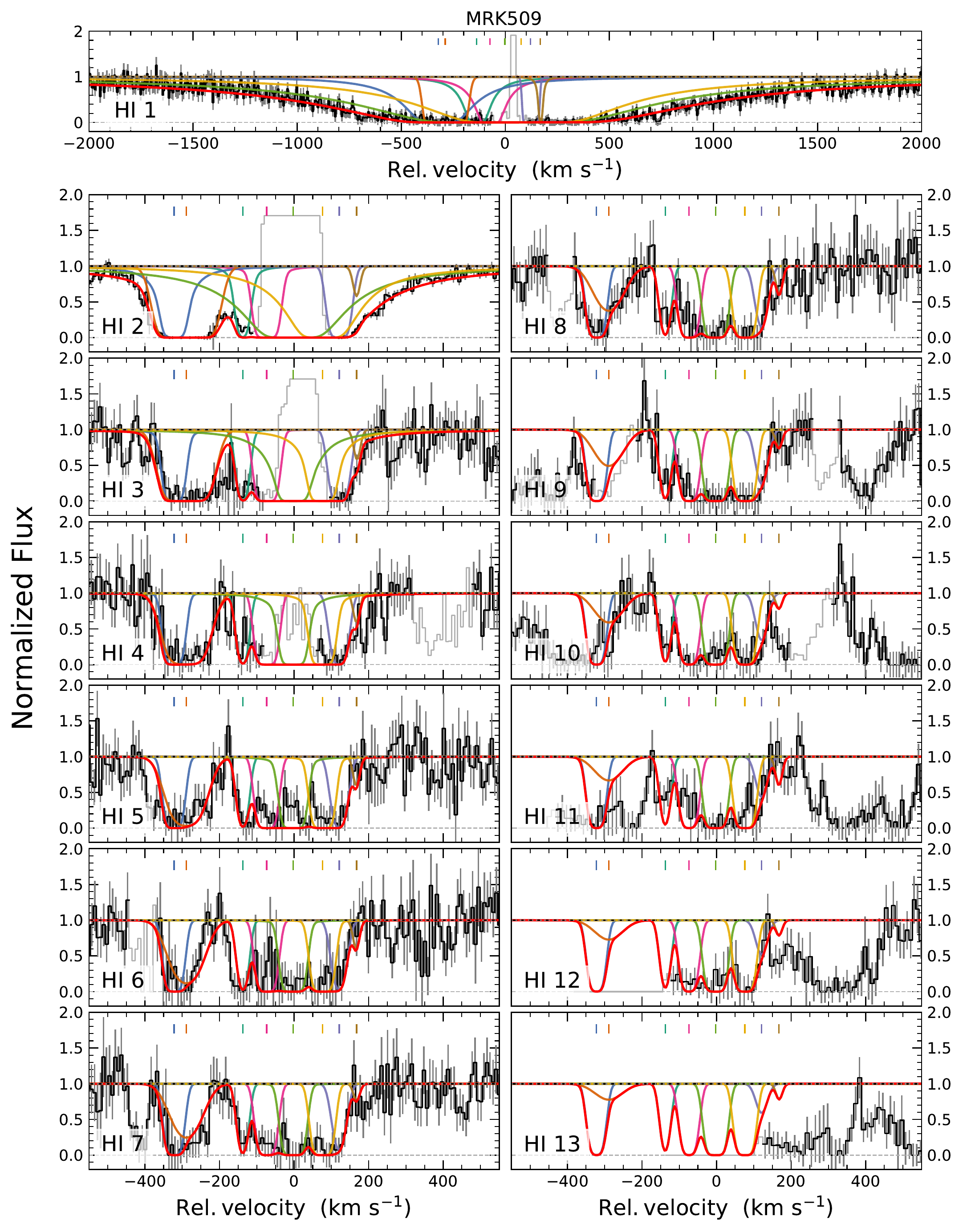}
  \caption{\small{Fits for MRK509. The data are shown in black, errors and/or masked regions in light grey, and the composite fit in red. Each contributing component is plotted with a unique color, and the matching tick marks in the top of each panel show the centroid velocity.}}
  \label{fig:MRK509}
  \vspace{10pt}
\end{figure*}

\begin{figure*}[ht!]
\centering
 \includegraphics[width=0.9\linewidth]{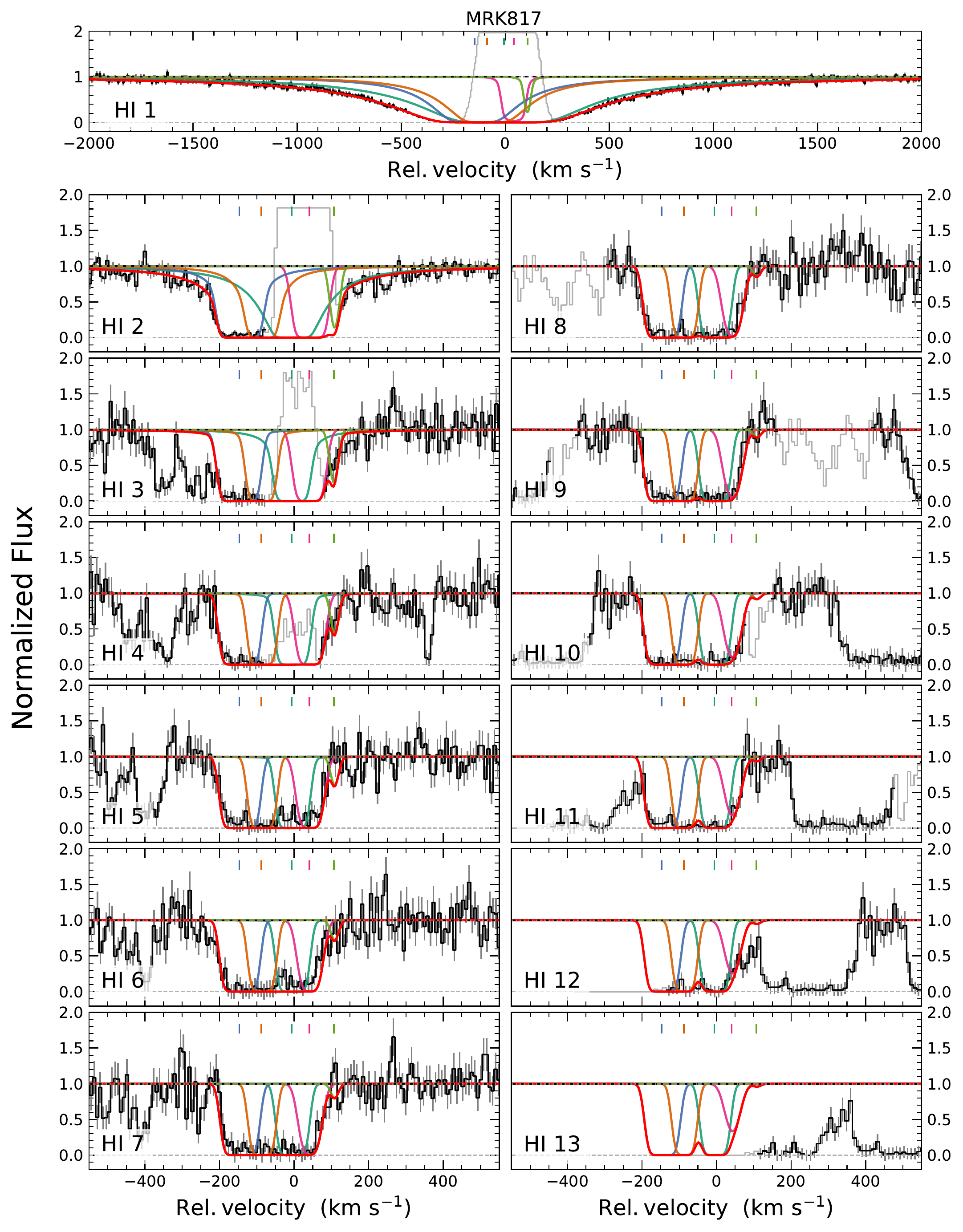}
  \caption{\small{Fits for MRK817. The data are shown in black, errors and/or masked regions in light grey, and the composite fit in red. Each contributing component is plotted with a unique color, and the matching tick marks in the top of each panel show the centroid velocity.}}
  \label{fig:MRK817}
  \vspace{10pt}
\end{figure*}

\begin{figure*}[ht!]
\centering
 \includegraphics[width=0.9\linewidth]{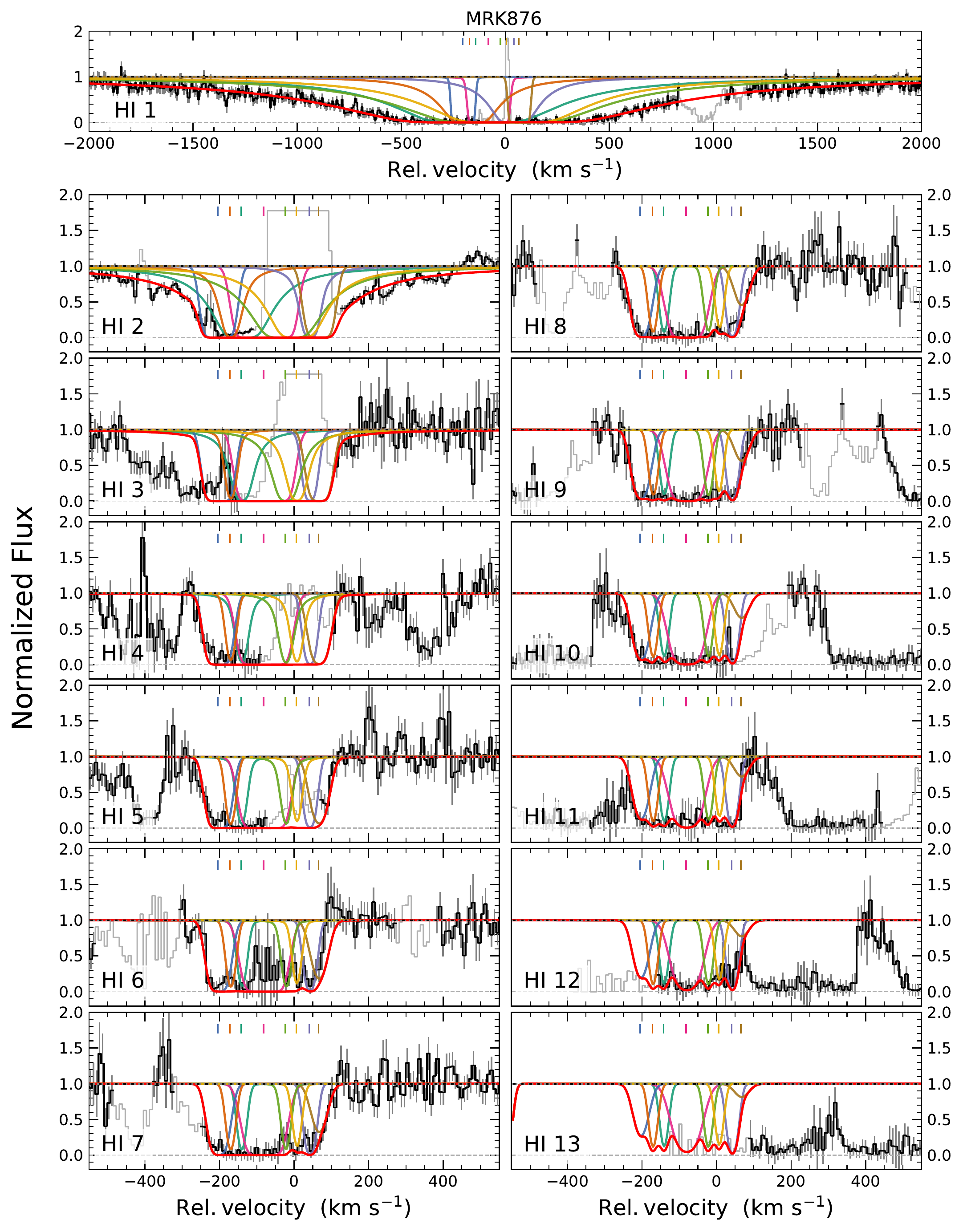}
  \caption{\small{Fits for MRK876. The data are shown in black, errors and/or masked regions in light grey, and the composite fit in red. Each contributing component is plotted with a unique color, and the matching tick marks in the top of each panel show the centroid velocity.}}
  \label{fig:MRK876}
  \vspace{10pt}
\end{figure*}

\begin{figure*}[ht!]
\centering
 \includegraphics[width=0.9\linewidth]{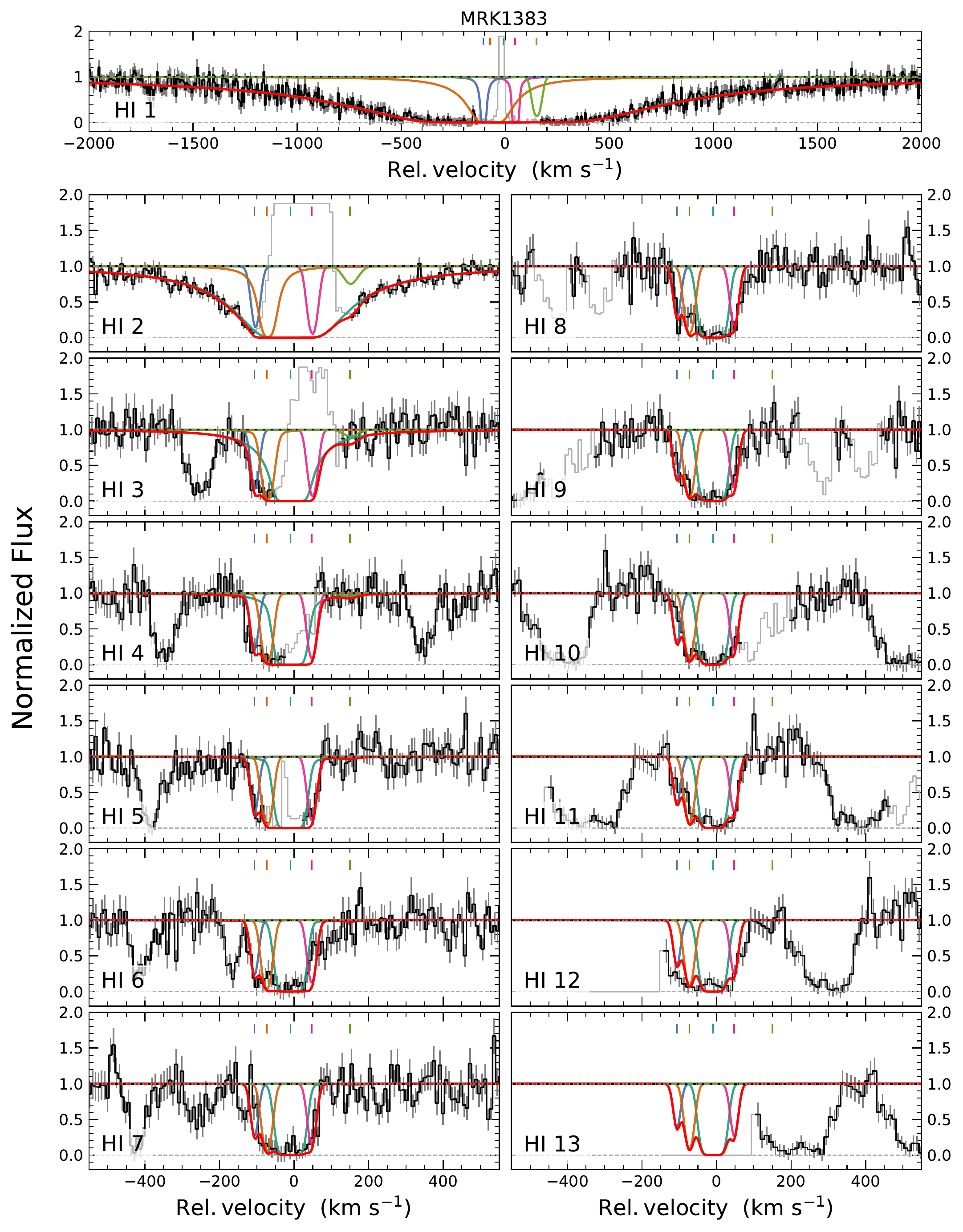}
  \caption{\small{Fits for MRK1383. The data are shown in black, errors and/or masked regions in light grey, and the composite fit in red. Each contributing component is plotted with a unique color, and the matching tick marks in the top of each panel show the centroid velocity.}}
  \label{fig:MRK1383}
  \vspace{10pt}
\end{figure*}

\begin{figure*}[ht!]
\centering
 \includegraphics[width=0.9\linewidth]{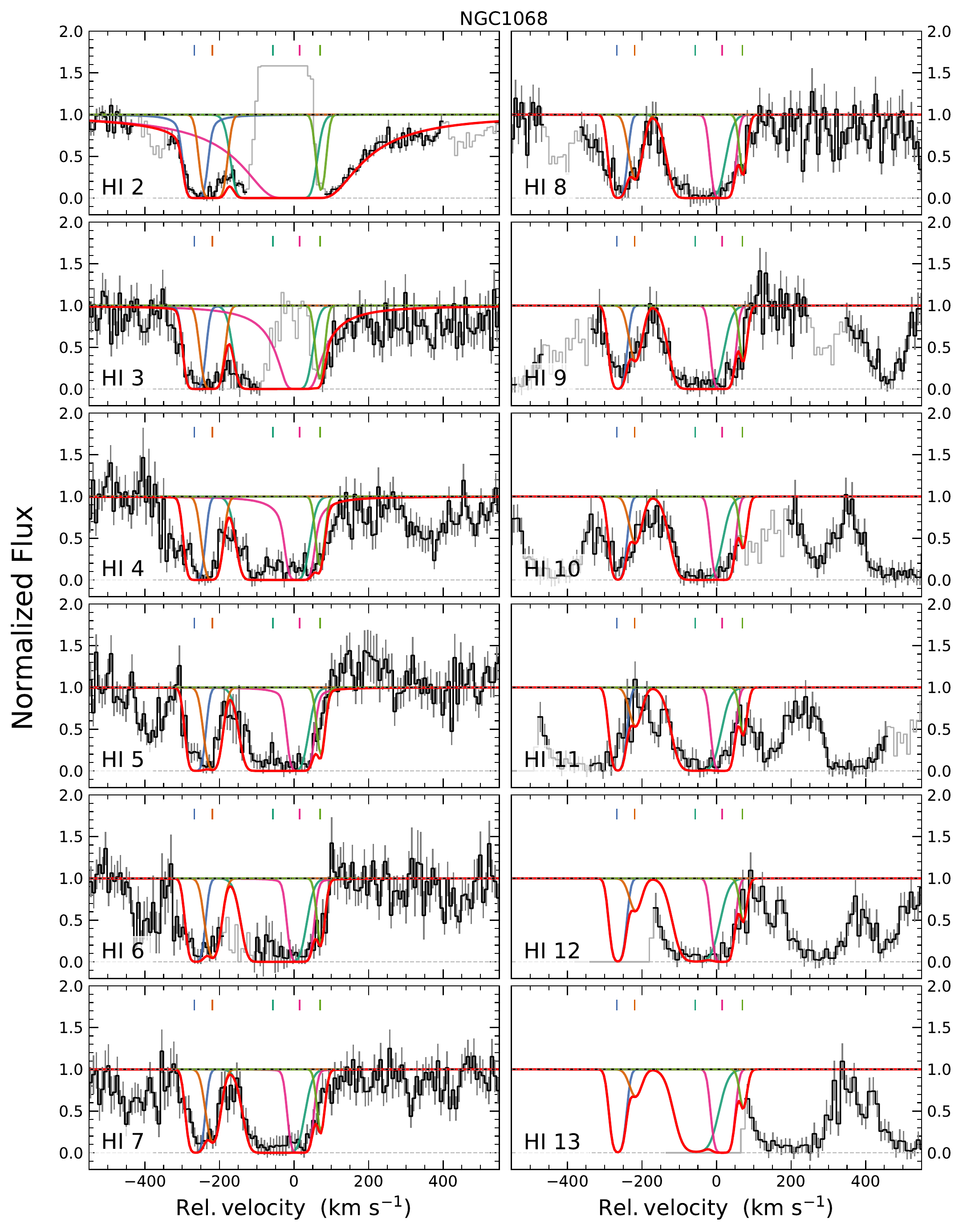}
  \caption{\small{Fits for NGC1068. The data are shown in black, errors and/or masked regions in light grey, and the composite fit in red. Each contributing component is plotted with a unique color, and the matching tick marks in the top of each panel show the centroid velocity.}}
  \label{fig:NGC1068}
  \vspace{10pt}
\end{figure*}

\begin{figure*}[ht!]
\centering
 \includegraphics[width=0.9\linewidth]{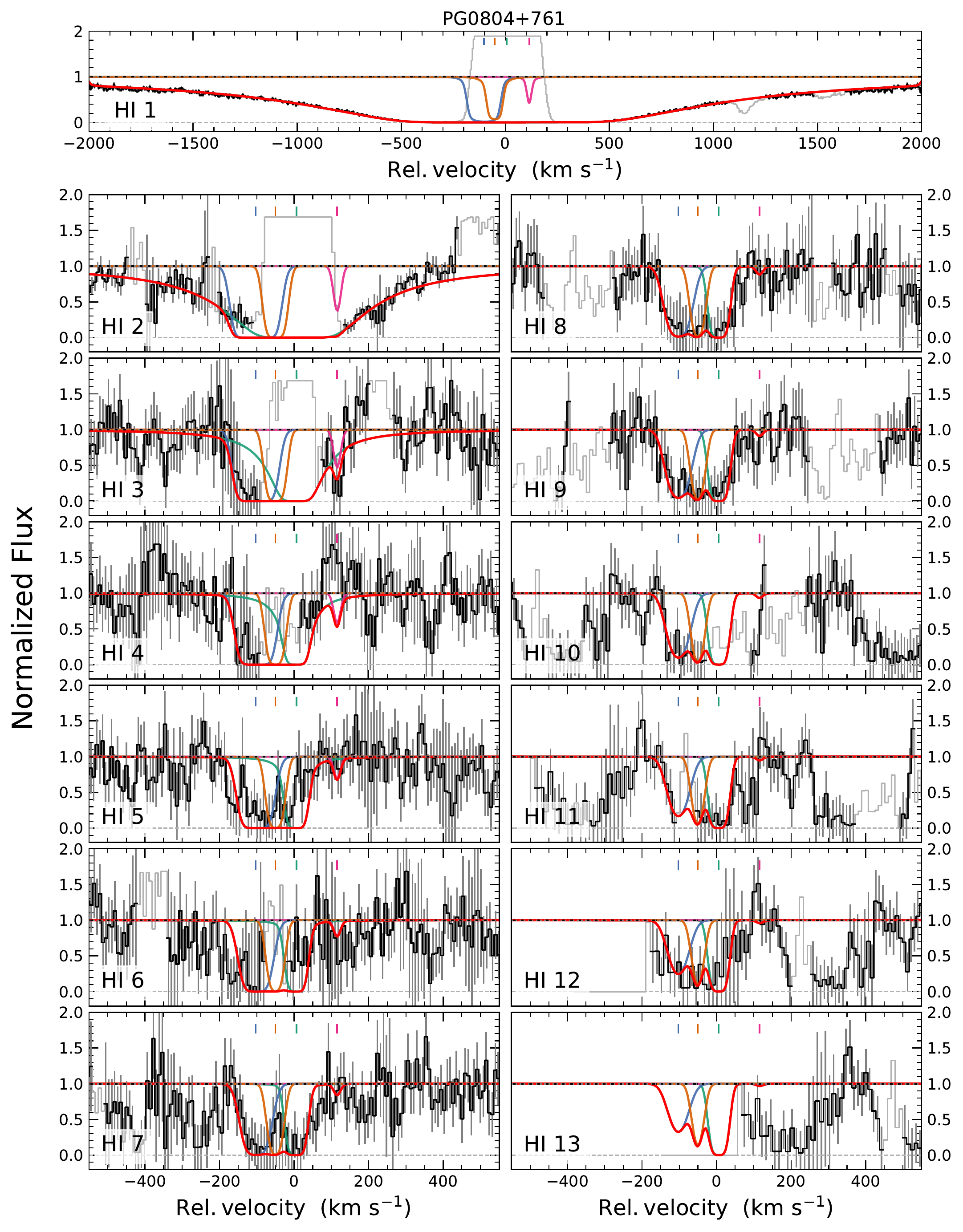} 
  \caption{\small{Fits for PG0804+761. The data are shown in black, errors and/or masked regions in light grey, and the composite fit in red. Each contributing component is plotted with a unique color, and the matching tick marks in the top of each panel show the centroid velocity.}}
  \label{fig:PG0804+761}
  \vspace{10pt}
\end{figure*}

\begin{figure*}[ht!]
\centering
 \includegraphics[width=0.9\linewidth]{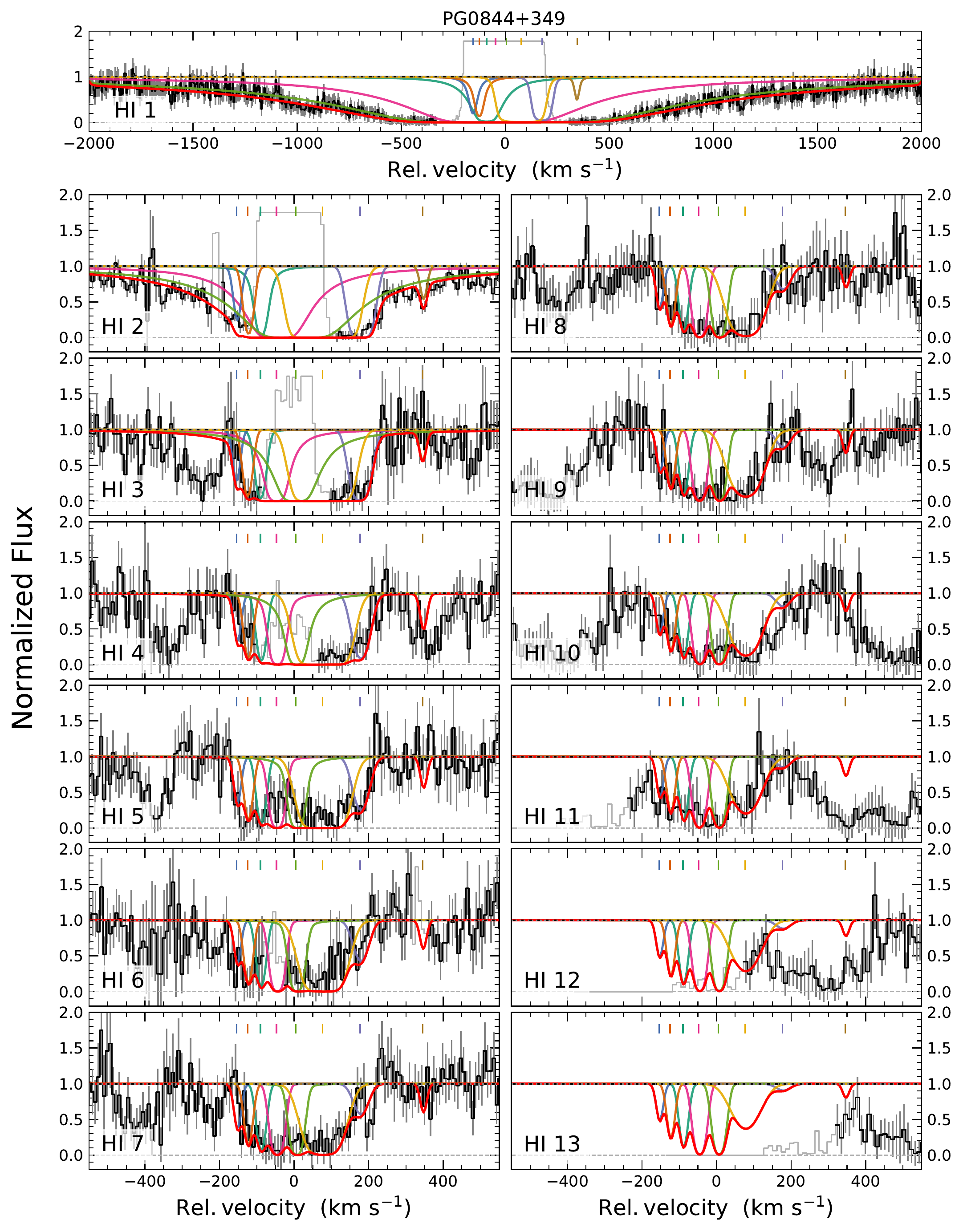}
  \caption{\small{Fits for PG0844+349. The data are shown in black, errors and/or masked regions in light grey, and the composite fit in red. Each contributing component is plotted with a unique color, and the matching tick marks in the top of each panel show the centroid velocity.}}
  \label{fig:PG0844+349}
  \vspace{10pt}
\end{figure*}

\begin{figure*}[ht!]
\centering
 \includegraphics[width=0.9\linewidth]{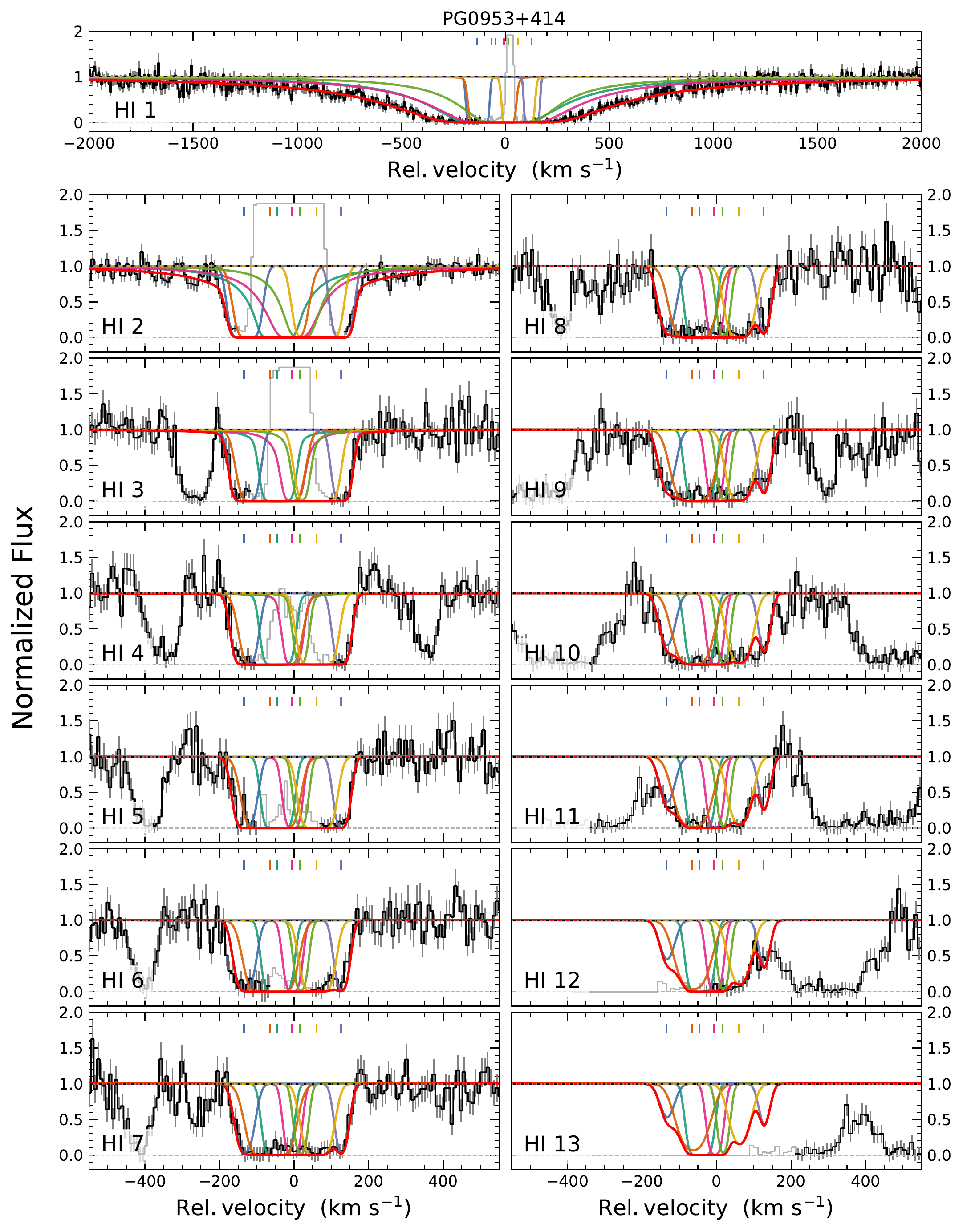}
  \caption{\small{Fits for PG0953+414. The data are shown in black, errors and/or masked regions in light grey, and the composite fit in red. Each contributing component is plotted with a unique color, and the matching tick marks in the top of each panel show the centroid velocity.}}
  \label{fig:PG0953+414}
  \vspace{10pt}
\end{figure*}

\begin{figure*}[ht!]
\centering
 \includegraphics[width=0.9\linewidth]{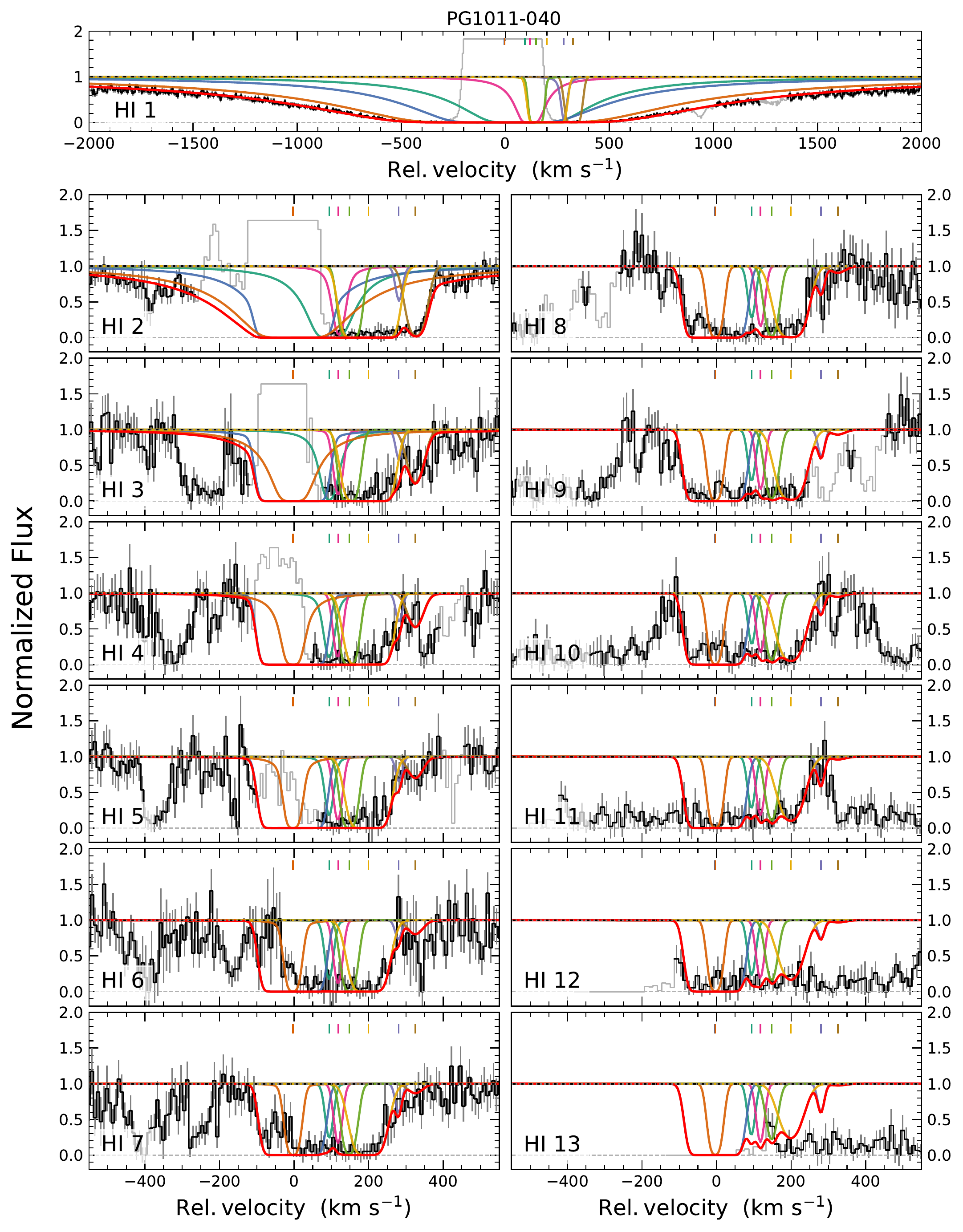}
  \caption{\small{Fits for PG1011-040. The data are shown in black, errors and/or masked regions in light grey, and the composite fit in red. Each contributing component is plotted with a unique color, and the matching tick marks in the top of each panel show the centroid velocity.}}
  \label{fig:PG1011-040}
  \vspace{10pt}
\end{figure*}

\begin{figure*}[ht!]
\centering
 \includegraphics[width=0.9\linewidth]{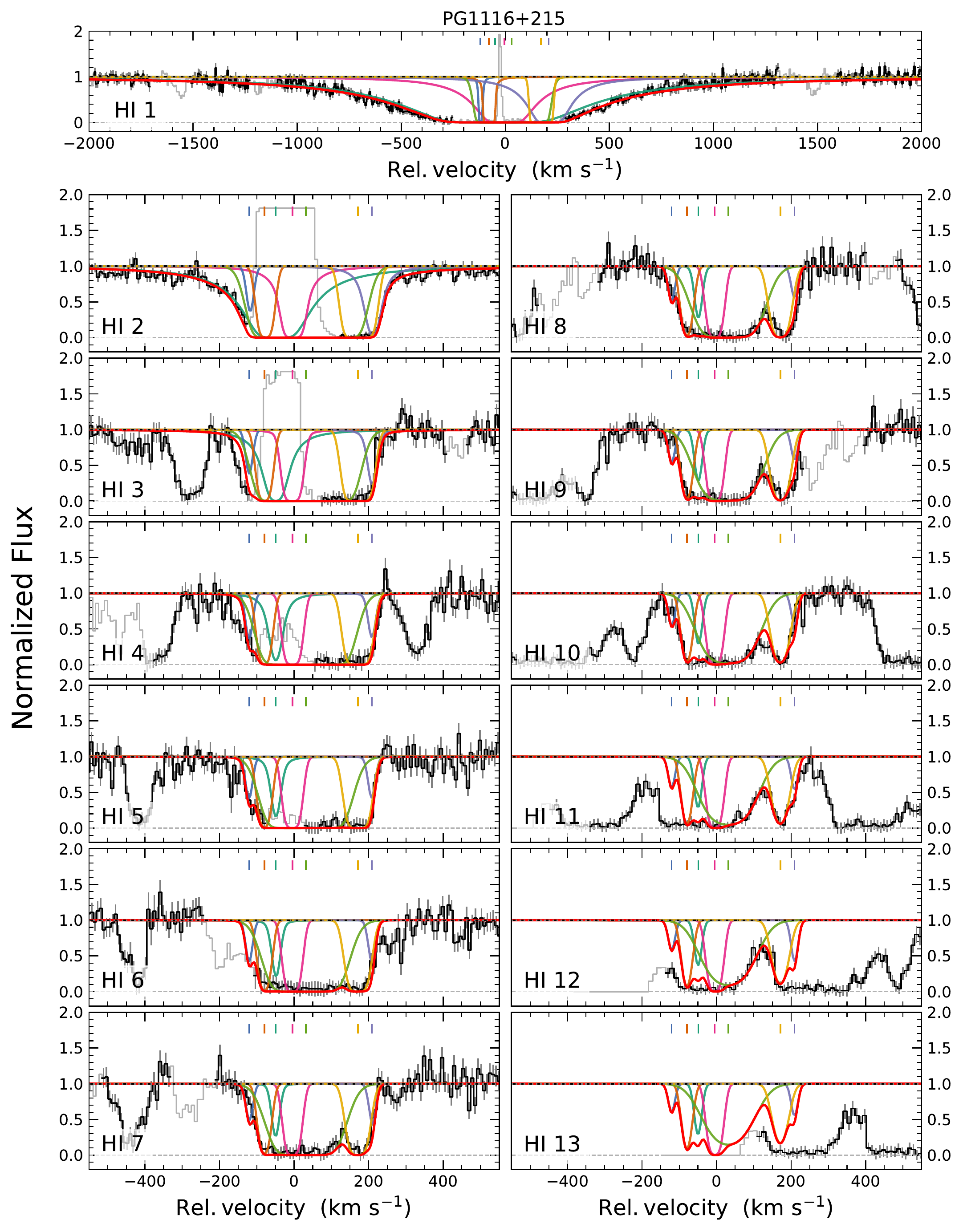}
  \caption{\small{Fits for PG1116+215. The data are shown in black, errors and/or masked regions in light grey, and the composite fit in red. Each contributing component is plotted with a unique color, and the matching tick marks in the top of each panel show the centroid velocity.}}
  \label{fig:PG1116+215}
  \vspace{10pt}
\end{figure*}

\begin{figure*}[ht!]
\centering
 \includegraphics[width=0.9\linewidth]{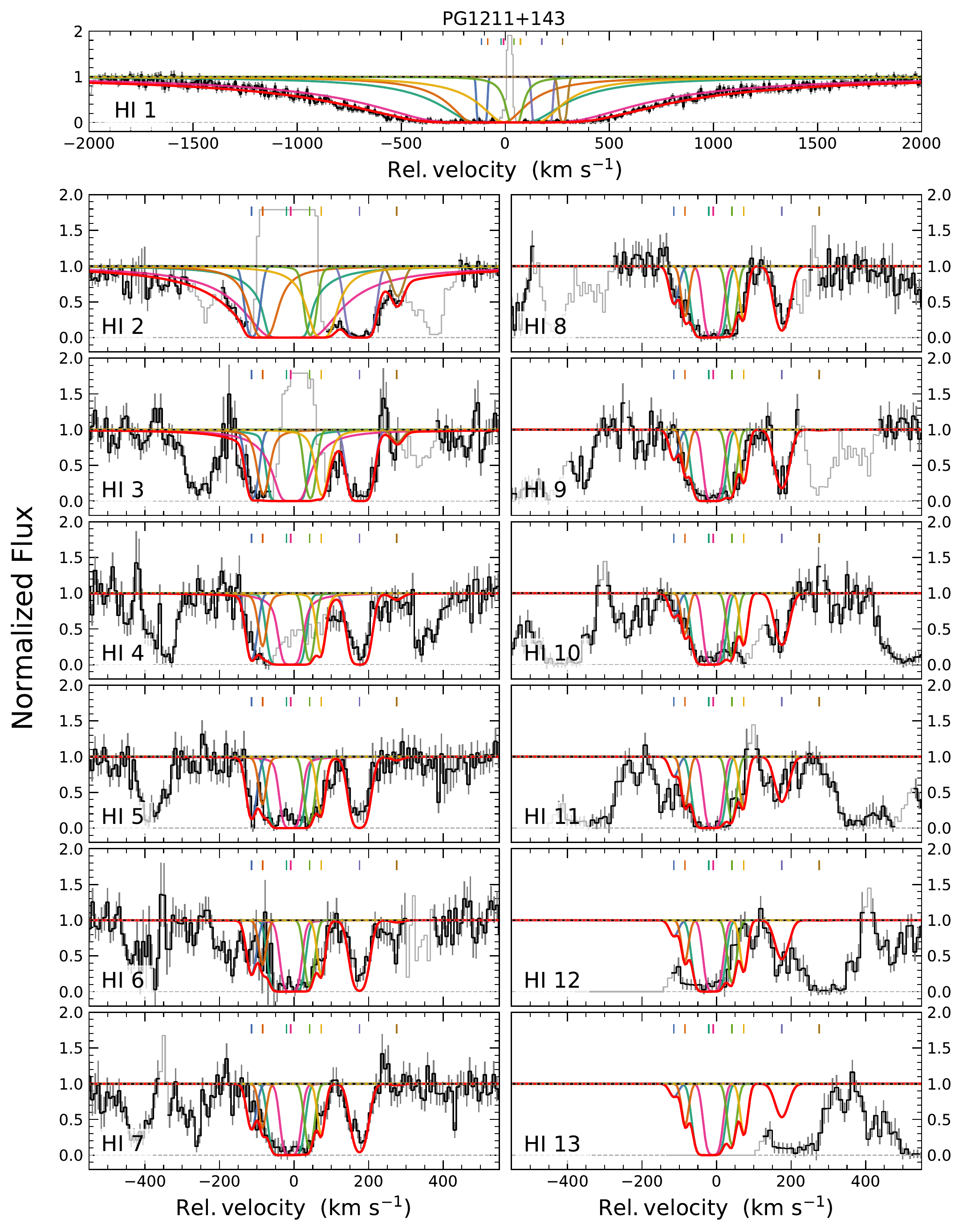}
  \caption{\small{Fits for PG1211+143. The data are shown in black, errors and/or masked regions in light grey, and the composite fit in red. Each contributing component is plotted with a unique color, and the matching tick marks in the top of each panel show the centroid velocity.}}
  \label{fig:PG1211+143}
  \vspace{10pt}
\end{figure*}

\begin{figure*}[ht!]
\centering
 \includegraphics[width=0.9\linewidth]{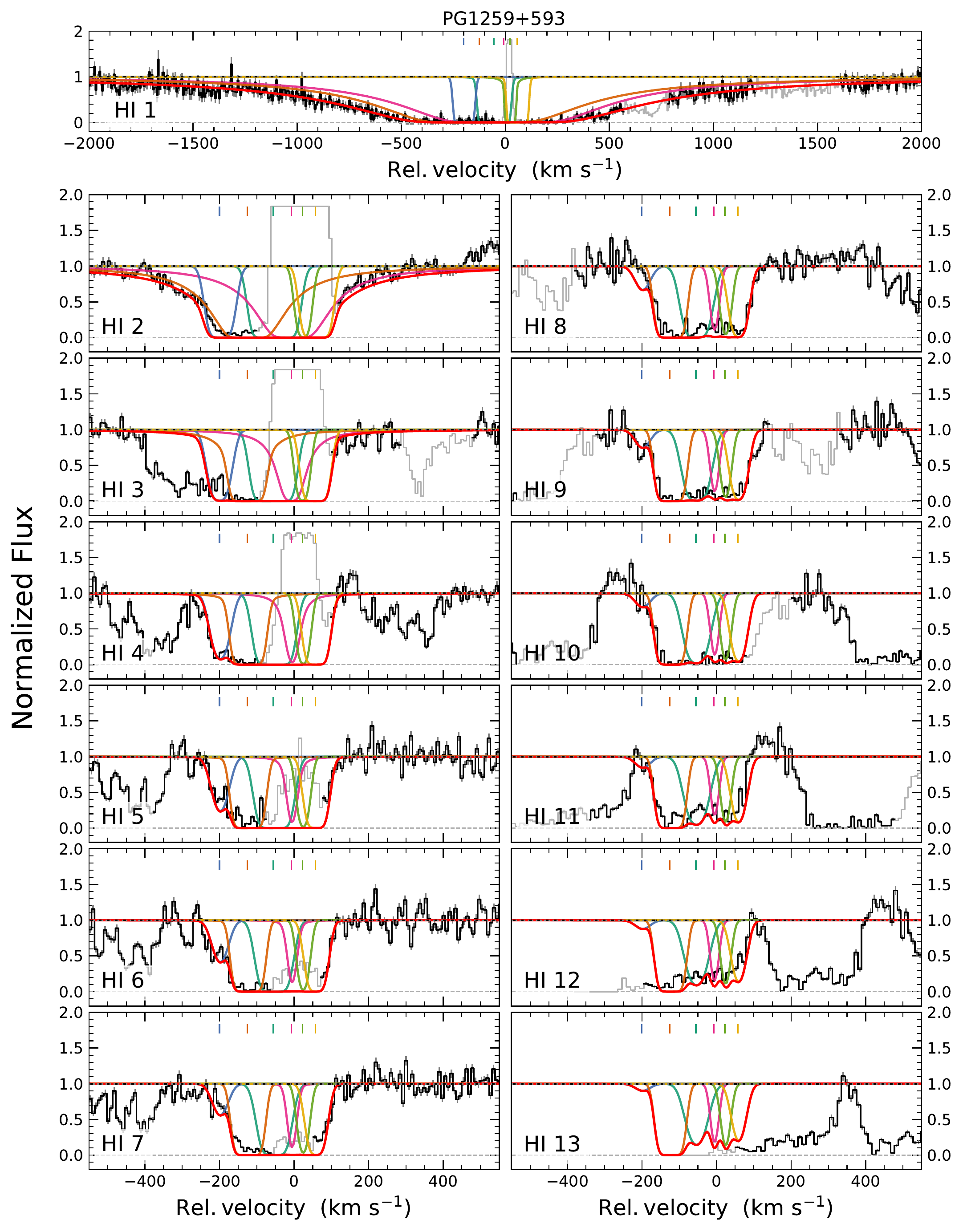}
  \caption{\small{Fits for PG1259+593. The data are shown in black, errors and/or masked regions in light grey, and the composite fit in red. Each contributing component is plotted with a unique color, and the matching tick marks in the top of each panel show the centroid velocity.}}
  \label{fig:PG1259+593}
  \vspace{10pt}
\end{figure*}

\begin{figure*}[ht!]
\centering
 \includegraphics[width=0.9\linewidth]{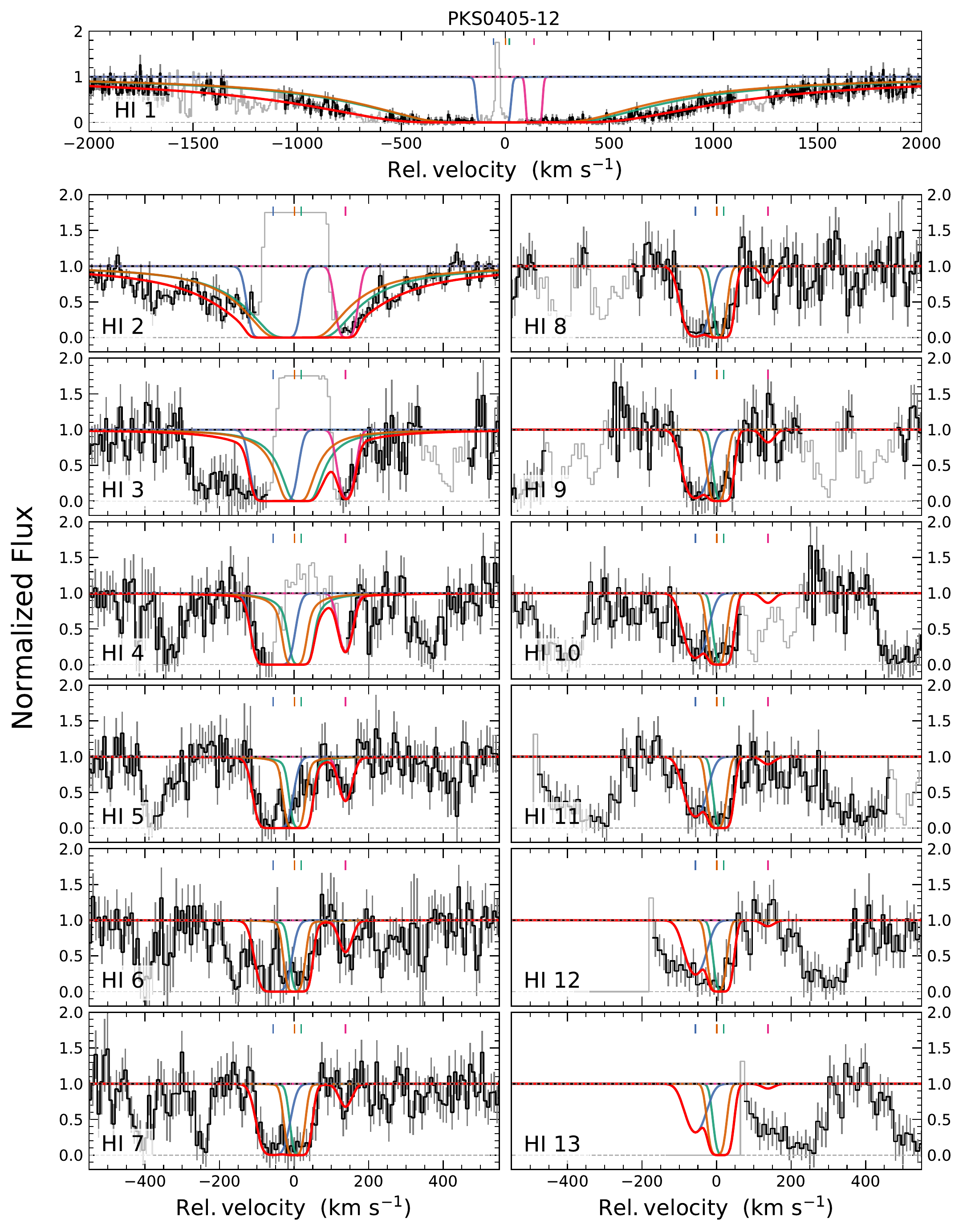}
  \caption{\small{Fits for PKS0405-12. The data are shown in black, errors and/or masked regions in light grey, and the composite fit in red. Each contributing component is plotted with a unique color, and the matching tick marks in the top of each panel show the centroid velocity.}}
  \label{fig:PKS0405-12}
  \vspace{10pt}
\end{figure*}

\begin{figure*}[ht!]
\centering
 \includegraphics[width=0.9\linewidth]{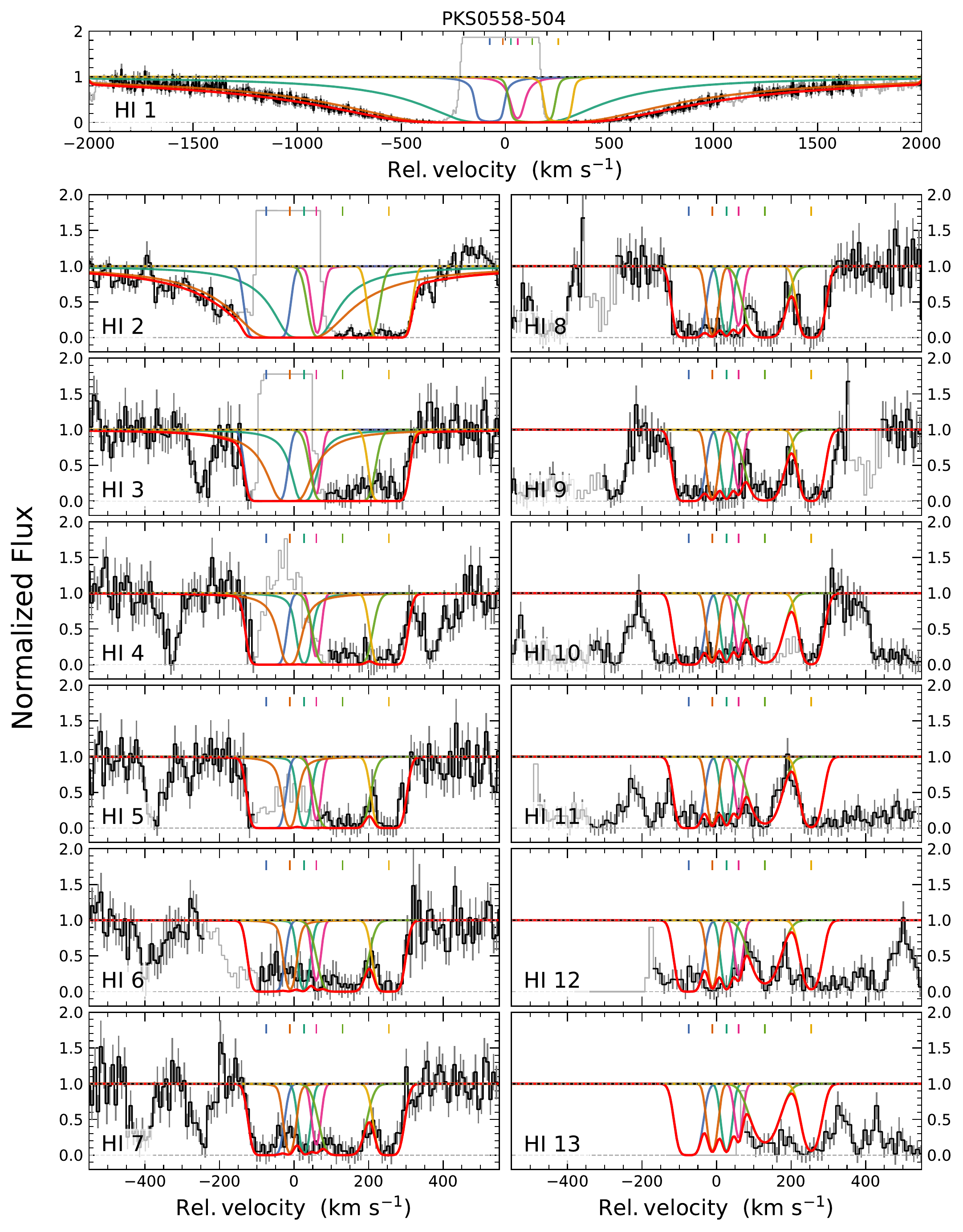}
  \caption{\small{Fits for PKS0558-504. The data are shown in black, errors and/or masked regions in light grey, and the composite fit in red. Each contributing component is plotted with a unique color, and the matching tick marks in the top of each panel show the centroid velocity.}}
  \label{fig:PKS0558-504}
  \vspace{10pt}
\end{figure*}

\begin{figure*}[ht!]
\centering
 \includegraphics[width=0.9\linewidth]{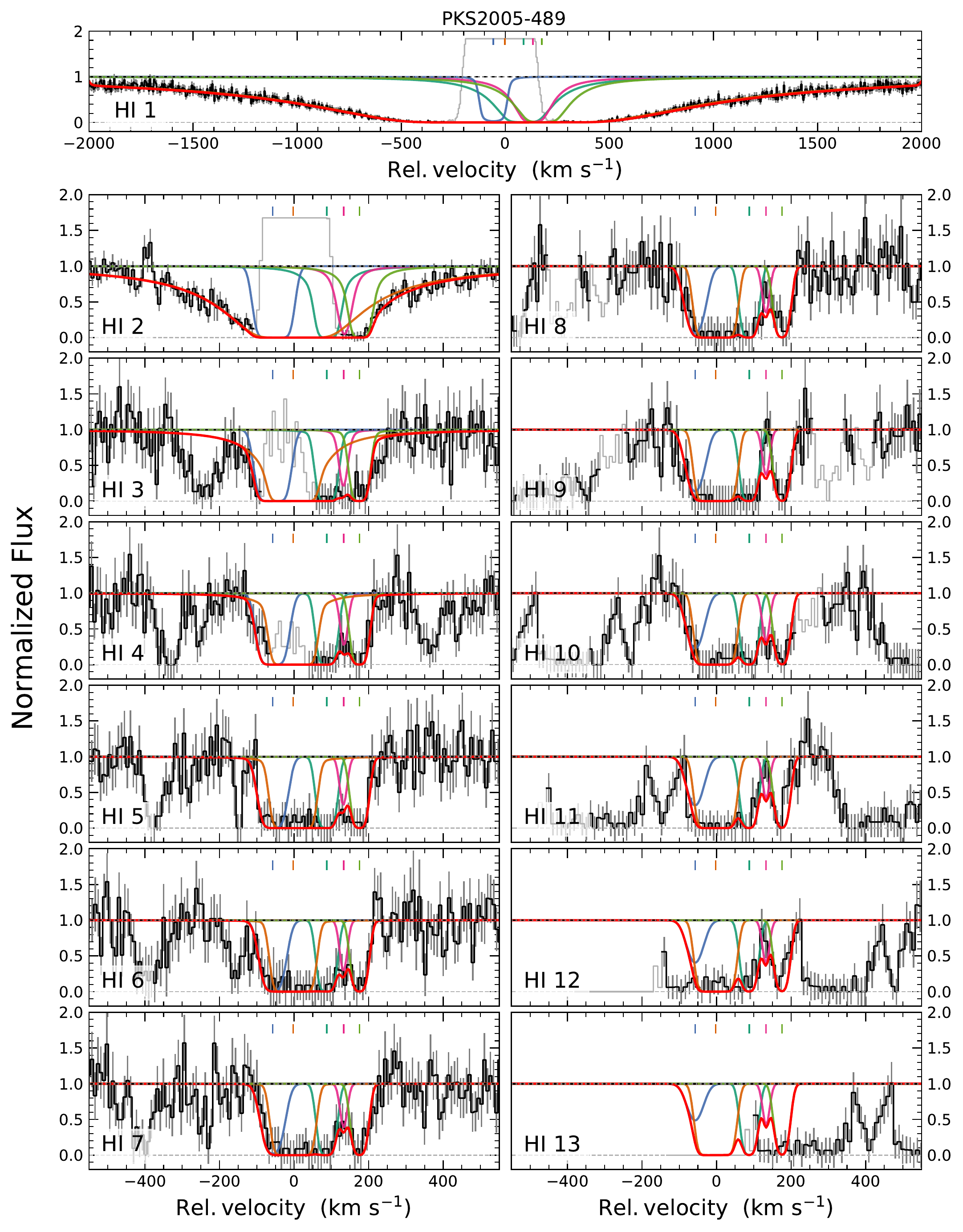}
  \caption{\small{Fits for PKS2005-489. The data are shown in black, errors and/or masked regions in light grey, and the composite fit in red. Each contributing component is plotted with a unique color, and the matching tick marks in the top of each panel show the centroid velocity.}}
  \label{fig:PKS2005-489}
  \vspace{10pt}
\end{figure*}

\begin{figure*}[ht!]
\centering
 \includegraphics[width=0.9\linewidth]{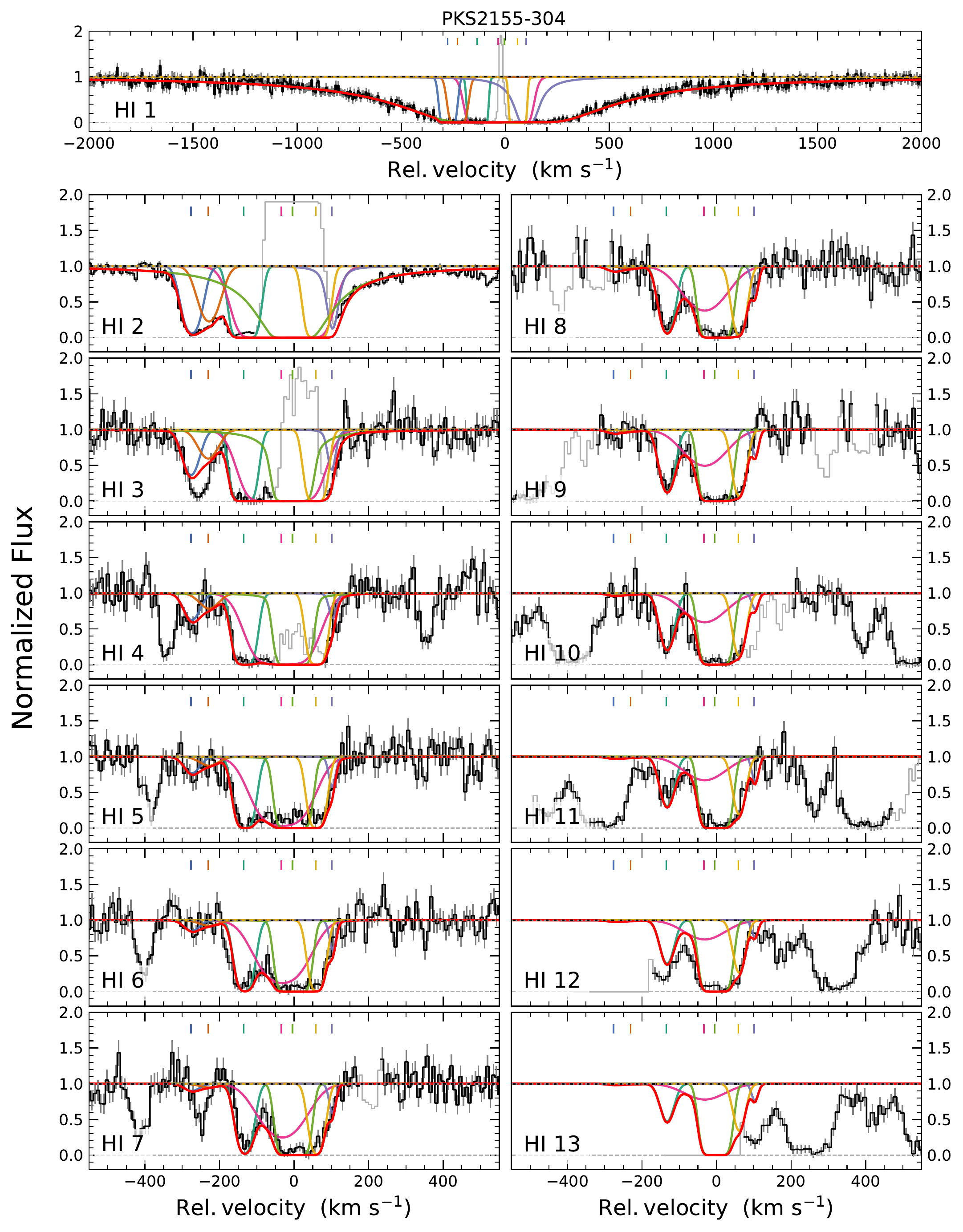}
  \caption{\small{Fits for PKS2155-304. The data are shown in black, errors and/or masked regions in light grey, and the composite fit in red. Each contributing component is plotted with a unique color, and the matching tick marks in the top of each panel show the centroid velocity.}}
  \label{fig:PKS2155-304}
  \vspace{10pt}
\end{figure*}

\begin{figure*}[ht!]
\centering
 \includegraphics[width=0.9\linewidth]{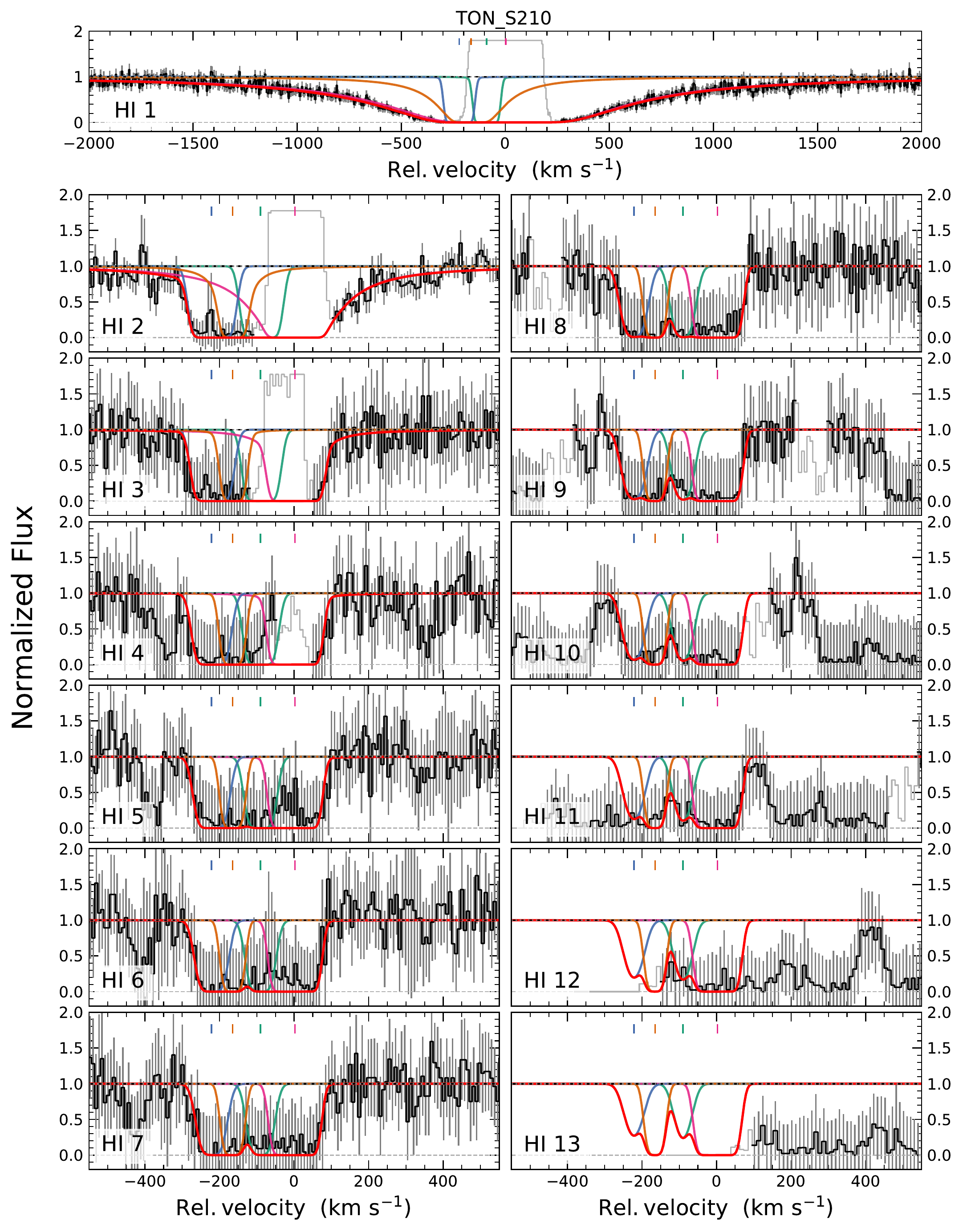}
  \caption{\small{Fits for TON\_S210. The data are shown in black, errors and/or masked regions in light grey, and the composite fit in red. Each contributing component is plotted with a unique color, and the matching tick marks in the top of each panel show the centroid velocity.}}
  \label{fig:TON_S210}
  \vspace{10pt}
\end{figure*}

Figures \ref{fig:1H0707-495N} - \ref{fig:TON_S210} present our best Voigt fits for each target and Lyman series transition. Ly$\alpha$ fits are included for all sightlines with data available (all but 1H0707-495 and NGC1068).

\end{document}